\documentclass[12pt]{article}

\usepackage{graphicx,natbib} 
\usepackage{enumerate}

\usepackage{url} 

\newcommand{\blind}{0}

\usepackage{tikz}
\usetikzlibrary{shapes.geometric, arrows}
\usepackage[doublespacing]{setspace} 
\usepackage{lineno} 
\usepackage{authblk} 
\usepackage{amsmath} 
\usepackage{bm} 
\usepackage{amssymb} 
\usepackage[inline]{enumitem} 
\usepackage{float}
\usepackage[labelsep=period,font=scriptsize]{caption} 
\usepackage[utf8]{inputenc} 
\usepackage[T1]{fontenc} 
\usepackage[title]{appendix}
\usepackage{multirow}
\usepackage{booktabs}
\usepackage{array}
\usepackage{booktabs,subcaption,amsfonts,dcolumn}
\usepackage{soul}
\newcolumntype{M}[1]{>{\centering\arraybackslash}m{#1}}
\usepackage{changes}
\newcommand{\stkout}[1]{\ifmmode\text{\sout{\ensuremath{#1}}}\else\sout{#1}\fi}
\setdeletedmarkup{\stkout{#1}}

\DeclareMathOperator{\Cov}{Cov}
\DeclareMathOperator{\E}{E}

\DeclareMathOperator*{\argmin}{argmin}

\DeclareMathOperator{\diag}{diag}

\tikzstyle{process} = [rectangle, minimum width=3cm, minimum height=1cm, text centered, draw=black ]
\tikzstyle{arrow} = [thick,->,>=stealth]
\begin{document}

\def\spacingset#1{\renewcommand{\baselinestretch}%
{#1}\small\normalsize} \spacingset{1}


\if0\blind
{
  \title{\bf An Adaptive Multivariate Functional Control Chart}

\author[1,2]{Fabio Centofanti\thanks{Corresponding author. e-mail: \texttt{fabio.centofanti@kuleuven.be}}}
\author[2]{Antonio Lepore}
\author[2]{Biagio Palumbo}

\affil[1]{Section of Statistics and Data Science, Department of Mathematics, KU Leuven, Belgium}
\affil[2]{Department of Industrial Engineering, University of Naples Federico II, Naples, Italy}

\setcounter{Maxaffil}{0}
\renewcommand\Affilfont{\itshape\small}
\date{}
\maketitle

} \fi

\if1\blind
{
  \begin{center}
    {\LARGE\bf An Adaptive Multivariate Functional Control Chart}
\end{center}
} \fi

\normalsize
\textbf{This is a preprint version of an article published by Taylor \& Francis in Technometrics available at: \url{https://doi.org/10.1080/00401706.2025.2491369}.}
\begin{abstract}

New data acquisition technologies allow one to gather huge amounts of data that are best represented as \textit{functional data}.
In this setting,  \textit{profile monitoring} assesses the stability over time of both univariate and multivariate functional quality characteristics.
The detection power of profile monitoring methods could heavily depend on parameter selection criteria, which usually do not take into account any information from the out-of-control (OC) state.
This work proposes a new framework, referred to as adaptive multivariate functional control chart (AMFCC), capable of adapting the monitoring of a multivariate functional quality characteristic to the unknown OC distribution, by combining $p$-values of the partial tests corresponding to  Hotelling $T^2$-type statistics calculated at different parameter combinations. 
Through an extensive Monte Carlo simulation study, the performance of AMFCC is compared with methods that have already appeared in the literature.
Finally, a case study is presented in which the proposed framework is used to monitor a resistance spot welding process in the automotive industry. 
AMFCC is implemented in the  \textsf{R} package \textsf{funcharts},  available on CRAN.
%
\end{abstract}
\noindent%
\vfill
\newpage

\spacingset{1.45}

\section{Introduction}
\label{sec_intro}
Nowadays, in many industrial settings,  the development of modern acquisition systems allows the collection of a large amount of data at a high rate, which is usually characterized by great complexity and high dimensionality. This development is the result of the rapid change in technology, industries, and societal patterns in the 21st century due to increasing interconnectivity and smart automation \citep{xu2018industry}.

Particularly relevant is the case where data are well represented as one or more functions defined on multidimensional domains, i.e.,  \textit{functional data} or \textit{profiles}  \citep{ramsay2005functional,ferraty2006nonparametric,horvath2012inference,kokoszka2017introduction}.
In this setting, the \textit{profile monitoring} \citep{noorossana2011statistical} problem refers to the 
 statistical process monitoring (SPM) \citep{montgomery2007introduction,qiu2013introduction}
through one or more quality characteristics of interest that are in the form of functional data, hereinafter referred to as functional quality characteristics.

 The main tool for SPM is the control chart that, through a plot over time of one or more statistics based on the quality characteristic of interest, aims at the process stability assessment. That is,   to assess whether the process is operating in the presence of a normal source of variations  (also known as common causes) only or, otherwise,  assignable
 sources of variations (also known as special causes) are present.
 In the former case, the process is said to be in control (IC), whereas in the latter the process is said to be out of control (OC).
 
 Because of its applicative relevance, profile monitoring has
 been extensively studied recently and several methods have been developed
 for monitoring linear and nonlinear functional quality characteristics based on both parametric and nonparametric models.
 For instance, the methods of \cite{mahmoud2004phase,wang2005using,zou2006control,zou2007monitoring, williams2007statistical} are based on multivariate
 control charts applied to the linear and nonlinear regression coefficients. \cite{qiu2010nonparametric,rajabi2017phase} presented methods
 based on  mixed-effect models  whereas the approaches of  \cite{chicken2009statistical,paynabar2011characterization}  rely on  wavelet decomposition, and that of \cite{qiu2010control} on non parametric regression.
 More recently, \cite{centofanti2021functional} introduced the functional regression control chart framework to take into account functional covariates, and \cite{centofanti2022real} presented a real-time monitoring method that accounts for both the phase and amplitude components.
It should be noted that all of the aforementioned profile monitoring methods refer to univariate functional quality characteristics alone. 

However, in many industrial settings, with measurements acquired by multichannel sensors, the process quality of each item is best characterized by a multivariate functional quality characteristic.  
In these cases, although the univariate methods could be separately applied to each component, a great amount of detection power is lost by neglecting the possible cross-correlation among profiles. 
For this reason, several methods have been recently proposed in the literature to fully exploit the multivariate nature of the quality characteristic. 
\cite{zou2012lasso} proposed a new methodology for monitoring general multivariate linear profiles via a LASSO-based multivariate SPM technique. 
\cite{chou2014monitoring} developed a
 process monitoring strategy for multiple correlated
 nonlinear profiles based on   B-spline coefficients extracted from each profile. Analogously, the method suggested by \cite{grasso2014profile} combines multichannel profiles into a  high-dimensional vector and applies principal component analysis to extract features and construct the monitoring statistic.
More recently, \cite{paynabar2016change} proposed a change-point model for Phase I analysis based on multichannel functional principal component analysis, and \cite{ren2019phase} developed a Phase II monitoring approach for multivariate profiles based on an exponentially weighted
moving average chart. Similar ideas are used in the methods presented by \cite{capezza2020control} and \cite{capezza2022robust}, where the multivariate functional quality characteristic is monitored via the Hotelling $T^2$ and squared prediction error (SPE) control charts applied to the
coefficients obtained from the multivariate functional principal component
decomposition \citep{chiou2014multivariate,happ2017mfpca}.

Each of these approaches is based on the choice of one or more parameters to be implemented. 
Throughout this paper, we use the term \textit{parameter choice or selection}   to indicate any decision needed to implement a method.
For example, multivariate monitoring strategies based on principal component analysis need to select the dimensionality of the space spanned by the retained principal components. 
Smoothing parameter selection is needed to recover multivariate functional data from noisy discrete measurements.
Parameter selection is usually based on criteria that take into account estimation or prediction errors. 
As an example, the number of principal components to be retained is chosen so that the spanned space explains a given proportion of the total variability, and the smoothing parameters are chosen by cross-validation or generalized cross-validation (GCV) \citep{ramsay2005functional}. 
Although these are reasonable criteria, it is widely recognized in the literature that the optimal parameter choice for model estimation is not necessarily optimal for testing \citep{ibragimov1981statistical,ingster1993asymptotically,hart1997classical}.
Many consistent  nonparametric tests, which are
constructed without reference to a particular class of parametric alternatives, are based on smoothing techniques \citep{hart1997classical} and thus, on the choice of the smoothing parameters.
In particular, adaptive tests select smoothing parameters through criteria that optimize testing performance by adapting to the unknown smoothness of the alternative hypothesis \citep{sperlich2014choice}.
 In this setting,  \cite{horowitz2001adaptive} and \cite{guerre2005data} proposed two relevant approaches to select optimal smoothing parameters for the lack-of-fit testing problem.
 In the profile monitoring literature itself, several works have pointed out that the detection power of a monitoring method could heavily depend on parameter choice, and parameter selection criteria optimized for model estimation may achieve unsatisfactory monitoring performance by overlooking any information about the OC condition \citep{zou2009nonparametric,qiu2010control,wang2018thresholded}.
For these reasons, some results presented in the nonparametric testing literature have been applied in the monitoring of univariate functional data
either to select the smoothing parameters for the monitoring of nonparametric profiles \citep{zou2009nonparametric,qiu2010control} or to deal with the choice of other regularization parameters \citep{zou2012lasso}.
However, no attempt has been made in the literature to deal with the issue that optimal parameters for model estimation are in general not optimal for monitoring a multivariate functional quality characteristic.

In the face of these considerations, we present a new framework, referred to as adaptive multivariate functional control chart (AMFCC), for the monitoring and diagnosis of a multivariate functional quality characteristic that is capable of adapting to the unknown distribution of the OC observations.  In particular, the AMFCC is designed for Phase II monitoring, which refers to the prospective monitoring of new observations, given a clean set of observations, which will be hereinafter referred to as IC or reference sample and used to characterize the IC operating conditions of the process (Phase I).
To this end,
 the AMFCC  monitors the multivariate functional quality characteristic by combining Hotelling $T^2$-type statistics corresponding to different parameter combinations. Inspired by the methods based on the functional principal component analysis \citep{ren2019phase,capezza2020control,capezza2022robust} each statistic relies on the coefficients obtained from the multivariate functional principal component decomposition \citep{chiou2014multivariate,happ2017mfpca} of the multivariate functional data. The latter is obtained from noisy discrete values through a data smoothing approach based on a roughness penalty \citep{ramsay2005functional} elaborated for multiple profiles. 
 The underlying idea is to obtain the monitoring statistics by combining $p$-values of the partial tests corresponding to the Hotelling $T^2$-type statistics according to the \textit{nonparametric combination methodology} \citep{salmaso2010permutation}.  In this way, the AMFCC method can adapt to OC conditions that are characterized by different optimal parameters for testing.
Furthermore, a diagnostic procedure based on the contribution plot approach \citep{kourti1996multivariate,kourti2005application} is developed to identify the set of functional variables responsible for the OC condition.

An extensive Monte Carlo simulation study was performed to quantify the monitoring
performance of the AMFCC in the identification of several OC conditions characterized by different mean shifts with respect to competing monitoring schemes that have already appeared in the literature.
 Finally, the practical applicability of the proposed method is
demonstrated through a case study on the SPM of a resistance spot welding (RSW) process in body-in-white automotive manufacturing. 

The paper is structured as follows. Section \ref{sec_method} introduces the AMFCC through the definition of its founding elements. Section \ref{se_perfo} contains the Monte Carlo simulation to assess and compare the performance of the AMFCC with competing methods, while Section \ref{se_casestudy} presents the case study in the automotive industry. 
Section \ref{se_conclusions} concludes the paper.
Supplementary materials for the article are available online.
All computations and plots have been obtained using the programming language \textsf{R} \citep{R2021}.
The  AMFCC is implemented in the  \textsf{R} package \textsf{funcharts},  available on CRAN \citep{capezza2023funcharts}.

\section{An Adaptive Multivariate Functional Control Chart}
\label{sec_method}
The AMFCC is a monitoring and diagnosis methodology that consists of several elements.
In Section \ref{sec_smoo}, a data smoothing approach via a roughness penalty is presented to obtain the multivariate functional data from noise discrete measurements. Section \ref{sec_mfcpa} describes the multivariate functional principal component analysis (MFPCA) that allows reducing the infinite dimensional nature of the functional data estimation problem to a finite number of
features to describe the structure of the data.
These elements are used in Section \ref{sec_mon} to construct and combine Hotelling $T^2$-type statistics.
In Section \ref{sec_dia}, a diagnostic approach is proposed to identify the set of functional variables responsible for the OC condition. 
Implementation details are presented in Section \ref{sec_impl}.

\subsection{Data Smoothing of Multivariate Functional Data}
\label{sec_smoo}
Let $L^2(\mathcal{T})$ be the Hilbert space of square integrable functions defined on the compact set $\mathcal T \in \mathbb{R}$, with the inner product of two functions $f,g \in L^{2}\left(\mathcal{T}\right)$ given by $\langle f,g\rangle=\int_{\mathcal{T}}f\left(t\right)g\left(t\right)dt$, and the norm $\lVert \cdot \rVert=\sqrt{\langle \cdot,\cdot\rangle}$.
Let us denote the multivariate functional quality characteristic by $\bm{X}$ which has $p$ components $\bm{X}=\left(X_1,\dots, X_p\right)^{T}$ and is a random vector with realizations in the Hilbert space $\mathbb{H}$ of $p$-dimensional vectors of $L^2(\mathcal{T})$ functions. The inner product of two function vectors $\mathbf{f}=\left(f_1,\dots,f_p\right)^{T}$ and $\mathbf{g}=\left(g_1,\dots,g_p\right)^{T}$ in $\mathbb{H}$ is given by $\langle \mathbf{f},\mathbf{g} \rangle _{\mathbb{H}}=\sum_{j=1}^{p}\langle f_j,g_j\rangle$ and the norm $\lVert \cdot \rVert_{\mathbb{H}}=\sqrt{\langle \cdot,\cdot\rangle_{\mathbb{H}}}$.
Data are usually collected by acquisition devices at given points, which means that any component of the generic realization $\bm{X}_i=\left(X_{i1},\dots, X_{ip}\right)^{T}$ of the multivariate functional quality characteristic $\bm{X}$ is observed through discrete values $\lbrace Y_{ik}\left(t_{ikj}\right), t_{ikj}\in\mathcal{T}, j=1,\dots,n_{ik} \rbrace_{k=1,\dots,p}$.
If these discrete data are assumed without any measurement error, multivariate functional data can be theoretically drawn up by merely connecting the set of points $ Y_{ik}\left(t_{ikj}\right)$, $j=1,\dots,n_{ik}$ for each $k$.  However, this does not represent the ordinary situation where observational error superimposes on the underlying signal $\bm{X}_i$, that is
\begin{equation}
	\label{eq_1}
	Y_{ik}\left(t_{ikj}\right)=X_{ik}\left(t_{ikj}\right)+\varepsilon_{ikj}, \quad t_{ikj}\in\mathcal{T}, j=1,\dots,n_{ik}, k=1,\dots,p,
\end{equation}
where $\bm{\varepsilon}_{ik}=\left(\varepsilon_{ik1},\dots,\varepsilon_{ikn_{ik}}\right)^{T}$ are zero mean random errors vectors with covariance and cross-covariance matrices $\bm{\Sigma}_{e,k_1k_2}=\Cov(\bm{\varepsilon}_{ik_1},\bm{\varepsilon}_{ik_2})$.
The aim is to recover the underlying signal represented by $\bm{X}_i$ using the information available in the discrete observations.
 To estimate $\bm{X}_i$  the following penalized weighted least-squares is considered
  \begin{equation}
  	\label{eq_smootspline}
  \min_{\bm{f}\in\mathbb{W}}\sum_{k_1=1}^p\sum_{k_2=1}^p\left(\bm{y}_{ik_1}-f_{k_1}\left(\bm{t}_{ik_1}\right)\right)^{T}\bm{W}_{k_1k_2}\left(\bm{y}_{ik_2}-f_{k_2}\left(\bm{t}_{ik_2}\right)\right)+\sum_{k=1}^p\lambda_k\int_{\mathcal{T}}f^{(m)}_k(t)^2dt,
\end{equation}
where $\mathbb{W}=\lbrace \bm{f}=\left(f_1,\dots,f_p\right)^{T}\in \mathbb{H}, \int_{\mathcal{T}}f^{(m)}_k(t)^2dt<\infty  \rbrace$, 
$\bm{y}_{ik}=\left(Y_{ik}(t_{k1}),\dots, Y_{ik}(t_{kn_{ik}})\right)^{T}$, $f_k\left(\bm{t}_{ik}\right)=\left(f_k(t_{ik1}),\dots,f_k(t_{ikn_{ik}})\right)^{T}$, $f^{(m)}$ is the $m$-derivative of $f$, and  $\bm{W}_{k_1k_2}$ are positive definite symmetric matrices that allow for unequal weighting of the squared residuals. The parameters $\lambda_k$, referred to as smoothing parameters control the trade-off between goodness-of-fit and the smoothness of the final estimates. 
To solve the optimization problem in Equation \eqref{eq_smootspline}, each component of $\bm{f}$ is represented as a linear combination of $K_k$ known basis functions $\phi_{k1},\dots,\phi_{kK_k}$ as follows
\begin{equation}
\label{eq_fk}
	f_k\left(t\right)=\sum_{m=1}^{K_k}c_{km}\phi_{km}\left(t\right)\quad  t\in\mathcal{T}, k=1,\dots,p.
\end{equation}
Then, the problem in Equation \eqref{eq_smootspline} reduces to the estimation of the unknown coefficient vectors $\bm{c}_{ik}=\left(c_{ik1},\dots,c_{ikK_k}\right)^{T}$ as 
\begin{multline}
	\label{eq_smootspline2}
\hat{\bm{c}}_{i1},\dots,\hat{\bm{c}}_{ik}=	\argmin_{\bm{c}_1\in\mathbb{R}^{K_1},\dots,\bm{c}_p\in\mathbb{R}^{K_p}}\sum_{k_1=1}^p\sum_{k_2=1}^p\left(\bm{y}_{ik_1}-\bm{c}_{k_1}^T\bm{T}_{k_1}\right)^{T}\bm{W}_{k_1k_2}\left(\bm{y}_{ik_2}-\bm{c}_{k_2}^T\bm{T}_{k_2}\right)\\+\sum_{k=1}^p\lambda_k\bm{c}_{k}^T\bm{R}_{k}^{(m)}\bm{c}_{k}dt,
\end{multline}
where $\bm{T}_{k} $ contains the  values of the $K_k$ basis functions evaluated at $t_{ik1}\dots,t_{ikn_{ik}}$, and   $\bm{R}^{(m)}_k $ is a matrix whose entries $i,j$ are $\langle \phi^{\left(m\right)}_{ki},\phi^{\left(m\right)}_{ki}\rangle$.
Finally, $\bm{X}_i$ is estimated through $\hat{\bm{X}}_i=\left(\hat{X}_{i1},\dots, \hat{X}_{ip}\right)^{T}$ with
\begin{equation}
	\label{eq_4}
	\hat{X}_{ik}\left(t\right)=\hat{\bm{c}}_{ik}^T\bm{\Phi}_k\left(t\right) \quad t\in\mathcal{T},  k=1,\dots,p,
\end{equation}
where  $\bm{\Phi}_k=\left(\phi_{k1},\dots,\phi_{kK_k}\right)^{T}$.

When $m=2$, it could be easily demonstrated that each component of the solution of Equation \eqref{eq_smootspline} is a cubic spline with knots at $t_{ikj}$ \citep{de1978practical,wood2006generalized}. 
For these reasons,  the B-spline basis functions are usually used in case of non-periodic functional data owing to good computational properties and great flexibility \citep{ramsay2005functional}. The penalty on the right-hand side of Equation \eqref{eq_smootspline2}  is computed by setting  $m=2$, i.e.,  by penalizing the function's second derivatives.
The values  $K_k$ in Equation \eqref{eq_fk}  are not crucial \citep{cardot2003spline}  until they are sufficiently large to capture the local behaviour of the multivariate functional data.
If the matrices $\bm{\Sigma}_{e,k_1k_2}$ are known, then a straightforward choice for the matrices $\bm{W}_{k_1k_2}$  would be $\bm{\Sigma}_{e,k_1k_2}^{-1}$ \citep{ramsay2005functional}. However, $\bm{\Sigma}_{e,k_1k_2}$ are rarely known and thus,  estimated from the data. Due to their large dimensions,  $\bm{\Sigma}_{e,k_1k_2}$ are usually assumed to be diagonal with equal diagonal entries.
The smoothing parameters $\lambda_k$ strongly affect the smoothing results and are considered the choice to be made in this step. They are usually chosen based on some estimation criteria \citep{wood2006generalized,ramsay2005functional} but, as highlighted in  Section \ref{sec_intro},  optimal parameters for estimation are not necessarily optimal for monitoring purposes. 
The monitoring scheme of the AMFCC, described in Section \ref{sec_mon}, relies on the combination of partial tests corresponding to each parameter combination. Therefore, if a different smoothing parameter $\lambda_k$ is considered for each variable, then the number of partial tests could become prohibitive when the multivariate functional quality characteristic comprises a large number of components.
For this reason, the smoothing parameters are set   as follows
\begin{equation}\label{eq_wei}
	\lambda_k=\lambda\frac{w_k}{\sum_{i=1}^pw_i}, \quad k=1,\dots,p,
\end{equation}
where $w_k=1/||\hat{f}^{0(m)}_k(t)||^2$, with $\hat{f}^{0(m)}_k$ the initial estimates of the $m$-derivative of $f_k$ that are obtained by solving the optimization problem in Equation \eqref{eq_smootspline2} with $\lambda_1=\dots=\lambda_p=\lambda$. Thus, the unique parameter of the data smoothing is $\lambda$. To make explicit the dependence on $\hat{\bm{X}}_i$, obtained by considering Equation    \eqref{eq_wei}, the latter   will be hereinafter indicated by $\hat{\bm{X}}_{i,\lambda}=\left(\hat{X}_{i1,\lambda},\dots, \hat{X}_{ip,\lambda}\right)^{T}$.

\subsection{Multivariate Functional Principal Component Analysis}
\label{sec_mfcpa}

We assume that $\bm{X}$ has mean $\bm{\mu}=\left(\mu_1,\dots,\mu_p\right)^T$, where $\mu_k(t)=\E(X_k(t))$, $k = 1,\dots, p$, $t\in\mathcal{T}$ and covariance $\bm{G}=\lbrace G_{k_1k_2}\rbrace_{1\leq k_1,k_2 \leq p}$, $G_{k_1k_2}(s,t)=\Cov(X_{k_1}(s),X_{k_2}(t))$, $s,t\in \mathcal{T}$.
In what follows, differences in variability and unit of measurements among $X_1,\dots, X_p$ are taken into account by using the transformation approach of \cite{chiou2014multivariate}. That is, we replace $\bm X$ with the vector of standardized  variables $\bm{Z}=\left(Z_1,\dots,Z_p\right)^T$, where $Z_k(t)=v_k(t)^{-1/2}(X_k(t)-\mu_k(t))$,  with $v_k(t)=G_{kk}(t,t)$, $k=1,\dots,p$, $t\in\mathcal{T}$.   From the multivariate Karhunen-Lo\`{e}ve's Theorem \citep{happ2018multivariate} it follows that
\begin{equation*}
	\bm{Z}(t)=\sum_{l=1}^{\infty} \xi_l\bm{\psi}_l(t),\quad t\in\mathcal{T},
\end{equation*}
where $\xi_l=\langle \bm{\psi}_l, \bm{Z}\rangle_{\mathbb{H}} $ are random variables, which are called \textit{principal component scores} or simply \textit{scores}, such that  $\E\left( \xi_l\right)=0$ and $\E\left(\xi_l \xi_m\right)=\eta_{l}\delta_{lm}$, being $\delta_{lm}$ the Kronecker delta.
The elements of the orthonormal set $\lbrace \bm{\psi}_l\rbrace $, $\bm{\psi}_l=\left(\psi_{l1},\dots,\psi_{lp}\right)^T$, with $\langle \bm{\psi}_l,\bm{\psi}_m\rangle_{\mathbb{H}}=\delta_{lm}$, are referred to as \textit{principal components}, and are the  eigenfunctions  of the covariance $\bm{C}$  of $\bm{Z}$ corresponding to the eigenvalues $\eta_1\geq\eta_2\geq \dots\geq 0$.

To estimate the eigenfunctions and eigenvalues of  $\bm{C}$, let us consider $n$ independent realizations $\bm{X}_1, \dots, \bm{X}_n$ of $\bm{X}$, and the corresponding estimates $\hat{\bm{X}}_{1,\lambda}, \dots, \hat{\bm{X}}_{n,\lambda}$ obtained through the data smoothing approach of Section \ref{sec_smoo}, for a given parameter $\lambda$. The corresponding  vectors of standardized  variables are $\hat{\bm{Z}}_{i,\lambda}=\left(\hat{Z}_{i1,\lambda},\dots,\hat{Z}_{ip,\lambda}\right)^T$, where  $\hat{Z}_{ik,\lambda}(t)=\hat{v}_{k,\lambda}(t)^{-1/2}(\hat{X}_{ik,\lambda}(t)-\hat{\mu}_{k,\lambda}(t))$, $k=1,\dots,p$, with  $\hat{\mu}_{k,\lambda}$ and $\hat{v}_{k,\lambda}$ denoting the sample mean and variance functions estimated from $\hat{\bm{X}}_{i,\lambda}$, $i=1\dots,n$, respectively.
Following the basis function expansion approach of \cite{ramsay2005functional},   we assume that each  component of the eigenfunction $\bm \psi_l$  is represented as linear combination of the $K_k$  basis functions $\phi_{k1},\dots,\phi_{kK_k}$ used to obtain the  estimates $\hat{\bm{X}}_{1,\lambda}, \dots, \hat{\bm{X}}_{n,\lambda}$ (Section \ref{sec_smoo}), that is
\begin{equation}
	\label{eq_appcov}
	\psi_{lk}(t)= \sum_{m=1}^{K_k} b_{lkm}\phi_{km}(t), \quad  t\in\mathcal{T}, k=1,\dots,p, l=1,2,\dots,
\end{equation}
where $\bm{b}_{lk}=\left(b_{lk1},\dots,b_{lkK_k}\right)^T$ are the eigenfunction coefficient vectors. 
With these assumptions,  multivariate functional principal component analysis \citep{ramsay2005functional,chiou2014multivariate} estimates eigenfunctions and eigenvalues of the covariance $\bm{C}$  by performing standard multivariate principal component analysis based on the random vectors $\bm{W}^{1/2}\tilde{\bm{c}}_{i,\lambda}$, where $\tilde{\bm{c}}_{i,\lambda}=\left(\tilde{\bm{c}}_{i1,\lambda}^{T},\dots,\tilde{\bm{c}}_{ip,\lambda}^{T}\right)^T$ is the coefficient vectors corresponding to $\hat{\bm{Z}}_{i,\lambda}$, and $\bm{W}$ is a block-diagonal matrix with diagonal blocks  $\bm{W}_k$, $k=1,\dots,p$, whose entries are $w_{k_1 k_2} = \langle\phi_{k_1},\phi_{k_2}\rangle$, $k_1,k_2=1,\dots, K$. 
Specifically, the estimated eigenvalues $\hat{\eta}_{l,\lambda}$ of $\bm{C}$ are the eigenvalues  of the sample  covariance matrix $\bm{S}$ obtained from $\bm{W}^{1/2}\tilde{\bm{c}}_{i,\lambda}$, whereas, each component of the estimated principal components $\hat{\bm{\psi}}_{l,\lambda}=\left(\hat{\psi}_{l1,\lambda},\dots,\hat{\psi}_{lp,\lambda}\right)^{T}$  is obtained through Equation \eqref{eq_appcov} with $\bm{b}_{lk}=\bm{W}^{-1/2}\bm{u}_{lk,\lambda}$ and $\bm{u}_{l,\lambda}=\left(\bm{u}_{l1,\lambda}^T,\dots,\bm{u}_{lp,\lambda}^T\right)^T$ representing the $l$-th eigenvectors of $\bm{S}$. 
In practice, as the eigenvalues decrease toward 0, it is assumed that  the leading eigenfunctions reflect the most important features of $\bm{X}$, which means that  $\hat{\bm{X}}_{i,\lambda}$ are approximated as $\hat{\bm{X}}_{i,\lambda}^L$ through the following truncated principal component decomposition
\begin{equation}
	\label{eq_appx}
	\hat{\bm{X}}_{i,\lambda}^L(t)= \hat{\bm{\mu}}_{\lambda}(t)+\hat{\bm{D}}_{\lambda}(t)\sum_{l=1}^{L} \hat{\xi}_{il,\lambda}\hat{\bm{\psi}}_{l,\lambda}(t) \quad t\in\mathcal{T}
\end{equation}
where $\hat{\bm{D}}_{\lambda}$ is a diagonal matrix whose diagonal entries are $\hat{v}_{k,\lambda}^{1/2}$, $\hat{\bm{\mu}}_{\lambda}=\left(\hat{\mu}_{k,\lambda},\dots,\hat{\mu}_{k,\lambda}\right)^T$,  and $\hat{\xi}_{il,\lambda}= \langle \hat{\bm{\psi}}_{l,\lambda}, \hat{\bm{Z}}_{i,\lambda}\rangle_{\mathbb{H}}$.
In the profile monitoring literature, the parameter $L$ is generally chosen such that the retained principal components explain at least a given percentage of the total variability, which is usually in the range 70-90$\%$ \citep{paynabar2016change,ren2019phase,centofanti2022real,capezza2022robust}, but more sophisticated methods are used as well \citep{capezza2020control}.
However, as already stated these approaches are based on criteria that are not optimized for monitoring purposes. For this reason, the AMFCC consider $L$ as the parameter of this step.
\subsection{The Monitoring Scheme}
\label{sec_mon}
Let $\bm{X}_i=\left(X_{i1},\dots,X_{ip}\right)^T$, $i=1,2,\dots$, be  the on-line
observations of the multivariate functional quality characteristic $\bm{X}$. The AMFCC aims to detect changes that may have occurred in the  mean function $\bm{\mu}_i$ of   $\bm{X}_i$ with respect to  the IC process mean  $\bm{\mu}$ by sequentially testing the hypothesis 
\begin{equation}\label{eq_hypo}
	H_0: \bm{\mu}_i=\bm{\mu}\quad  versus \quad H_1:\bm{\mu}_i\neq\bm{\mu}.
\end{equation}
The proposed method relies on the following  Hotelling $T^2$-type statistics
\begin{multline}\label{eq_monts}
	T^2_{i;\lambda, L}=\sum_{k_1=1}^p\sum_{k_2=1}^p\int_{\mathcal{T}}\int_{\mathcal{T}}\hat{v}_{k_1,\lambda}(s)^{-1/2}(\hat{X}_{ik_1,\lambda}(s)-\hat{\mu}_{k_1,\lambda}(s))\hat{K}_{k_1k_2;\lambda, L}^*(s,t)\\\hat{v}_{k_2,\lambda}(t)^{-1/2}(\hat{X}_{ik_2,\lambda}(t)-\hat{\mu}_{k_1,\lambda}(t))dsdt,
\end{multline}
with 
\begin{equation}\label{eq_t21}
	\hat{K}_{k_1k_2;\lambda, L}^*(s,t)=\sum_{l=1}^{L}\frac{1}{\hat{\eta}_{l,\lambda}}\hat{\psi}_{lk_1,\lambda}(s)\hat{\psi}_{lk_2,\lambda}(t), \quad s,t\in \mathcal{T},
\end{equation}
where $\hat{v}_{k,\lambda}$, $\hat{\mu}_{k,\lambda}$, $\hat{\eta}_{l,\lambda}$ $\hat{\psi}_{lk,\lambda}$, $k=1,\dots,p$, $l=1,\dots,L$, are estimated as described in Section \ref{sec_mfcpa} using the   observations in the IC sample.
Note that, Equation \eqref{eq_t21} is the functional extension of the likelihood ratio test statistic for the mean of multivariate normal data \citep{johnson1992applied}.

In standard profile monitoring methods the monitoring statistic in Equation \eqref{eq_monts}  would have been obtained by setting $\lambda$ and $L$  to some optimal values identified via estimation-based criteria, whereas the AMFCC combines the values of 	$T^2_{i;\lambda, L} $ at different  $\lambda$ and $L$ into a single monitoring statistic that optimizes power in testing the hypothesis in Equation \eqref{eq_hypo}.
The combination method of the Hotelling $T^2$-type statistics is inspired by the nonparametric combination methodology of \cite{salmaso2010permutation}. 
The main idea is that  $H_0$ may be decomposed down into a set of null sub-hypothesis $H_{0t}$, $t=1,\dots,T$, such that $H_0$ is true if all the $H_{0t}$ are simultaneously true.
Let rewrite	$T^2_{i;\lambda, L} $ as 
\begin{equation}\label{eq_stat2}
	T^2_{i;\lambda, L}=\hat{\bm{\xi}}_{i;\lambda, L}^{T}\hat{\bm{H}}_{\lambda, L}\hat{\bm{\xi}}_{i;\lambda, L}
\end{equation}
where $\hat{\bm{\xi}}_{i;\lambda, L}=\left(\hat{\xi}_{i1,\lambda},\dots,\hat{\xi}_{iL,\lambda}\right)^{T}$, and $\hat{\bm{H}}_{\lambda,L}=\diag\left(\hat{\eta}_{1,\lambda},\dots,\hat{\eta}_{L,\lambda}\right)$.
Then, the  null $H_{0t}$ and the alternative  $H_{1t}$ sub-hypothesis  are
\begin{equation}\label{key}
	H_{0t}: \bm{\xi}_{i;\lambda, L}=\bm{0} \quad H_{1t}:\bm{\xi}_{i;\lambda, L}\neq\bm{0}, \quad t=1,\dots,T,
\end{equation}
with $\bm{\xi}_{i;\lambda, L}=\left(\xi_{i1,\lambda},\dots,\xi_{iL,\lambda}\right)^{T}$, $\xi_{il,\lambda}= \langle \bm{\psi}_{l}, \hat{\bm{Z}}_{i,\lambda}\rangle_{\mathbb{H}}$, and $T$ is the number of parameter combinations that are considered. Note that, although, in principle, the parameter $\lambda$ could take infinite values in $\mathbb{R}^{+}$, in the following we consider only a finite number of smoothing parameters. Moreover, the partial tests $T^2_{i;\lambda, L}$ corresponding to each null sub-hypothesis $H_{0t}$ are indicated as $T^2_{i,t} $.

Following the nonparametric combination methodology, the partial tests are combined  in an overall monitoring statistic $T_i^2$ by considering their associated $p$-values $p_1,\dots,p_{T}$, defined as
\begin{equation*}
	p_t(x)=\Pr(T^2_{i,t}\geq x|H_0),  \quad x\overset{d}{=}T^2_{i,t}, t=1,\dots,T.
\end{equation*} 
This ensures the combination takes place on a common scale. 
Then, the overall monitoring statistic $T_i^2$ is obtained through  a combining function $\Theta$ of $p_1,\dots,p_{T}$ via, that is
\begin{equation}\label{eq_theta}
	T_i^2=\Theta(p_1,\dots,p_{T}),
\end{equation} 
where $\Theta:(0,1)^{T}\rightarrow \mathbb{R}$ is a continuous, non-increasing, univariate, measurable, and non-degenerate function that satisfies several reasonable properties (see  \cite{salmaso2010permutation} for details).
Then, the AMFCC triggers a signal if $T_i^2>C$, where $C>0$ is the chart control limit and is set as the quantile $(1-\alpha)$  of the IC distribution of  $T_i$, with $\alpha$ denoting the overall type I error probability \citep{lehmann2006testing}. Note that the suggested procedure avoids the multiple comparison issue because $C$ is chosen to ensure a given Type I error probability based on the IC distribution.

The rationale of the proposed procedure is based on the ability of the overall statistics to blend the information from partial tests corresponding to a wide range of parameter combinations. 
The proposed procedure is then expected to achieve a larger detection power with respect to procedures that use a fixed parameter combination optimized for model estimation. 

Several choices for the combining function  $\Theta$ are available in the literature  \citep{salmaso2010permutation,loughin2004systematic}.
The most popular one    is the \textit{Fisher omnibus} \citep{fisher1992statistical} that  combines the $p$-values $p_1,\dots,p_{T}$ as 
\begin{equation}\label{key}
	T^2_{i,F}=-2\frac{\sum_{t=1}^{T}\log(p_t)}{T},
\end{equation}
 which is derived from the so-called multiplicative rule and enjoys asymptomatic optimality \citep{littell1971asymptotic,littell1973asymptotic}.
 Another well-known choice is the \textit{Tippett} \citep{tippett1931methods} combining function that produces the following  statistic 
 \begin{equation}\label{key}
 	T^2_{i,T}=-2\log \left(\min_{1\le t \le T}p_t\right).
 \end{equation}
Note that an optimal choice of the combining function depends on the data and the true alternative hypothesis and no single combining function proves to be the best under all circumstances. However, the Fisher omnibus and Tippett combining functions have shown adequate detection power in several contexts \citep{loughin2004systematic} and thus, will be used to implement particular instances of the AMFCC. Their favorable performance is further confirmed by the Monte Carlo simulation study of Section \ref{se_perfo} and the case study presented in Section \ref{se_casestudy}.

Since the IC distributions of $T^2_{i,t}$ are rarely known, $p_1,\dots,p_{T}$ cannot be usually obtained in a straight way. To this aim, we propose  the use of the following estimators \citep{salmaso2010permutation,edgington2007randomization}
\begin{equation}\label{eq_papp}
	\hat{p}_t(x)=\frac{1+\sum_{i=1}^{n_{IC}}I\left(T^2_{i,t}\geq  x\right)}{n_{IC}+1},\quad t=1,\dots,T,
\end{equation}
where $T^2_{1,t},\dots, T^2_{n_{IC},t}$,  are the partial test values calculated at each  IC observation. Implementation details are provided in Section \ref{sec_impl}.

\subsection{Post-Signal Diagnostics}
\label{sec_dia}
After detecting a change in the functional mean of the multivariate functional quality characteristic, it is often critical to identify the set of components that are responsible for the OC condition, i.e., whose functional means have changed significantly from the IC counterparts.
The proposed method addresses this problem through a diagnostic procedure based on a contribution plot approach that explicitly takes into account the possibility that optimal parameters for estimation purposes are generally not optimal for hypothesis testing.

Then, upon observing that   Equation \eqref{eq_stat2} can be rewritten as
\begin{equation}\label{key}
	T^2_{i;\lambda, L}=\sum_{l=1}^{L}\frac{\hat{\xi}_{il,\lambda}}{\hat{\eta}_{l,\lambda}}\hat{\xi}_{il,\lambda}=\sum_{k=1}^{p}\sum_{l=1}^{L}\frac{\hat{\xi}_{il,\lambda}}{\hat{\eta}_{l,\lambda}}\langle \hat{\psi}_{lp,\lambda},\hat{Z}_{i1,\lambda}\rangle,
\end{equation}
the contributions  to the statistics 	$T^2_{i;\lambda, L} $ can be defined as
\begin{equation}\label{key}
	c_{ik;\lambda, L}^{T^2}=\sum_{l=1}^{L}\frac{\hat{\xi}_{il,\lambda}}{\hat{\eta}_{l,\lambda}}\langle \hat{\psi}_{lp,\lambda},\hat{Z}_{i1,\lambda}\rangle,\quad k=1,\dots,p.
\end{equation}
Similarly to Section \ref{sec_mon}, we propose a diagnostic procedure to combine contributions $c_{ik;\lambda, L}^{T^2}$ at different values of the parameters $\lambda$ and $L$, into a single monitoring  statistic for each component of the multivariate functional quality characteristic.
To this aim, we define, the overall contribution to the monitoring statistic $T^2_i$ for each component $k$ as
\begin{equation}\label{key}
	c_{ik}^{T^2}=\Theta_{c}(p^c_1,\dots,p^c_{T}),\quad k=1,\dots,p,
\end{equation}
where $\Theta_{c}$ is a combining function applied to the $p$-values $p^c_1,\dots,p^c_{T}$ are the  $p$-values associated to the $T$ partial tests corresponding to the contributions $c_{ik,\lambda L}^{T^2}$ for each considered combination of parameters $\lambda$ and $L$.
Following the idea of \cite{westerhuis2000generalized},    control limits for the
overall contributions are introduced to support the identification of the anomalous components. Specifically, the $k$-th variable is labelled as anomalous when $c_{ik}^{T^2}>C_k$, with $C_k>0$ representing the $(1-\alpha_k)$ quantile of the IC distribution of  $c_{ik}^{T^2}$ and $\alpha_k$ being the overall type I error probability corresponding to $X_{ik}$, $k=1,\dots,p$, $i=1,2,\dots$.

As for $\Theta$ in \eqref{eq_theta}, the Fisher omnibus or the Tippett combining functions are reasonable choices for $\Theta_{c}$, and the $p$-values $p^c_1,\dots,p^c_{T}$ could be estimated by using the IC observations through  Equation \eqref{eq_papp}.
\subsection{Implementation Details}
\label{sec_impl}
 A reference or Phase I sample of $n_{IC}$ discrete values realizations \sloppy $\lbrace\bm{y}_{1,k}\rbrace_{k=1,\dots,p},\dots,\lbrace\bm{y}_{n_{IC},k}\rbrace_{k=1,\dots,p}$ is assumed available to characterize the IC operating conditions of the process.
At a given value of the smoothing parameter $\lambda$, the multivariate functional observations, denoted by $\hat{\bm{X}}_{1,\lambda}, \dots, \hat{\bm{X}}_{n_{IC},\lambda}$, are obtained from the reference sample through the data smoothing method described in Section \ref{sec_smoo}. Then, $\hat{\bm{X}}_{1,\lambda}, \dots, \hat{\bm{X}}_{n_{IC},\lambda}$ are used to estimate the MFPCA model as described in Section \ref{sec_mfcpa},  namely the principal components $\lbrace \bm{\psi}_l\rbrace $, the eigenvalues $\lbrace \eta_l\rbrace $, the mean $\mu_k$ and the variance $v_k$  functions, $k=1,\dots,p$.
Thereafter, the values of the statistics $T_{i,\lambda L}^2$ are obtained for different combinations of $\lambda$ and $L$.
These combinations should be reasonably set in a range where the AMFCC monitoring statistic can best adapt to the unknown distribution of the possible OC state of the process.
To this end, we suggest considering a finite number of  $\lambda$ values in the interval $\left[\lambda_{min},\lambda_{max}\right]$, where $\lambda_{min}$ and  $\lambda_{max}$  allow a very wide range of smoothness level,  combined with the values   $L$  in the set $\lbrace L_{\delta_{i}}\rbrace$, where $L_{\delta_{i}}$  the number of the retained principal components that explain at least a given percentage $\delta_{i}\%$ of the total variability, with $\delta_{i}$  in the interval $\left[\delta_{min},\delta_{max}\right]$. Reasonable values of $\delta_{min}$ and $\delta_{max}$ are 0.4 and 0.99, respectively.
 The monitoring statistics $T_{i}^2$ are obtained by combining the $p$-values corresponding to $T_{i,\lambda L}^2$, which are estimated as in Equation \eqref{eq_papp}, and are used to calculate the control limit $C$ for a given type I error probability  $\alpha$. 
To reduce possible overfitting issues \citep{ramaker2004effect,kruger2012statistical} and increase the monitoring performance of the AMFCC, the reference sample is customarily partitioned into training and tuning sets, to be separately used for the MFPCA model and control limit estimation, respectively.
Analogous steps follow to estimate the control limits  $C_k$ for the overall contribution $c_{ik}^{T^2}$.

The AMFCC is designed to be used in Phase II,  the current observation  $\bm{X}_{new}$ is projected onto the  MFPCA model, estimated on the training sample,  to compute the monitoring statistic $T_{new}^2$  by combining the $p$-values corresponding to $T_{new,\lambda L}^2$, which are estimated as in Equation \eqref{eq_papp}.
An alarm signal is issued if  $ T_{new}^2$ violates the control limit $C$.
In that case, the components whose overall contribution $c_{newk}^{T^2}$ violates the relative control limit $C_k$ are deemed responsible for the OC condition.
Note that $ T_{new}^2>C$ dos not necessarily imply that $\exists \ k=1,\dots,p$ such that $c_{new   k}^{T^2}>C_k$.

\section{Simulation Study}
\label{se_perfo}
An extensive Monte Carlo simulation study is performed to assess the performance of the AMFCC in identifying mean shifts of the multivariate functional quality characteristic. 
The data generation process is inspired by the work of \cite{centofanti2021functional} and is detailed in  Supplementary Materials A.
Two scenarios are considered where data are generated with different covariance structures with $p=5$. Specifically,  in Scenario 1 the covariance structure is based on the \textit{Bessel} correlation function of
the first kind, whereas in Scenario 2 it is based on the  \textit{Gaussian} correlation function \citep{abramowitz1964handbook}. As results in both scenarios are very similar, in this section, we present only Scenario 1,
whereas results for Scenario 2 are reported in Supplementary Materials B.
Each scenario explores three models with decreasing dependence levels within and between the components of the multivariate functional characteristic, which are denoted by D1, D2, and D3.
To assess the performance of the AMFCC over a wide variety of OC conditions, four different types of mean shifts are considered and named Shift A, Shift B, Shift C, and Shift D. Shift A is characterized by a mean function increase in the central part of the domain, in Shift B the mean function linearly decreases in the second half of the domain, in Shift C the mean function shift is characterized by a sinusoidal behaviour, whereas the mean function quadratically changes in Shift D.
For each shift, four different severity levels $ d\in\lbrace 1,2,3,4\rbrace $ are explored.

The AMFCC is compared with several competing approaches.
The first method considered is the multivariate functional control chart (MFCC) presented in \cite{capezza2020control,capezza2022funcharts}, where the multivariate functional data are monitored via the Hotelling $T^2$ and squared prediction error (SPE) control charts applied to the
coefficients obtained from the MFPCA. Specifically, we consider three instances of this method are considered, which are referred to as MFCC\textsubscript{07}, MFCC\textsubscript{08}, and  MFCC\textsubscript{09}  by varying the dimensionality of the space spanned by the retained principal components to explain 70\%, 80\%, and 90\% of the total variability, respectively. These values have been selected in the typical range considered in the literature \citep{jolliffe2011principal}. 
The second and third competing methods do not consider the functional nature of the data and are based on the classical Hotelling’s $T^2$ control chart built on either the $p$-dimensional    vector of means obtained by averaging each component profile or directly from the discrete values realizations of the multivariate functional quality characteristic without the data smoothing and MFPCA. These are, respectively, referred to as multivariate control chart (MCC) and discrete control chart (DCC).
The AMFCC is implemented as outlined in Section \ref{sec_impl} based on the Fisher omnibus and  Tippett combining functions, referred to as AMFCC\textsubscript{F} and  AMFCC\textsubscript{T}, respectively, and the parameter $\lambda$ and $L$ respectively selected over an equally spaced grid on the log-scale between $\lambda_{min}=10^{-6}$ and $\lambda_{max}=10^2$, and a grid of $\delta_i$ in the interval  $\left[0.4,0.99\right]$. The number of basis functions $K_k$ is set for each component equal to 20. 

For each scenario, dependence level, OC condition, and severity level, 50 simulation runs are performed. 
Each run considers a Phase I random sample of 2000  observations that are equally split into training and tuning sets, and  500 randomly generated OC observations.
The AMFCC and the competing method performance are assessed by means of the true detection rate (TDR) and the false alarm rate (FAR), which are defined as the proportion of points that fall outside the control limits whilst the process is, respectively, OC or IC.
The FAR should be as similar as possible to the overall type I error probability $\alpha$ considered to obtain the control limits, which is set equal to 0.05, whereas the TDR should be as close to one as possible.

\begin{figure}[h]
	\caption{Mean FAR ($ d=0 $) or TDR ($ d \neq 0 $) achieved by AMFCC\textsubscript{F},  AMFCC\textsubscript{T}, MFCC\textsubscript{09}, MFCC\textsubscript{08},  MFCC\textsubscript{07}, MCC, and DCC for each dependence level (D1, D2 and D3), OC condition (Shift A, Shift B, Shift C, and Shift D) as a function of the severity level $d$  in Scenario 1.}
	
	\label{fi_results_1}
	
	\centering
	\hspace{-2.1cm}
	\begin{tabular}{cM{0.24\textwidth}M{0.24\textwidth}M{0.24\textwidth}M{0.24\textwidth}}
			\textbf{\footnotesize{D1}}&\includegraphics[width=.25\textwidth]{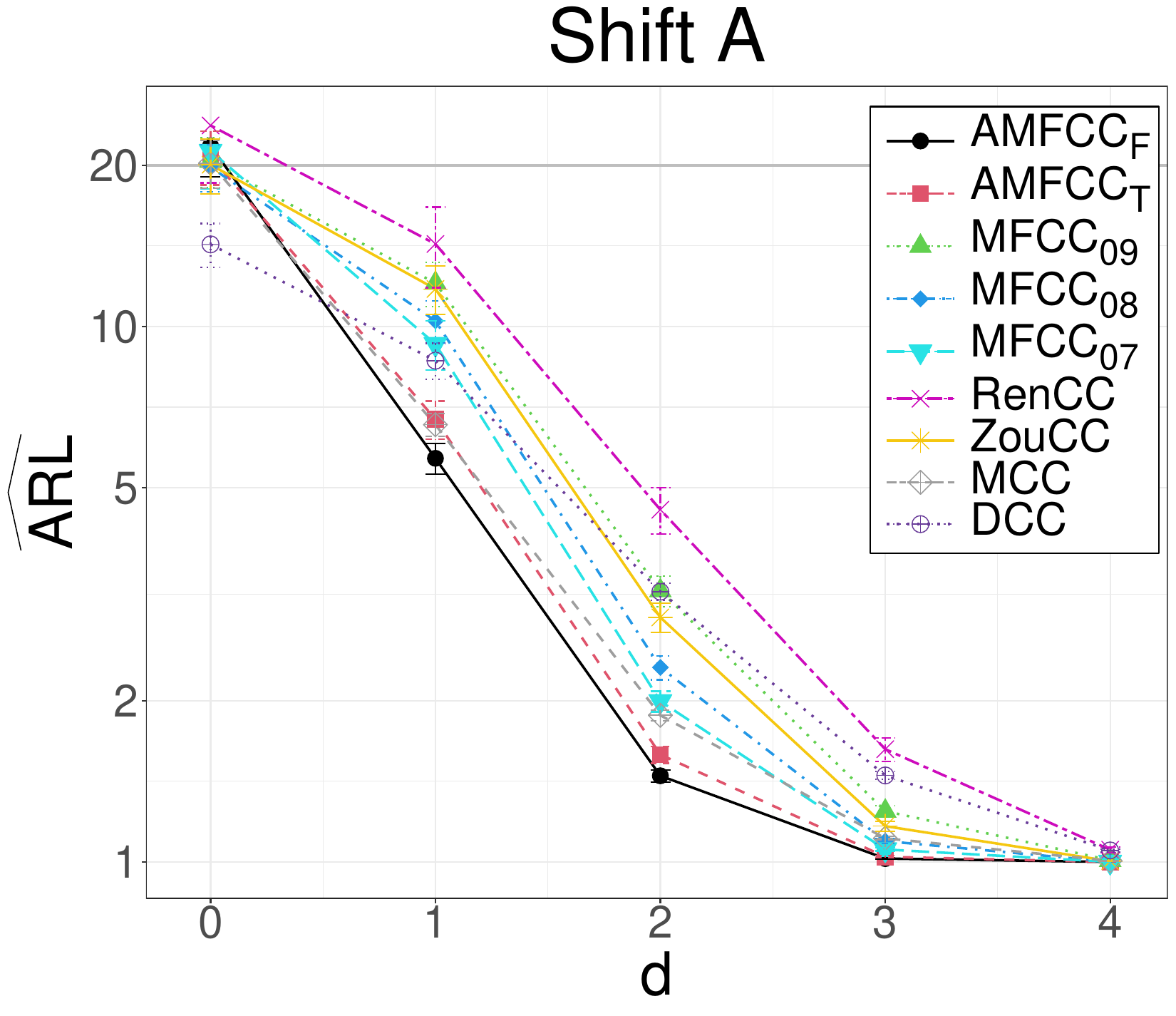}&\includegraphics[width=.25\textwidth]{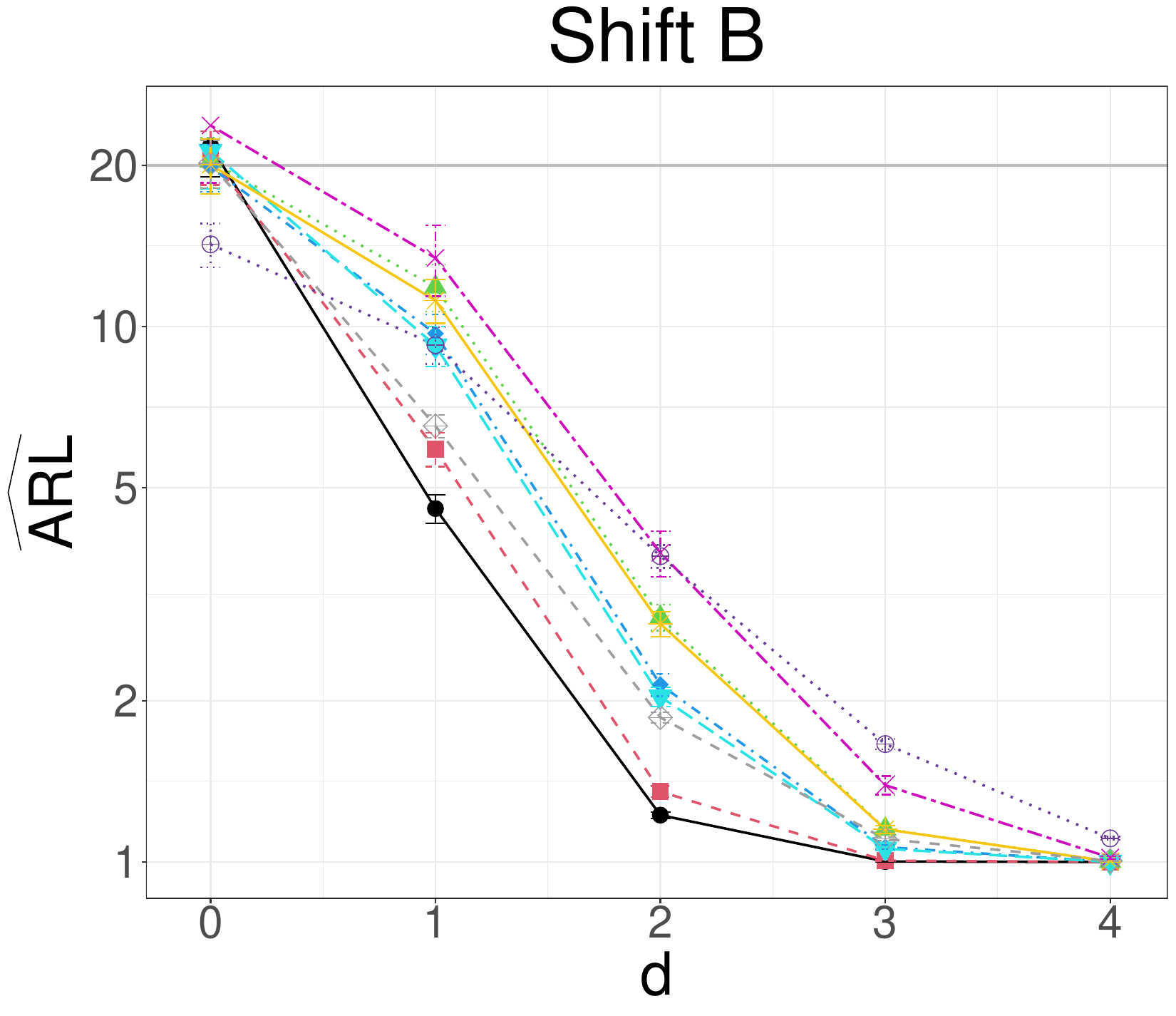}&\includegraphics[width=.25\textwidth]{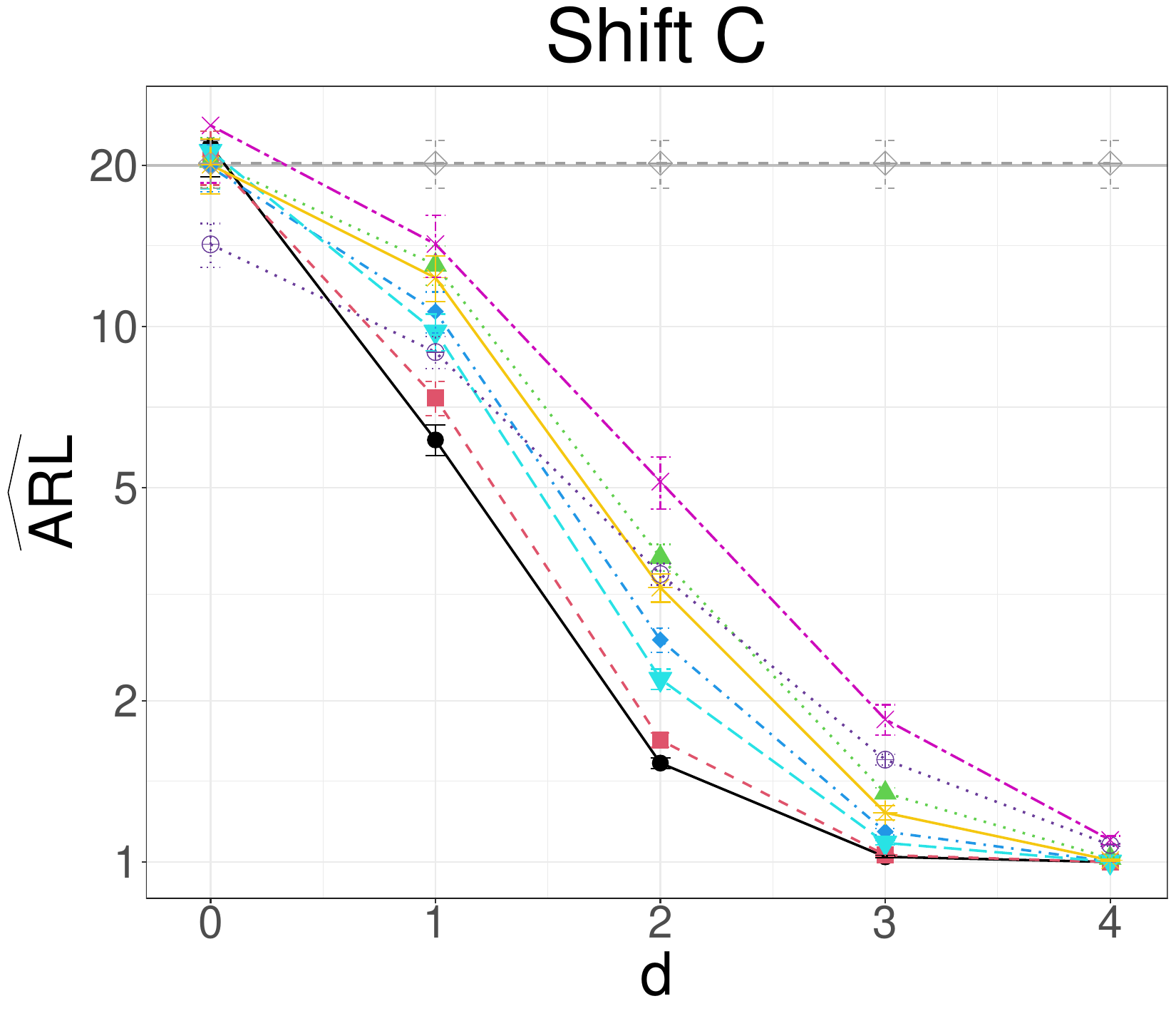}&\includegraphics[width=.25\textwidth]{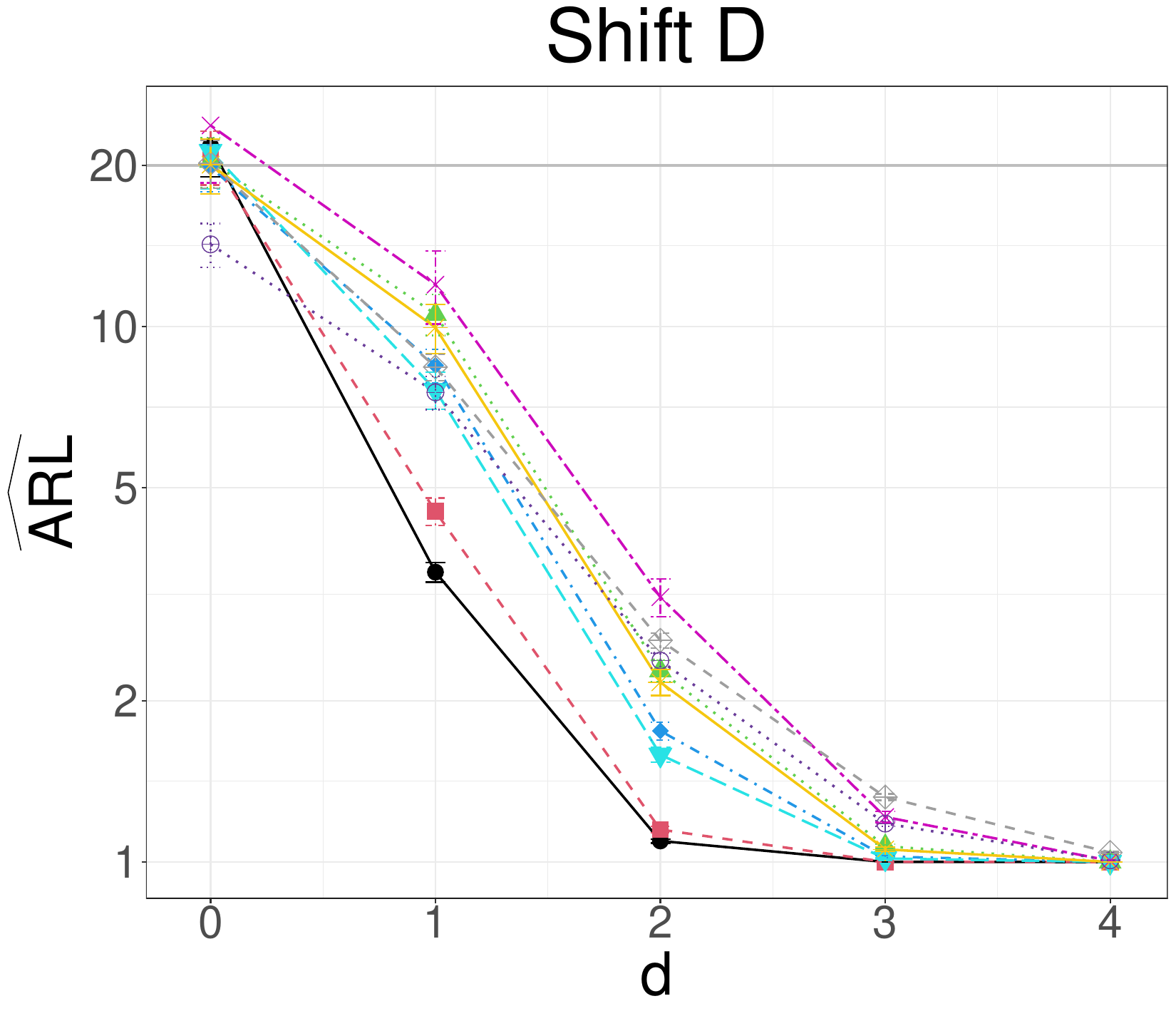}\\
		\textbf{\footnotesize{D2}}&\includegraphics[width=.25\textwidth]{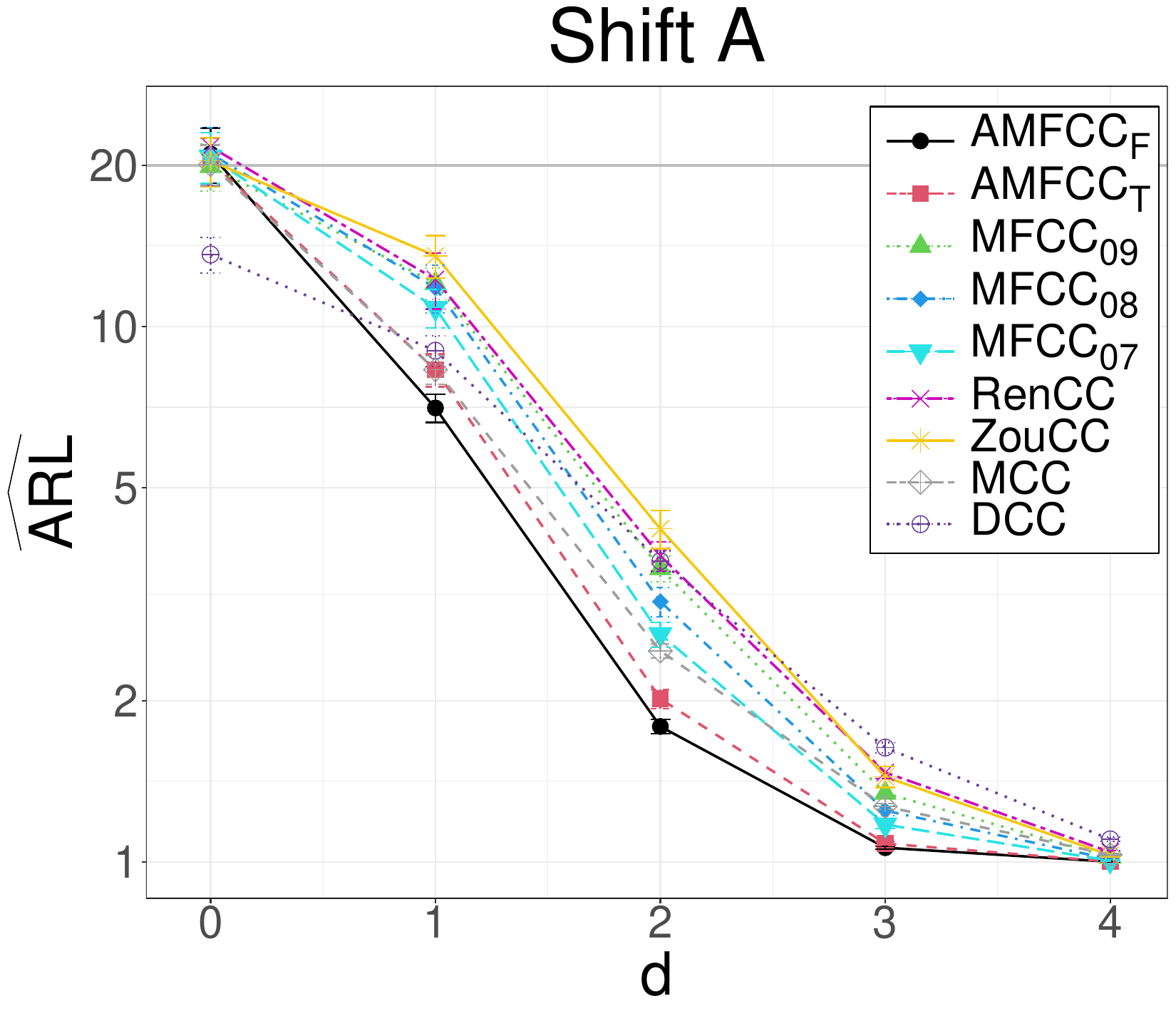}&\includegraphics[width=.25\textwidth]{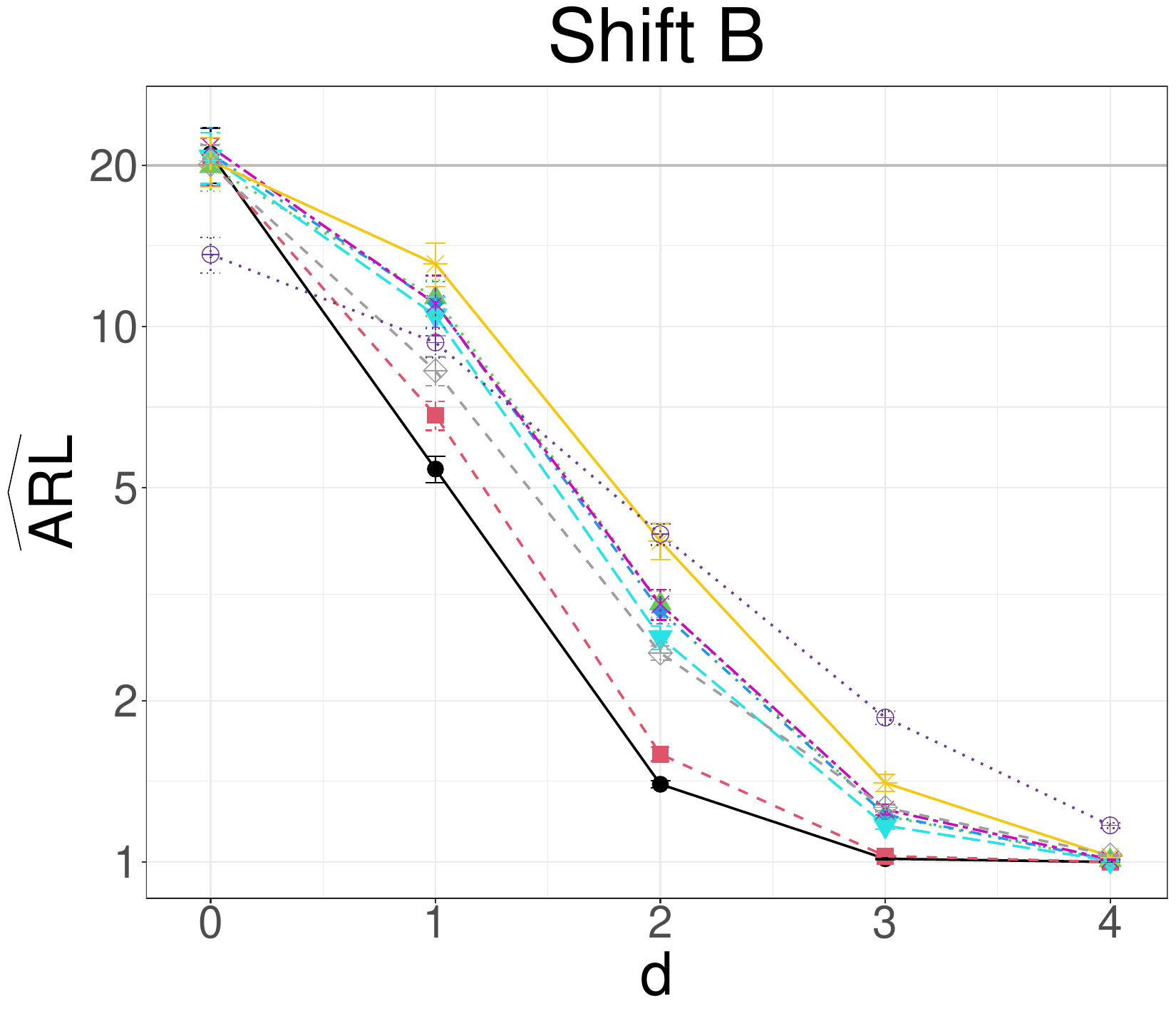}&\includegraphics[width=.25\textwidth]{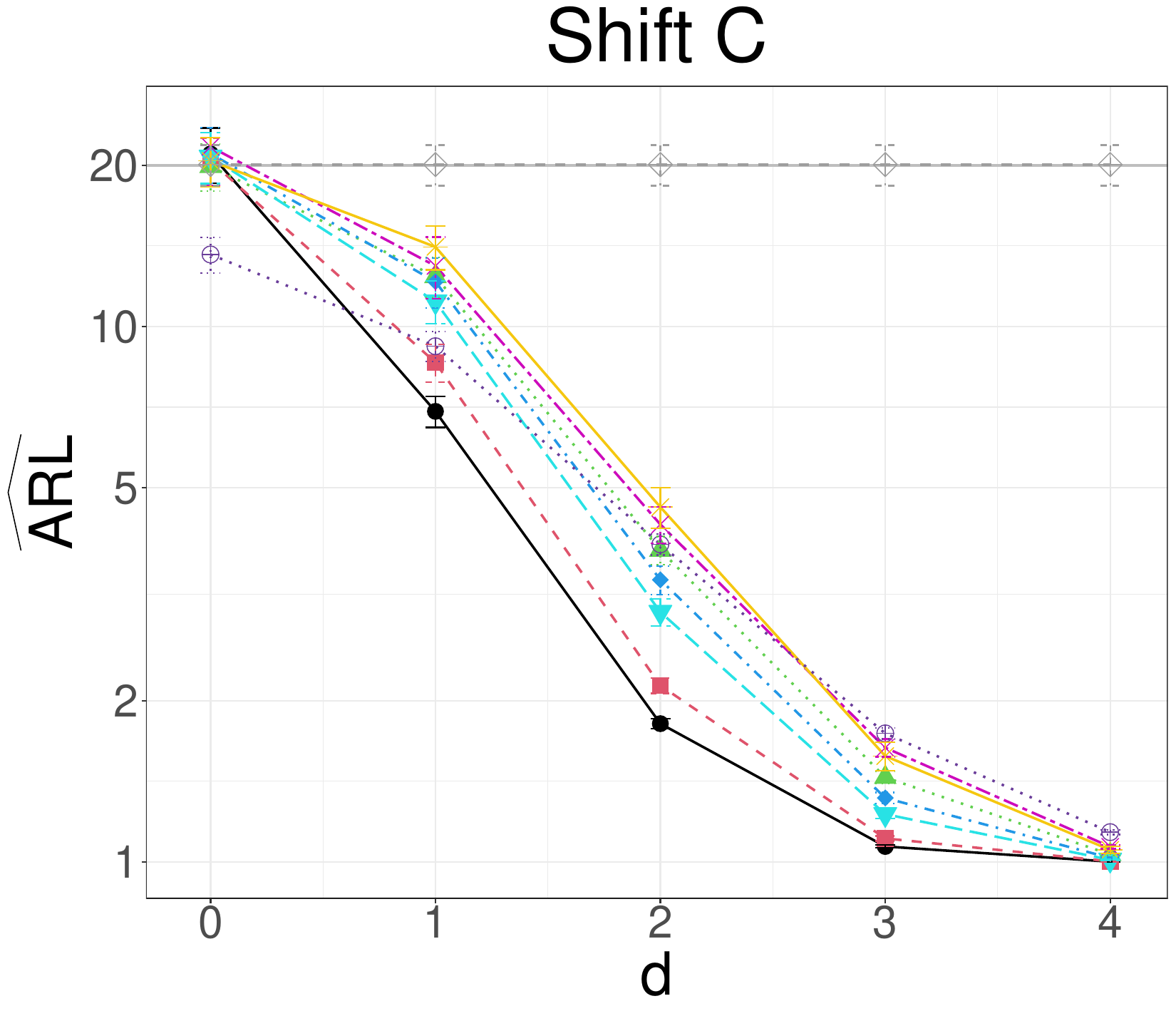}&\includegraphics[width=.25\textwidth]{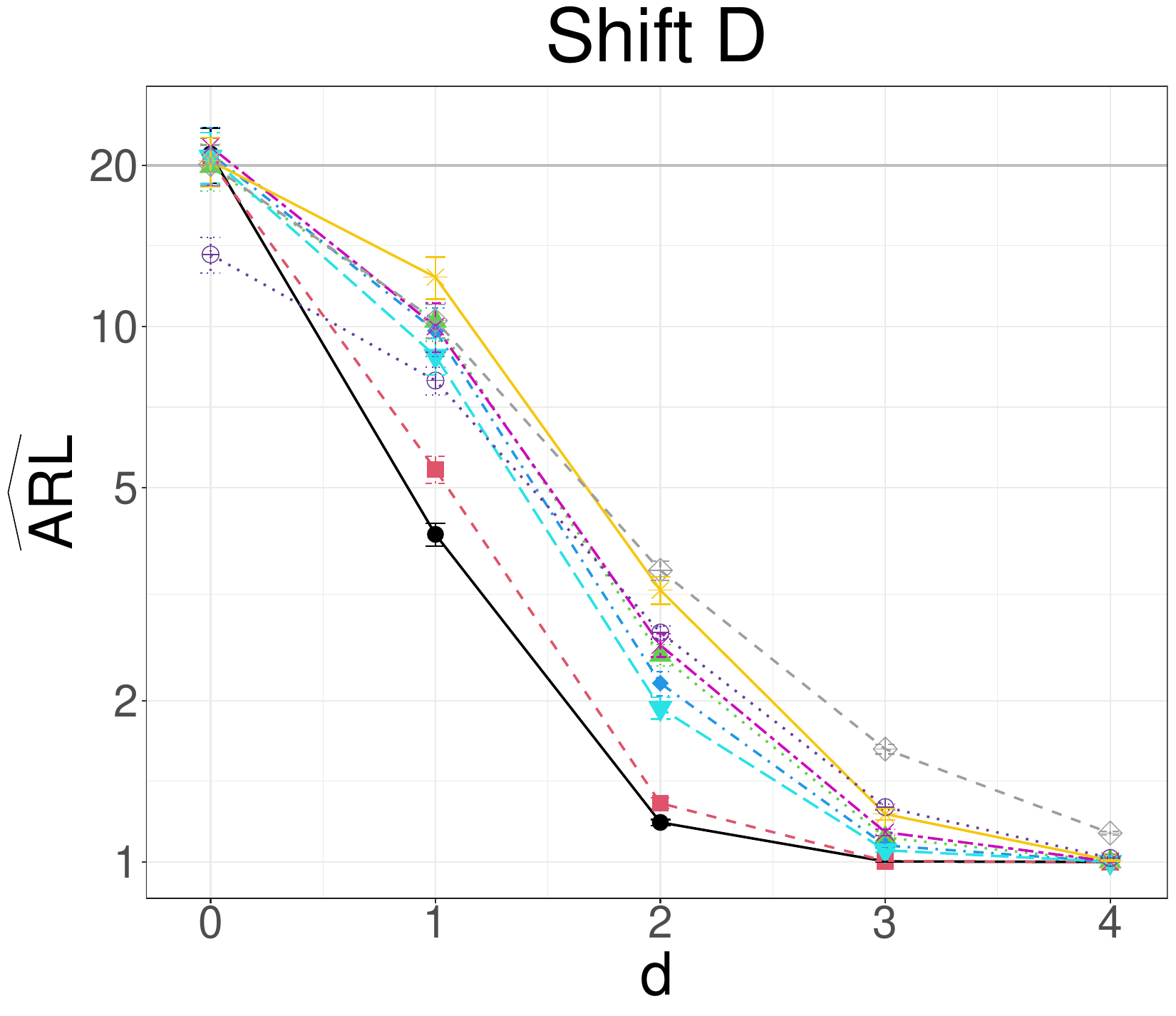}\\
		\textbf{\footnotesize{D3}}&\includegraphics[width=.25\textwidth]{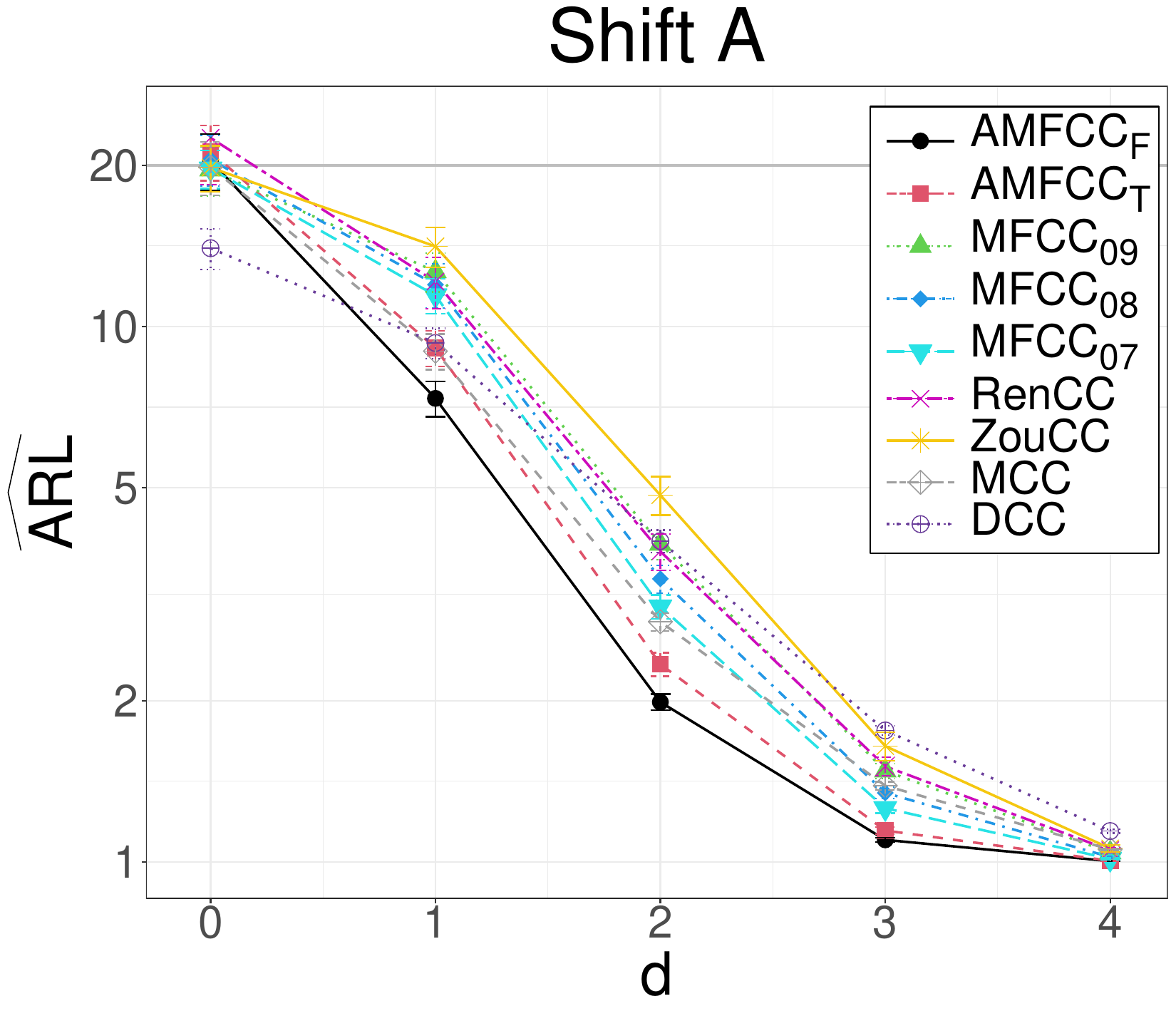}&\includegraphics[width=.25\textwidth]{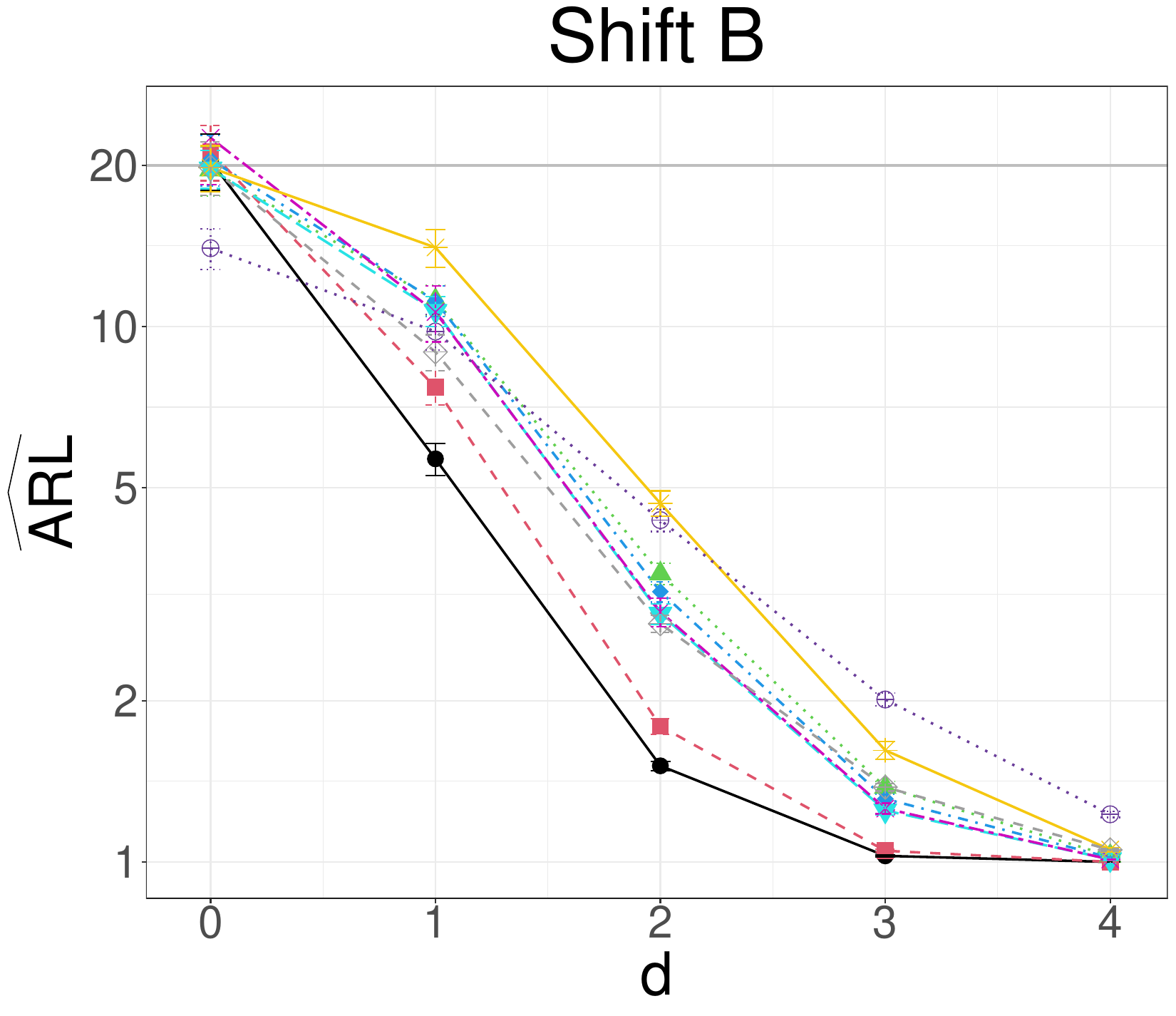}&\includegraphics[width=.25\textwidth]{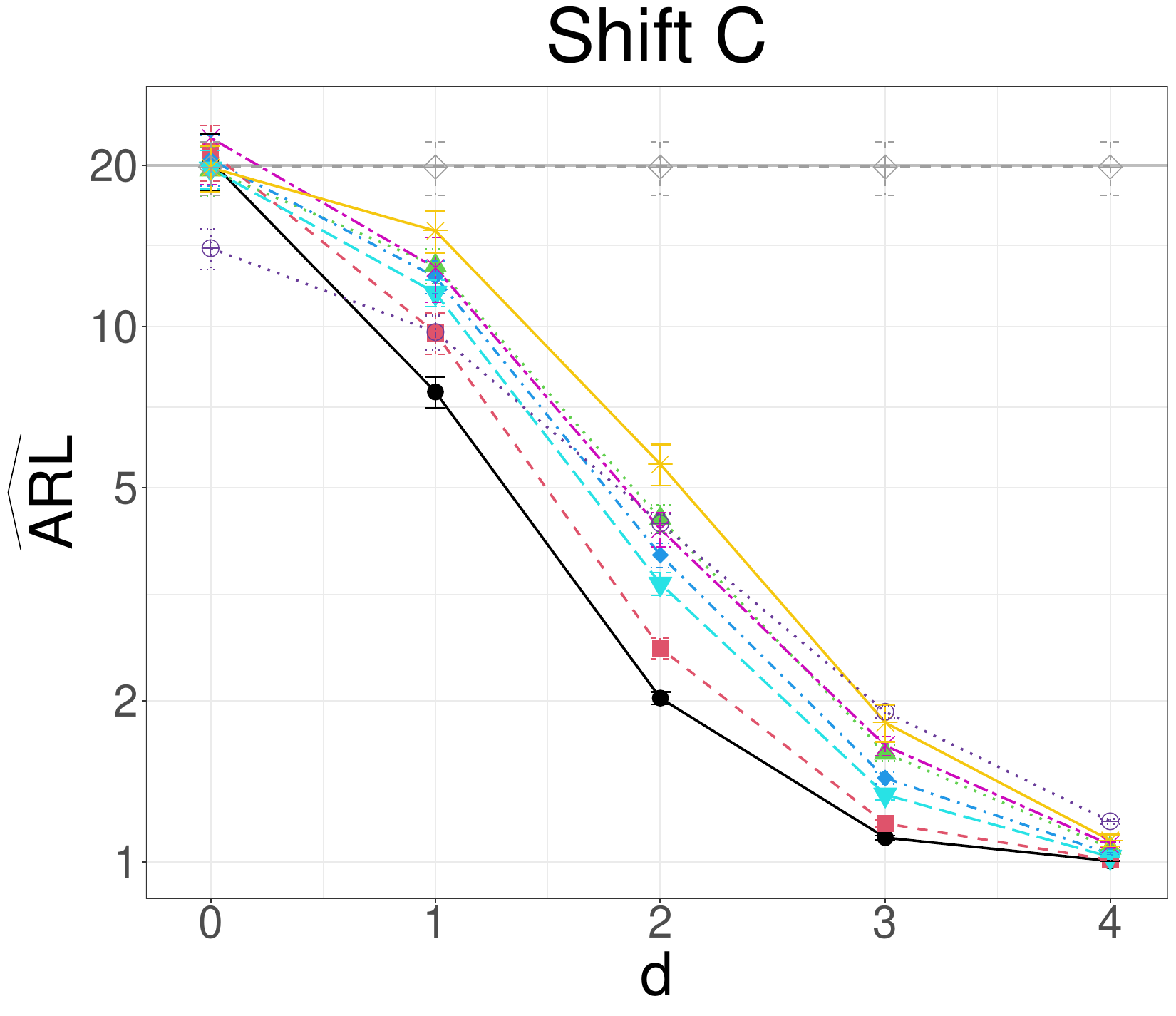}&\includegraphics[width=.25\textwidth]{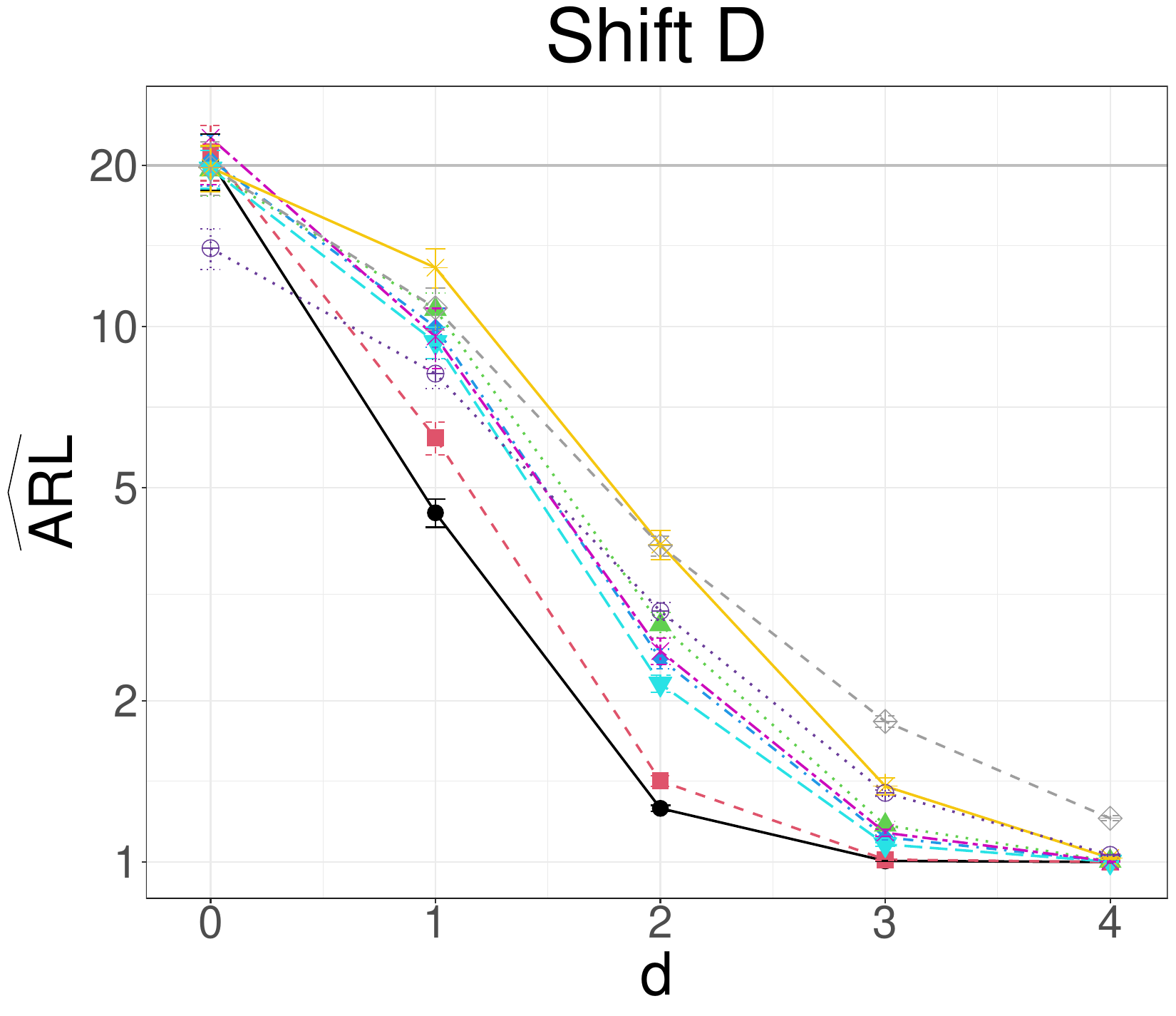}\\
	
	\end{tabular}
	\vspace{-.5cm}
\end{figure}

Figure \ref{fi_results_1}  displays the mean FAR ($ d=0 $) or TDR ($ d \neq 0 $) for Scenario 1  as a function of the severity level $d$ for each dependence level (D1, D2 and D3), OC condition (Shift A, Shift B, Shift C, and Shift D).
The proposed AMFCC\textsubscript{F} and  AMFCC\textsubscript{T} outperform all the competing methods for each dependence level and OC condition.
Indeed, the proposed methods better adapt to the unknown distribution of the OC process by considering a suitable range of parameter combinations. 
The performance of the AMFCC\textsubscript{F} appears overall slightly better than that of the AMFCC\textsubscript{T}. We motivate this by recognizing that the OC conditions of this simulation study allow a large number of partial tests to be significant. It is known indeed that the  Fisher omnibus is more appropriate when any strength of evidence is available from most of the partial tests, whereas the Tippett combining function is better suited when one or very few of the partial tests are significant \citep{loughin2004systematic}.
It is worth noting that the former function takes into account the $p$-values corresponding to all the partial tests by placing greater emphasis on smaller $p$-values than on larger ones, whereas the latter is based on the minimum $p$-value alone.
The most intriguing conclusion from this study comes from the poor performance of the MFCC\textsubscript{09}, MFCC\textsubscript{08}, and MFCC\textsubscript{07} versus the AMFCC that evidences the clear inadequacy of the popular criteria based on the total variability explained by the retained principal components.
The MFCC methods show overall similar performance even though MFCC\textsubscript{07} seems to perform slightly better than MFCC\textsubscript{09} and MFCC\textsubscript{08}. 
Among the non-functional approaches, the MCC shows good performance for Shift A and Shift B whose detection power is comparable to that of the MFCC approaches. This is not surprising because these shifts are characterized by an overall increase in the mean function that is better captured by the average of each component profile. On the contrary, MCC shows bad performance for Shift D and, especially, for Shift C, where shifts mainly affect the shape of the mean profile.
As expected, the DCC achieves very low detection power overall because it suffers from the high dimensionality of the problem.
%

\begin{figure}[h]
	\caption{Mean cFAR ($ d=0 $) or cTDR ($ d \neq 0 $) achieved by AMFCC\textsubscript{F},  AMFCC\textsubscript{T}, MFCC\textsubscript{09}, MFCC\textsubscript{08}, and  MFCC\textsubscript{07} for each dependence level (D1, D2 and D3), OC condition (Shift A, Shift B, Shift C, and Shift D) as a function of the severity level $d$  in Scenario 1.}
	
	\label{fi_results_vs}
	
	\centering
	\hspace{-2.1cm}
	\begin{tabular}{cM{0.24\textwidth}M{0.24\textwidth}M{0.24\textwidth}M{0.24\textwidth}}
		\textbf{\footnotesize{D1}}&\includegraphics[width=.25\textwidth]{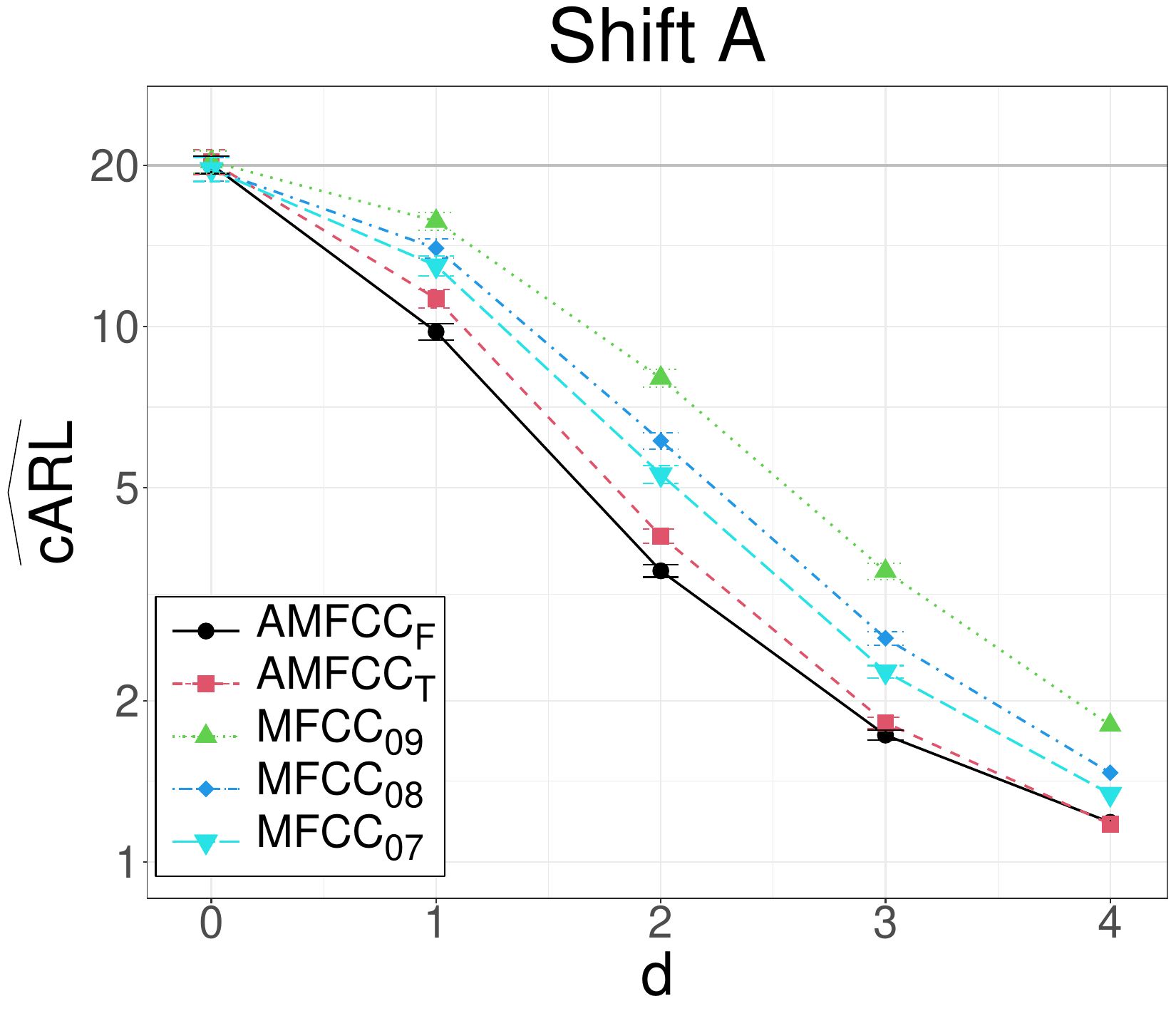}&\includegraphics[width=.25\textwidth]{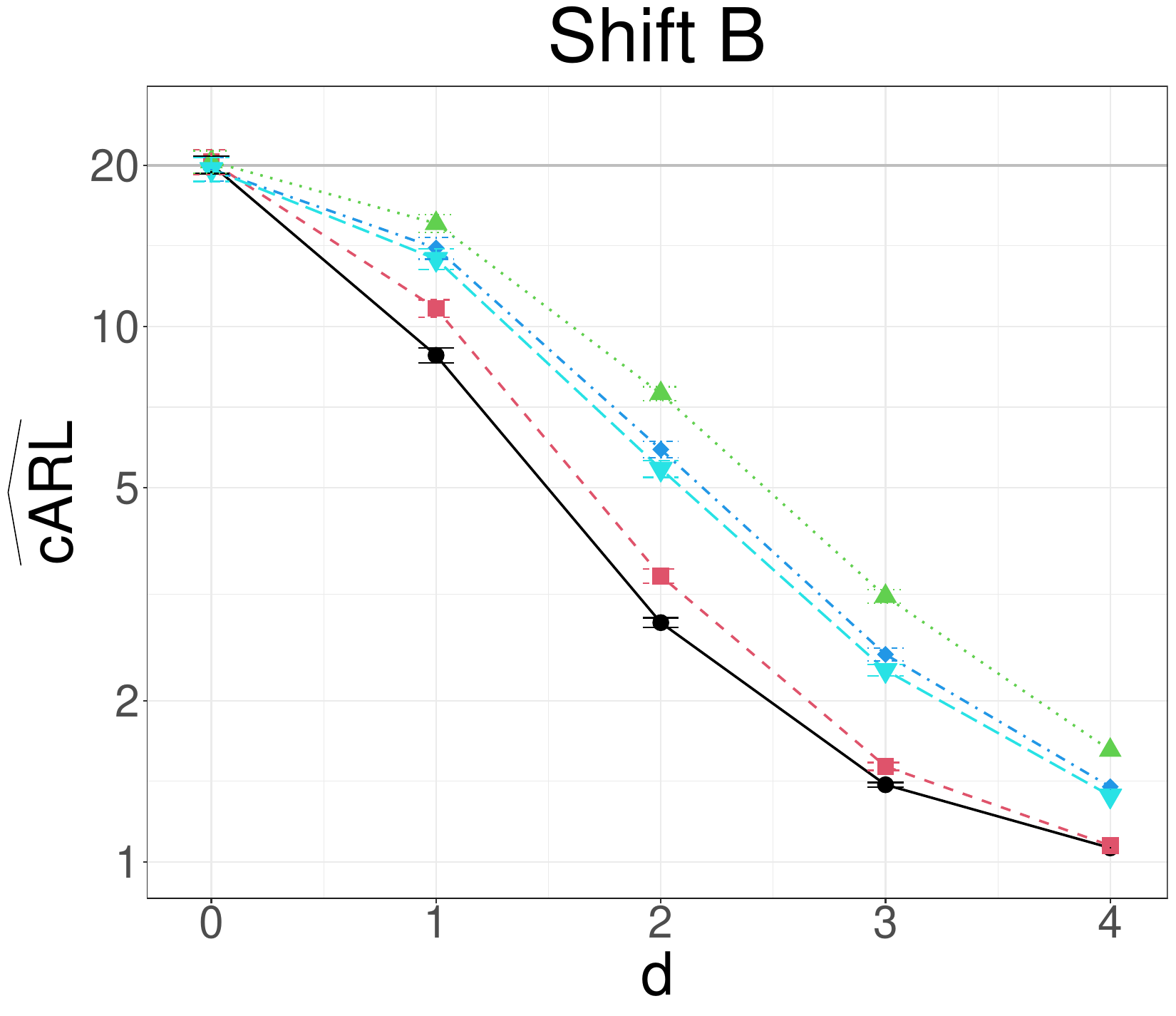}&\includegraphics[width=.25\textwidth]{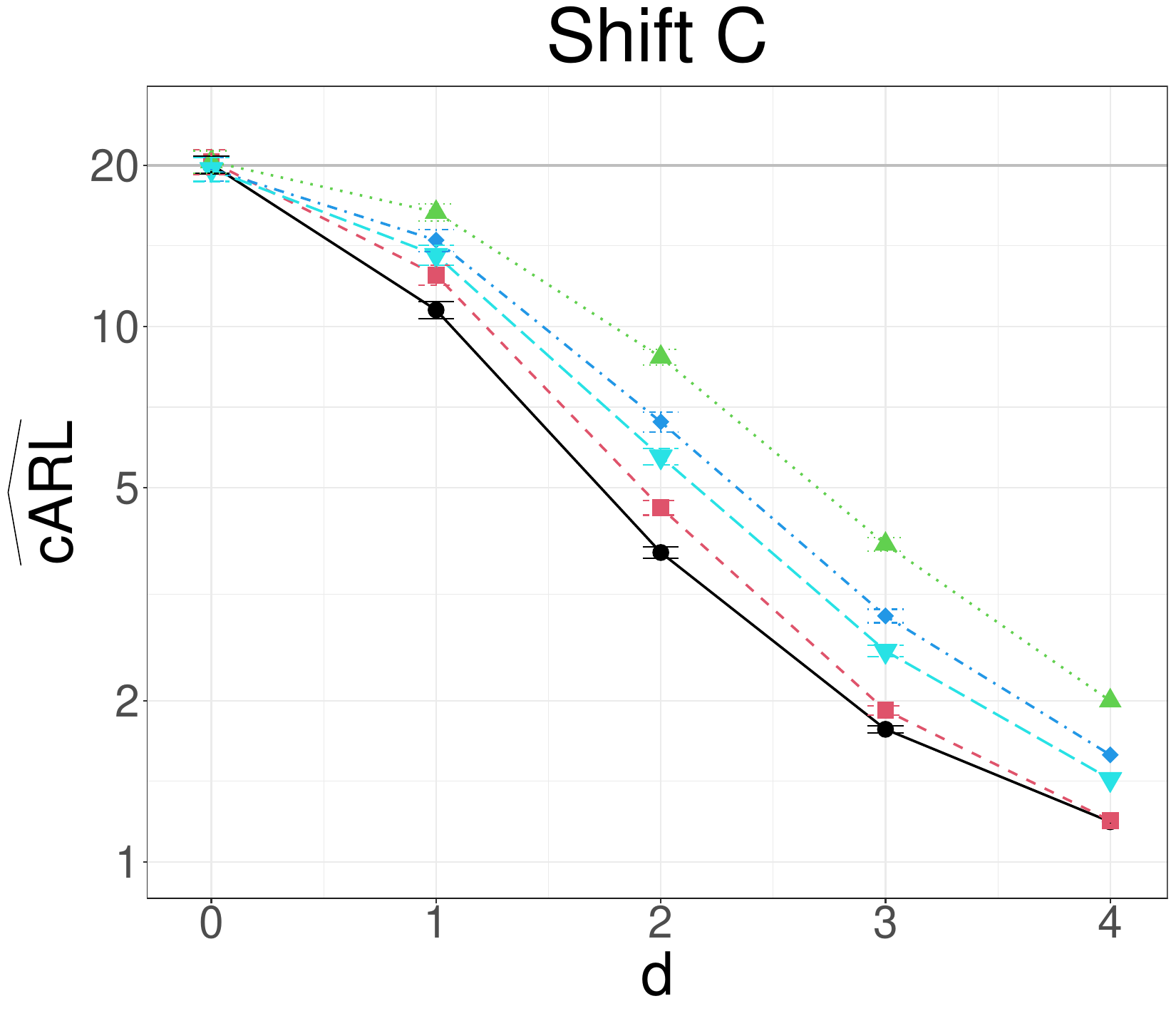}&\includegraphics[width=.25\textwidth]{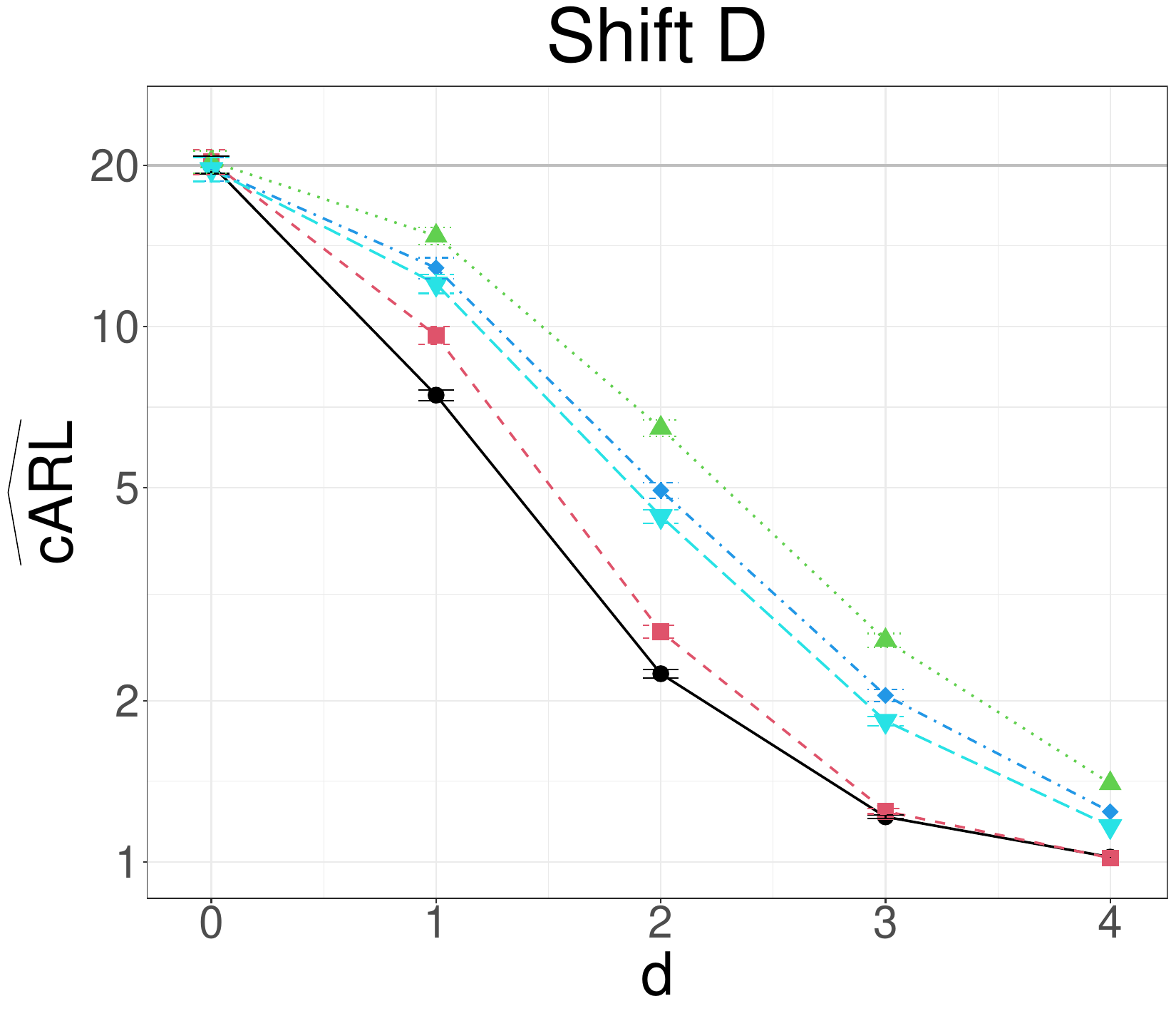}\\
		\textbf{\footnotesize{D2}}&\includegraphics[width=.25\textwidth]{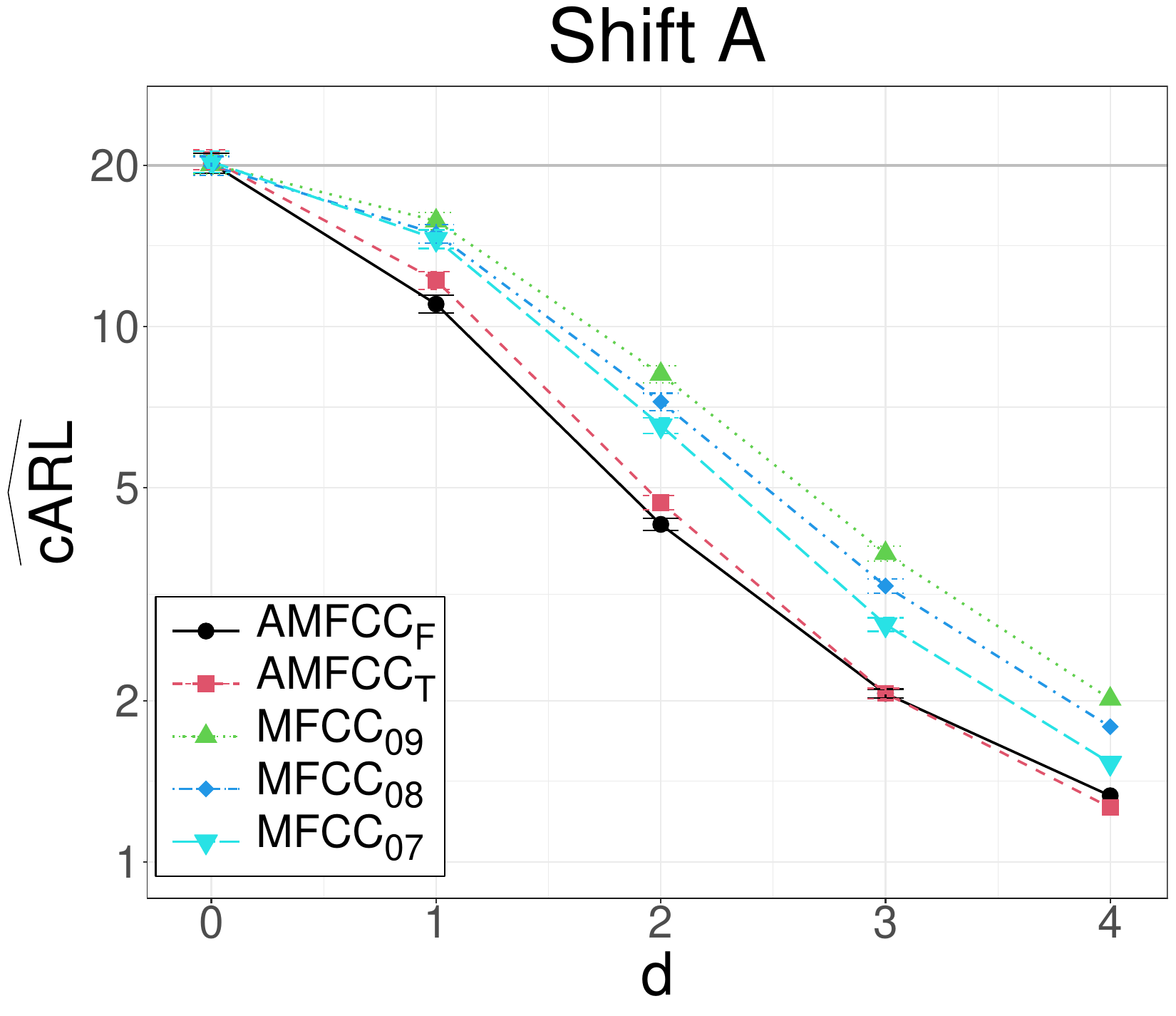}&\includegraphics[width=.25\textwidth]{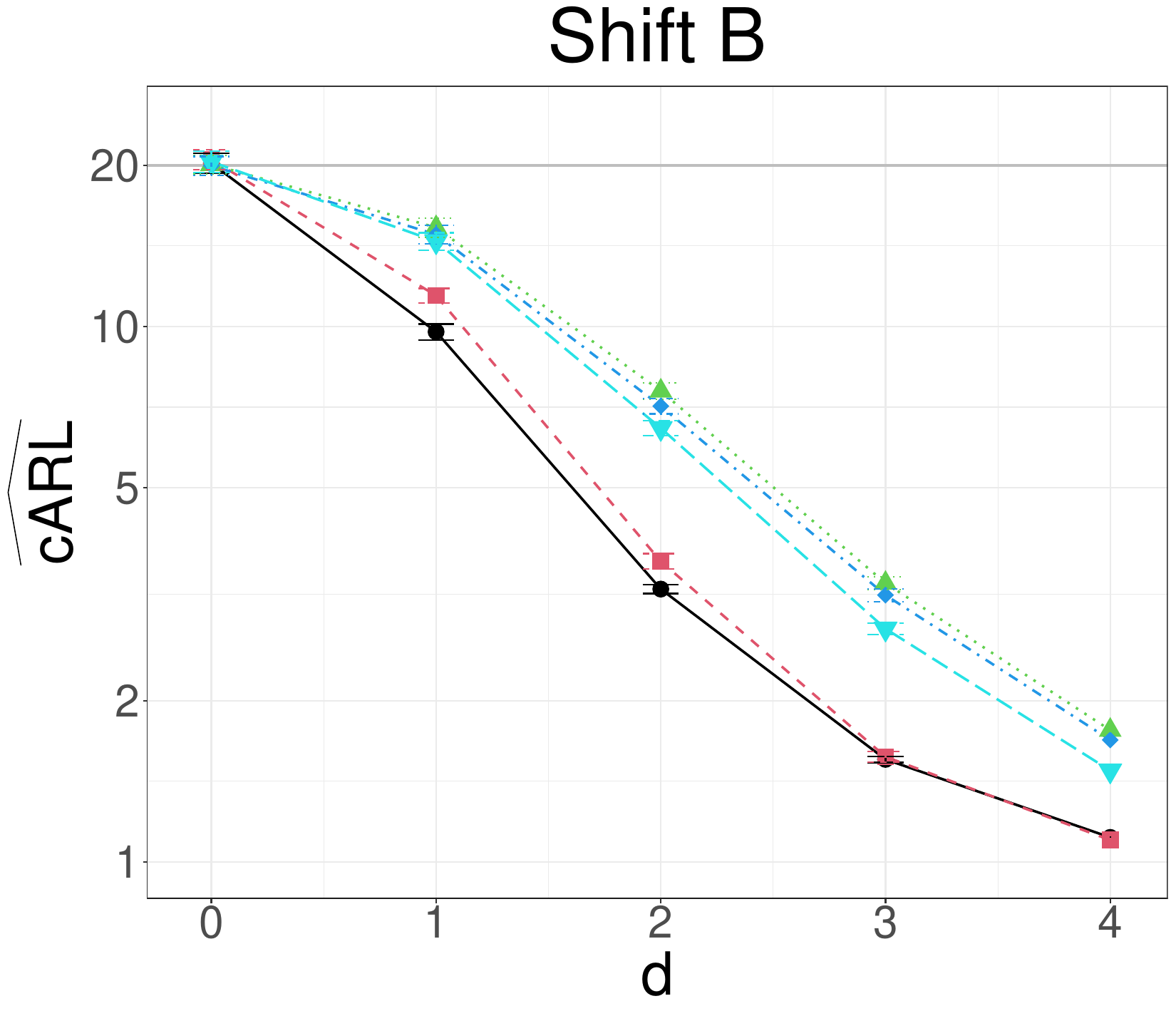}&\includegraphics[width=.25\textwidth]{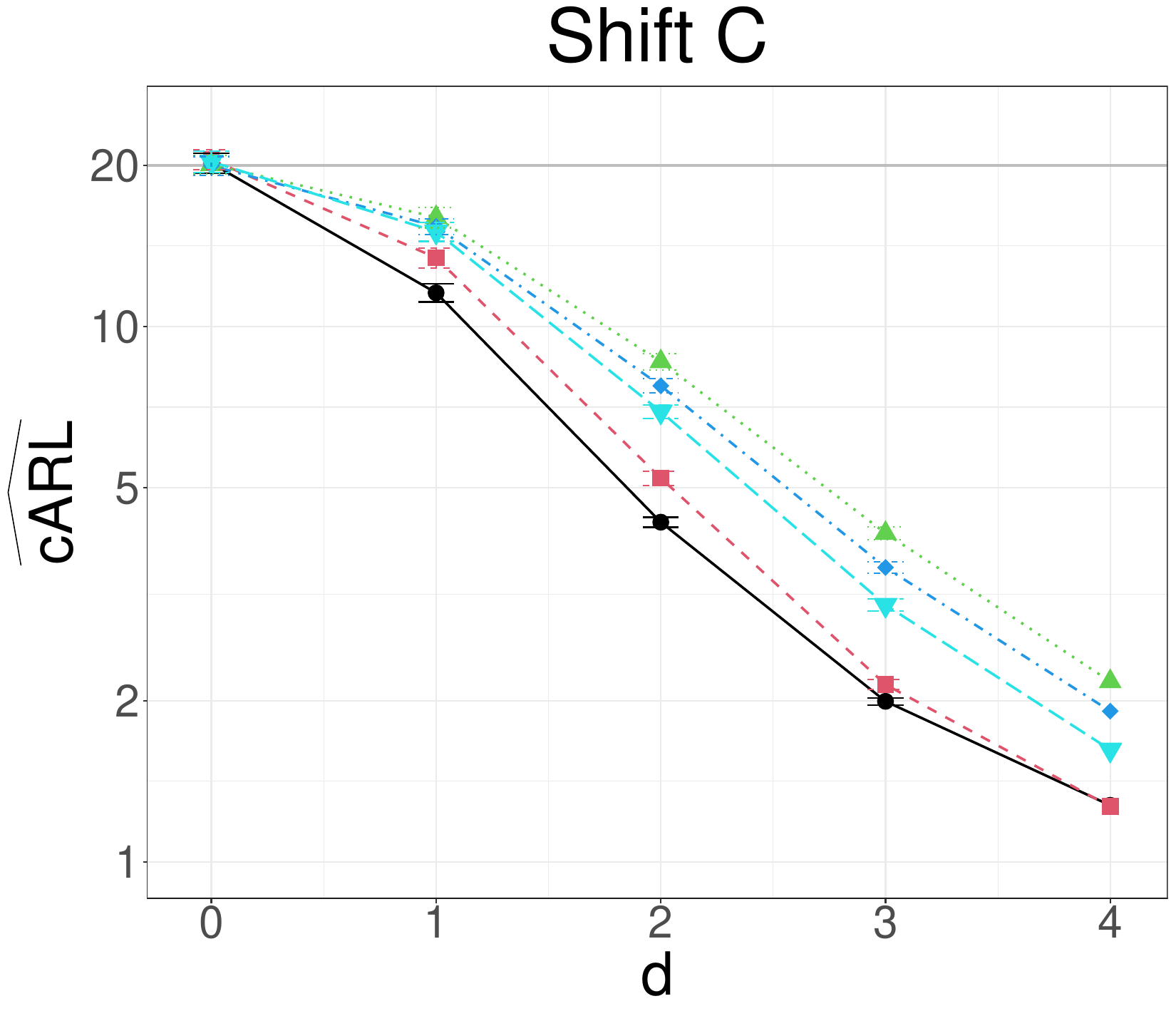}&\includegraphics[width=.25\textwidth]{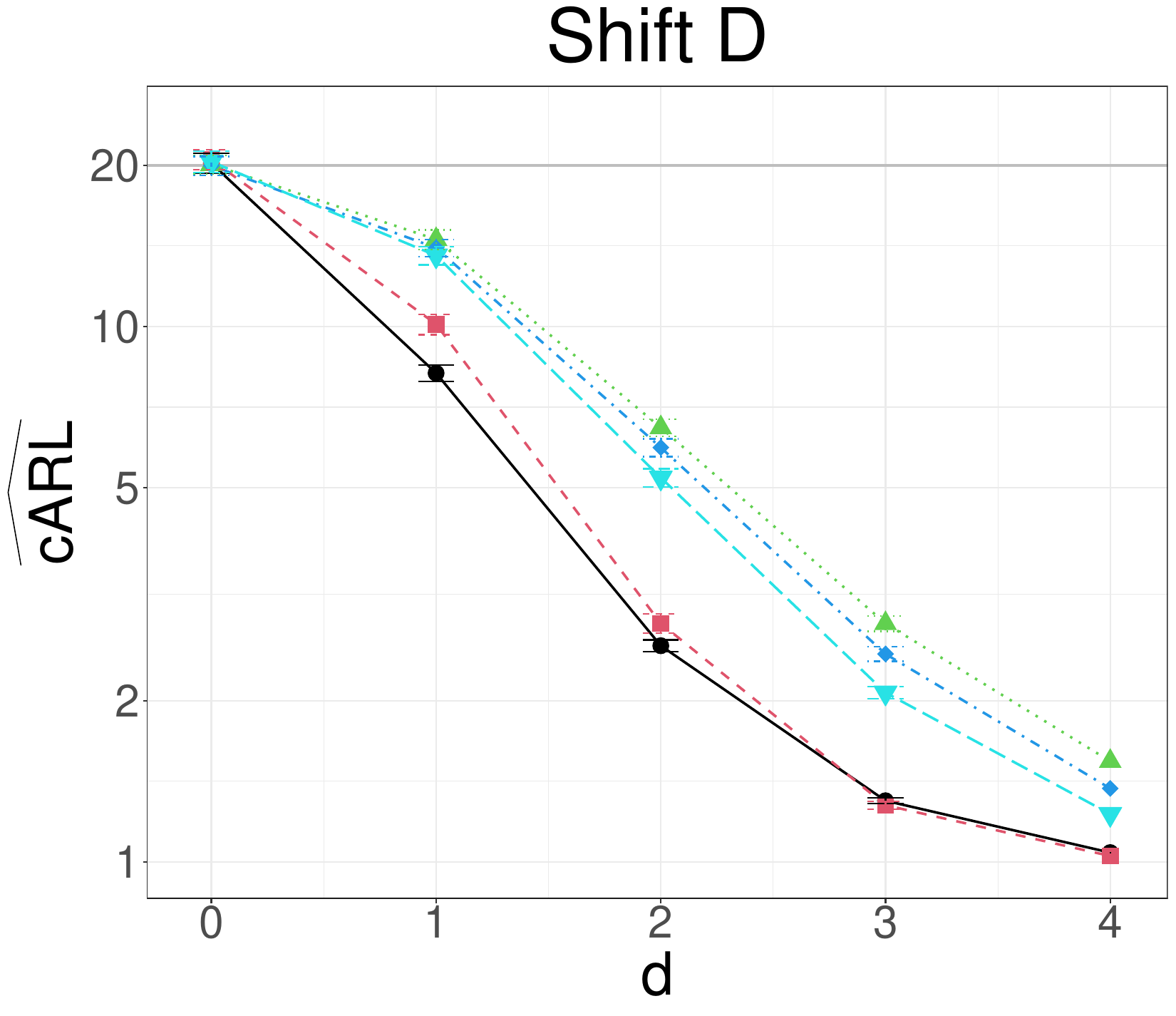}\\
		\textbf{\footnotesize{D3}}&\includegraphics[width=.25\textwidth]{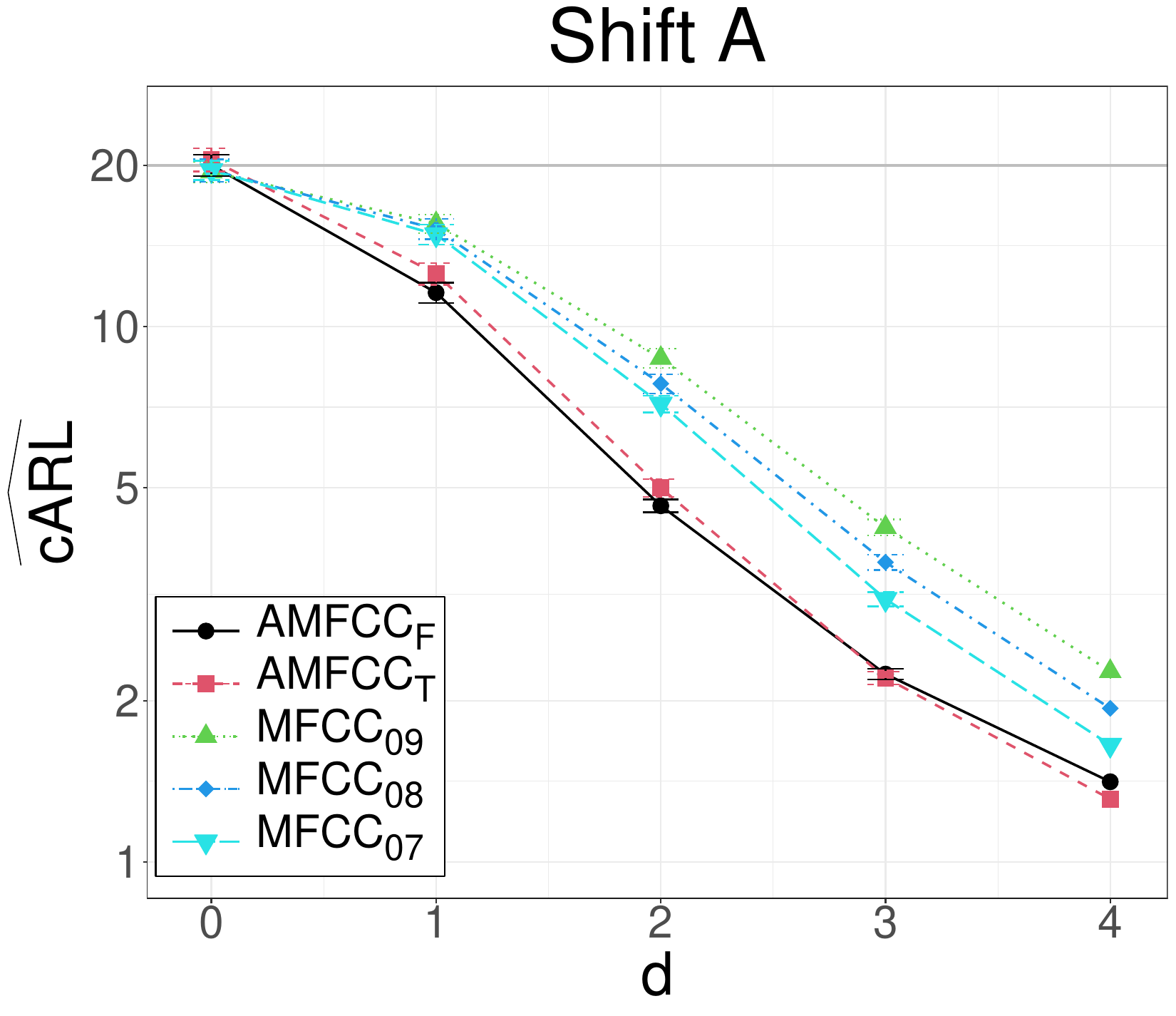}&\includegraphics[width=.25\textwidth]{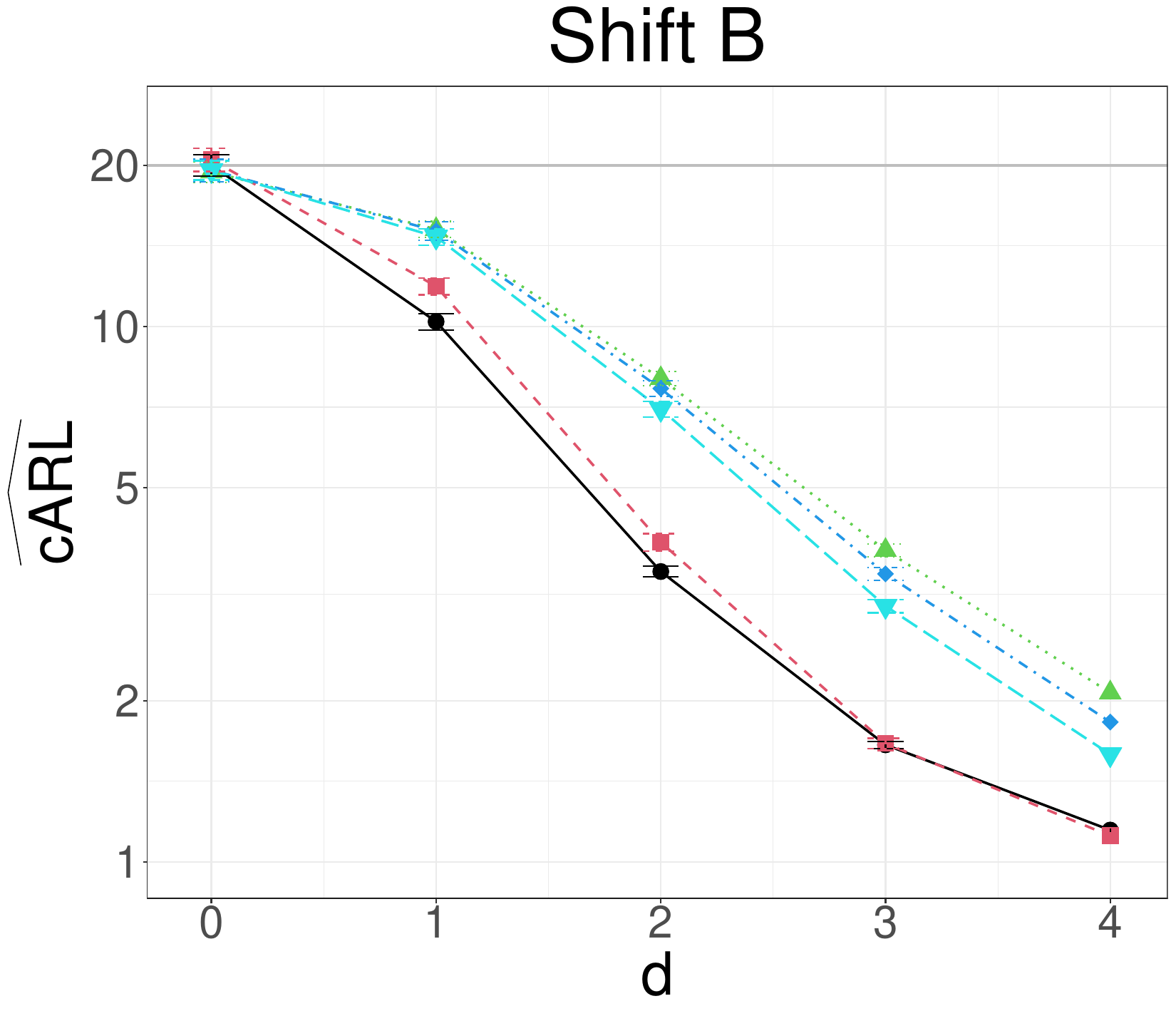}&\includegraphics[width=.25\textwidth]{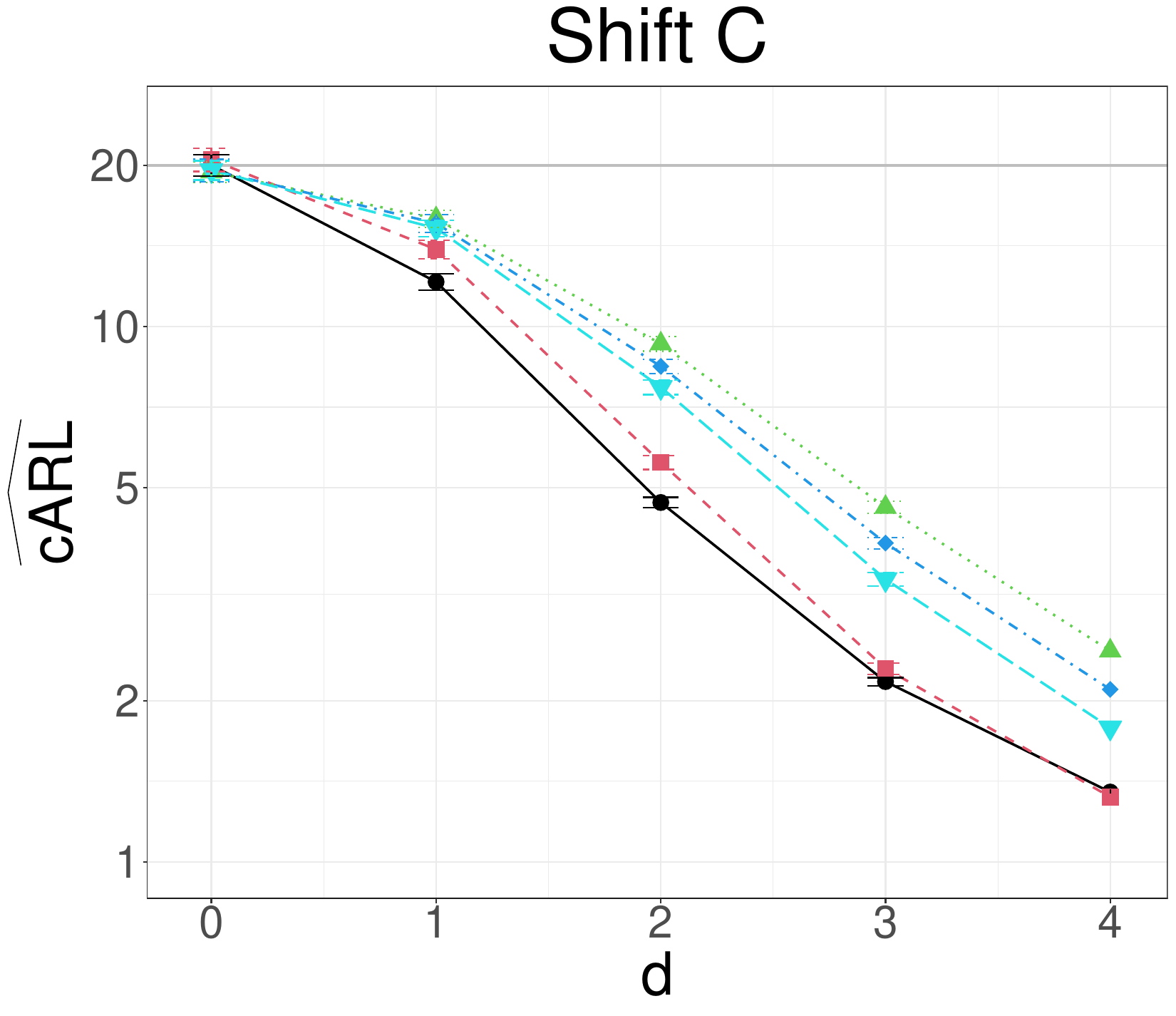}&\includegraphics[width=.25\textwidth]{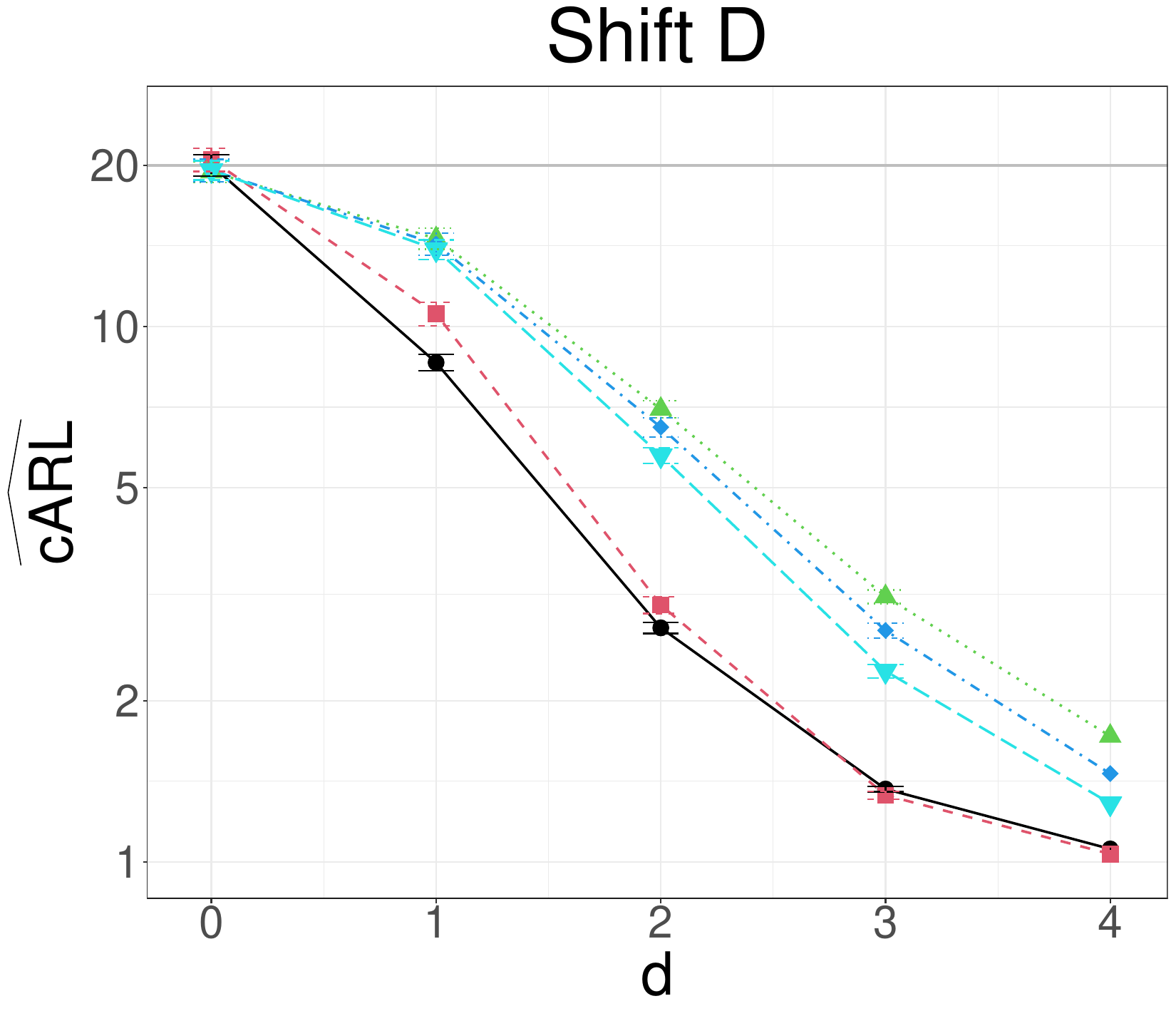}\\
		
	\end{tabular}
	\vspace{-.5cm}
\end{figure}

The effectiveness of the proposed diagnostic procedure described in Section \ref{sec_dia} is evaluated by means of the component true detection rate (cTDR) and the component false alarm rate (cFAR), which are defined as the proportion of times a component of the multivariate functional quality characteristic is, respectively,  rightly or wrongly signaled as anomalous.
For each component, $k=1,\dots,p$, the cFAR should be as similar as possible to the overall type I error probabilities $\alpha_k$ considered to obtain the control limits $C_k$, which is set equal to 0.05, whereas the cTDR should be as close to one as possible.
Post-signal diagnostics of AMFCC\textsubscript{F} and  AMFCC\textsubscript{T} are compared with those of MFCC\textsubscript{07}, MFCC\textsubscript{08}, and  MFCC\textsubscript{09}, which identify the components responsible for the OC condition through their overall contribution to the Hotelling $T^2$ and SPE statistics.
Figure \ref{fi_results_vs}  displays the mean cFAR ($ d=0 $) or cTDR ($ d \neq 0 $) for Scenario 1 as a function of the severity level $d$ for each dependence level (D1, D2 and D3), and OC condition (Shift A, Shift B, Shift C, and Shift D). Results for Scenario 2 are reported in Supplementary Materials B.

The proposed diagnostic approach with both the Fisher omnibus and  Tippett combining functions largely outperforms MFCC\textsubscript{09}, MFCC\textsubscript{08}, and  MFCC\textsubscript{07}. Differently from Figure \ref{fi_results_1}, the AMFCC\textsubscript{T} slightly outperforms AMFCC\textsubscript{F}.
Although performance differences are very small, this result is an example that, as stated in Section \ref{sec_mon}, no single combining function proves to be the best under all circumstances and its performance depends on the data and the true alternative hypothesis.
Finally, among the competing approaches, MFCC\textsubscript{07} stands out as the best method in accordance with the results of Figure \ref{fi_results_1}.

\section{Case Study: Resistance Spot Welding  Process Monitoring in the Automotive Industry}
\label{se_casestudy}
The performance in practical situations of the AMFCC  is demonstrated through a case study in the automotive industry.
It addresses the issue of monitoring the quality of the RSW process to guarantee the structural integrity and solidity of welded assemblies in each vehicle \citep{martin}.
The RSW process \citep{zhang2011resistance} is an autogenous welding process in which two overlapping conventional steel galvanized sheets are joined together, without the use of any filler material. 
Joints are formed by applying pressure to the weld area from two opposite sides through two copper electrodes to which a voltage is applied and generates a current flowing through the material.
Heat is generated by the resistance offered by the metal and increases the metal temperature at the faying surfaces of the workpieces up to the melting point.
Due to the mechanical pressure of the electrodes, the molten metal of the jointed metal sheets cools,  and solidifies, and forms the so-called weld nugget \citep{raoelison}.
The modern automotive Industry 4.0 framework allows the automatic acquisition of a large volume of RSW process variables.
Among online measurements, the \textit{dynamic resistance curve} (DRC) is recognized as the most informative technological signature of the metallurgical development of a spot weld \citep{dickinson,capezza2021functional_clustering} and is customarily regarded as an online low-cost proxy of the RSW process quality in contrast with  destructive off-line tests, which cannot be obviously performed
on each welded spot.
Further details on how the typical behaviour of a DRC is related to the physical and metallurgical development of a spot weld are provided by \cite{capezza2021functional_clustering}.
The RSW process quality is directly affected by electrode wear that leads to changes in electrical, thermal, and mechanical contact conditions at electrode and metal sheet interfaces \citep{manladan2017review}.
To counteract the wear issue, electrodes go through periodical dressing operations.
In this setting, the aim of this case study is the monitoring of spot welds to swiftly identify  DRC mean shifts caused by electrode wear, which could be considered as a further criterion to address the electrode dressing. 

The data analyzed are courtesy of Centro Ricerche Fiat and are recorded at the Mirafiori Factory during lab tests.
Specifically,  the multivariate functional quality characteristic is the vector of ten DRCs, collected on the same item and pertaining to ten spot welding points of interest made by the same welding machine.
The dataset contains 1260 observations of the multivariate functional quality characteristic where each component, i.e., DRC, is obtained by resistance measurements collected at a regular grid of points equally spaced by 1 ms. Without loss of generality, the latter are normalized onto time domain $[0, 1]$.
The Phase I sample is formed by 708 observations corresponding to spot welds made immediately after electrode renewal, whereas  552 additional observations, gathered immediately before electrode renewal, are used in Phase II to evaluate online monitoring performance of the AMFCC and competing methods in detecting the DRC mean shift caused by electrode wear. Moreover, resistance measurements were collected at a regular grid of points equally spaced by 1 ms. 
The proposed AMFCC\textsubscript{F} and  AMFCC\textsubscript{T} are  implemented  as  in Section \ref{se_perfo} with the Phase I  sample  equally split into training and tuning sets.
The type I error probabilities  $\alpha$ and $\alpha_k$ are set equal to 0.05.
The 354 IC observations forming the tuning set are plotted in  Supplementary Materials C.

The estimated TDR values, denoted as $\widehat{TDR}$, of the  AMFCC\textsubscript{F},  AMFCC\textsubscript{T} and the competing methods on the Phase II sample are shown in Table \ref{tab_arlreal}.
The uncertainty of $\widehat{TDR}$ is quantified through a bootstrap analysis \citep{efron1994introduction} as in \cite{centofanti2021functional,capezza2022robust}.
Table \ref{tab_arlreal} reports the mean of the empirical bootstrap distribution of $\widehat{TDR}$, denoted by $\overline{TDR}$,  and the corresponding bootstrap 95\% confidence interval (CI)  for each method. 
\begin{table}[]
	\caption{Estimated TDR values, denoted as $\widehat{TDR}$,  mean of the empirical bootstrap distribution of $\widehat{TDR}$, denoted by $\overline{TDR}$,  and the corresponding bootstrap 95\% confidence interval (CI)  for each monitoring method in the case study. }
	\centering
	\resizebox{.5\columnwidth}{!}{
		\begin{tabular}{ccccc}
			\toprule
			& $\widehat{TDR}$ & $\overline{TDR}$ & CI\\
			\midrule
			AMFCC\textsubscript{F}& 0.788 & 0.787 & [0.759,0.817]\\
			AMFCC\textsubscript{T} & 0.779 & 0.777 & [0.746,0.806]\\
			MFCC\textsubscript{09} & 0.462 & 0.460 & [0.421,0.498]\\
			MFCC\textsubscript{08} & 0.612 & 0.606 & [0.568,0.640]\\
			MFCC\textsubscript{07} & 0.712 & 0.710 & [0.671,0.746]\\
			MCC& 0.520 & 0.519 & [0.476,0.553]\\
				DCC& 0.275 & 0.278 & [0.240,0.315]\\
			\bottomrule
		\end{tabular}
	}
	
	\label{tab_arlreal}
\end{table}
The proposed  AMFCC\textsubscript{F} and  AMFCC\textsubscript{T} achieve larger  $\overline{TDR}$ than the competing methods and corresponding bootstrap 95\% confidence intervals are strictly above those of the competing approaches.
These results are in accordance with those presented in Section \ref{se_perfo}, and further demonstrate the inability of the competing methods either to adapt to the unknown distribution of the OC observations (the MFCC\textsubscript{09}, MFCC\textsubscript{08}, and  MFCC\textsubscript{07}) or to capture the functional nature of the data (the MCC and DCC).

\begin{figure}
	\caption{AMFCC\textsubscript{F} and $T_{2}$ and $SPE$ control charts for the MFCC\textsubscript{07} control charts in the case study. The vertical line  separates the monitoring statistics calculated for the tuning set, on the left,  and the Phase II sample on the right, while the horizontal lines define the control limits.}
	\label{fi_cs_exa}
	\resizebox{\textwidth}{!}{
		\begin{tabular}{cM{0.45\textwidth}M{0.45\textwidth}}
			\footnotesize{AMFCC\textsubscript{F}}&\multicolumn{2}{M{\textwidth}}{\includegraphics[width=0.6\textwidth]{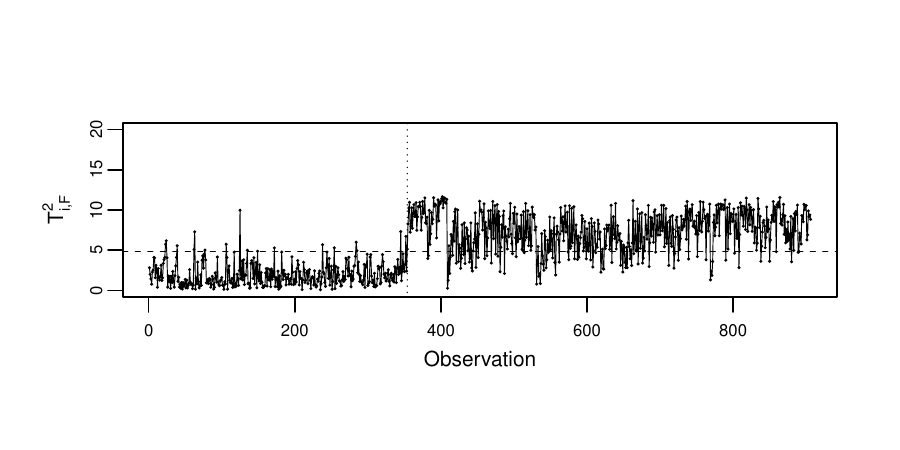}}\\[-1.5cm]
			\footnotesize{MFCC\textsubscript{07}}&\includegraphics[width=0.6\textwidth]{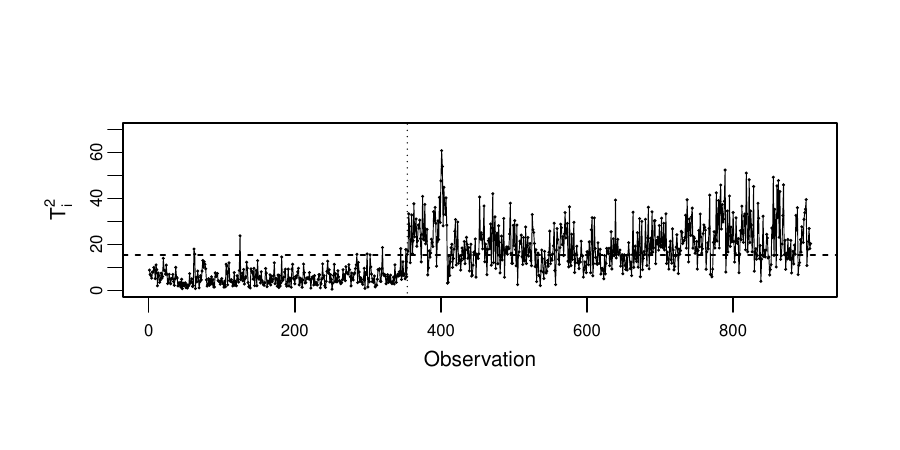}&\includegraphics[width=0.6\textwidth]{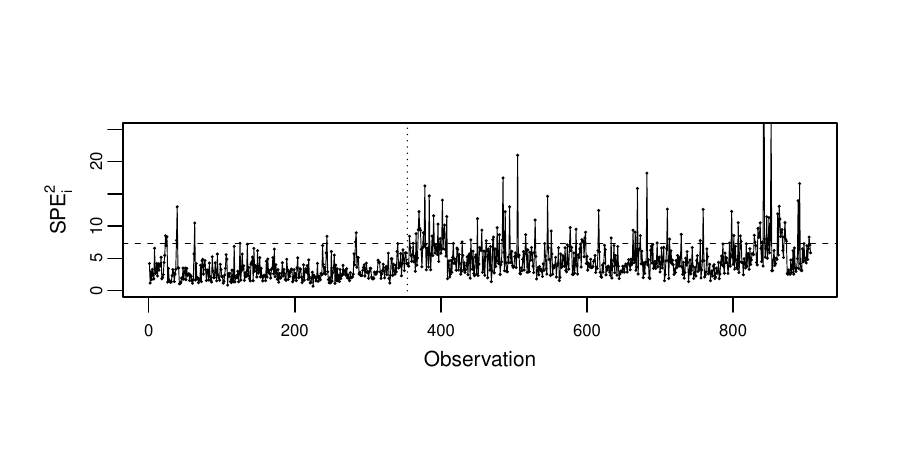}\\
	\end{tabular}}
\end{figure}

For illustrative purposes, Figure \ref{fi_cs_exa} reports the Phase II observations monitored through the  AMFCC\textsubscript{F} and the MFCC\textsubscript{07}. As a direct consequence of  the results in Table \ref{tab_arlreal}, the latter signals a  much lower number of OC observations than the former.
%
%

\begin{figure}
	\caption{Overall contributions $c_{ik}^{T_F^2}$ to  $T^2_{i,F}$  of each component corresponding to the 30th observation of the Phase II sample in the case study. Contributions exceeding upper control limits, represented by the horizontal lines, are plotted in red.}
	\label{fi_cs_exa2}
	%
	\centering
		\includegraphics[width=0.5\textwidth]{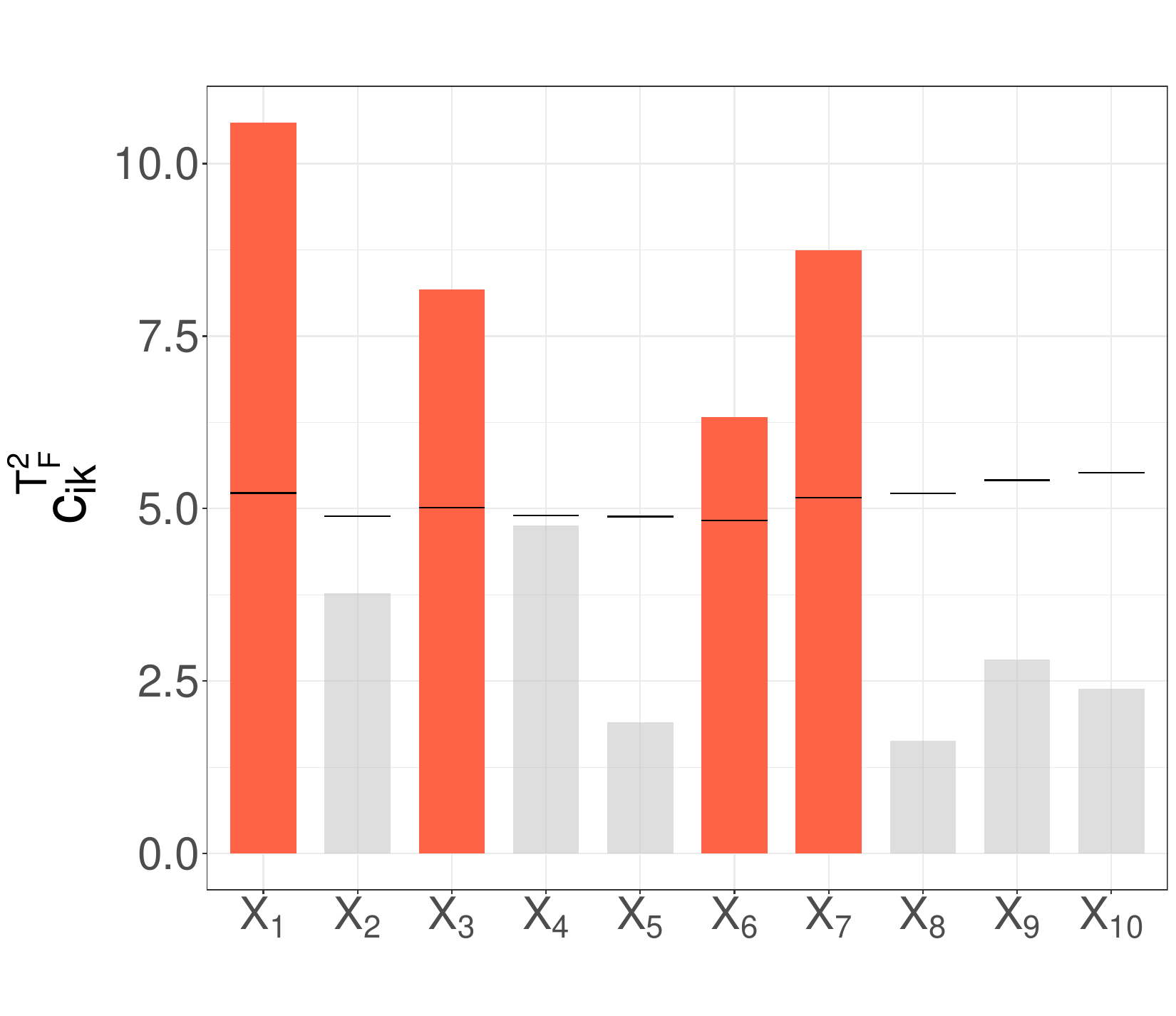}
			
\end{figure}

The advantages  of the proposed diagnostic procedure are further illustrated 
 through Table \ref{tab_arlreal2}  that reports the estimated cTDR values, denoted as $\widehat{cTDR}$,  the mean of the empirical bootstrap distribution of $\widehat{cTDR}$, denoted by $\overline{cTDR}$,  and the corresponding bootstrap 95\% confidence interval (CI)  for  the AMFCC\textsubscript{F}, AMFCC\textsubscript{T}, MFCC\textsubscript{09}, MFCC\textsubscript{08}, and MFCC\textsubscript{07}.
\begin{table}[]
	\caption{Estimated cTDR values, denoted as $\widehat{cTDR}$,  mean of the empirical bootstrap distribution of $\widehat{cTDR}$, denoted by $\overline{cTDR}$,  and the corresponding bootstrap 95\% confidence interval (CI\textsubscript{c}) for  the AMFCC\textsubscript{F}, AMFCC\textsubscript{T}, MFCC\textsubscript{09}, MFCC\textsubscript{08}, and MFCC\textsubscript{07} in the case study. }
	\centering
	\resizebox{.5\columnwidth}{!}{
		\begin{tabular}{ccccc}
			\toprule
			& $\widehat{cTDR}$ & $\overline{cTDR}$ & CI\textsubscript{c}\\
			\midrule
			AMFCC\textsubscript{F}& 0.392 & 0.391 & [0.373,0.407]\\
			AMFCC\textsubscript{T} & 0.445 & 0.442 & [0.423,0.462]\\
			MFCC\textsubscript{09} & 0.221 & 0.221 & [0.209,0.234]\\
			MFCC\textsubscript{08} & 0.281 & 0.279 & [0.261,0.295]\\
			MFCC\textsubscript{07} & 0.318 & 0.318 & [0.299,0.333]\\
			\bottomrule
		\end{tabular}
	}
	
	\label{tab_arlreal2}
\end{table}
All values clearly point out the proposed diagnostic procedure outperforms the competitors. Bootstrap 95\% confidence intervals for the proposed method are strictly above those achieved by the competing approaches and support the previous statement.
As an example of its practical  applicability,  Figure \ref{fi_cs_exa2} reports  the overall contributions $c_{30k}^{T_F^2}$ of each component to the monitoring statistic  $T^2_{30,F}$   corresponding to the Phase II observation $i=30$, for which $T^2_{30,F}>C$. 
The overall contribution values corresponding to $X_1$, $X_3$, $X_6$, and $X_7$ appear as exceeding the corresponding contribution control limits $C_1$, $C_4$, $C_6$, and $C_7$, and thus, are deemed a responsible for the  OC signal.
\section{Conclusions}
\label{se_conclusions}
In this paper, we propose a new approach for Phase II monitoring and diagnosis of a multivariate functional quality characteristic, referred to as \textit{adaptive multivariate functional control chart} (AMFCC).
The AMFCC relies on the twofold notions that (i) all parameters chosen to implement any profile monitoring scheme could heavily influence the ability to identify anomalous behaviors of the process; (ii)   the usual parameter selection criteria that are optimized for model estimation are not necessarily optimal for testing and may poorly perform in terms of detection power of the SPM scheme because they do not take into account any information under the OC condition.
The AMFCC is the first monitoring scheme for a multivariate functional quality characteristic able to adapt to the unknown distribution of the OC observations.
Specifically, this feature is obtained by combining $p$-values of the partial tests corresponding to  Hotelling $T^2$-type statistics calculated at different parameter combinations.
 Each statistic is obtained from the multivariate functional principal component decomposition of the multivariate functional data, which, in turn, are estimated from noisy discrete values through a data smoothing approach based on a roughness penalty specifically designed for multiple profiles. 
 Furthermore, a diagnostic procedure based on the contribution plot approach is proposed to identify the components of the quality characteristic mainly responsible for the OC condition.

The performance of the AMFCC is assessed through an extensive Monte Carlo simulation study where it is compared with several approaches already present in the literature both for functional and multivariate data. The proposed method has proved to outperform the competing methods in terms of both OC monitoring and diagnosis. The latter turned out to be unable to adapt to the unknown OC conditions. 
The practical applicability of the proposed method is further illustrated through a case study in the monitoring of a resistance spot-welding process in the automotive industry through the multivariate functional quality characteristic formed by dynamic resistance curves on interest on each item. Also in this case, the AMFCC stands out as the best method in the identification as well as the diagnosis of OC conditions mainly caused by excessive electrode wear.

The AMFCC presents a practical way to translate the notion that optimal parameters for estimation purposes are in general not optimal for testing the IC versus OC state of the process through the monitoring of multivariate functional data. Indeed, its implementation relies on the idea of combining in a single test different statistics that aim at capturing a wide range of OC conditions.
In future research, this idea could be extended to the functional real-time monitoring \citep{centofanti2022real} or when additional covariate information is available \citep{centofanti2021functional,capezza2020control}.
Nevertheless, in many settings, prior knowledge about some OC conditions could be explicitly available or easily elicited. Additional research could be addressed to integrate this information and further improve the proposed AMFCC.

\section*{Supplementary Materials}
The Supplementary Materials contain additional details about the data generation process in the simulation study (A), additional simulation results (B), and additional plots for the data example (C), as well as the \textsf{R} code to reproduce graphics and results over competing methods in the simulation study.

\bibliographystyle{apalike}
\setlength{\bibsep}{5pt plus 0.2ex}
{\small
\spacingset{1}
\bibliography{References}
}

\end{document}




\if0\blind
{
  
\title{Supplementary Materials to ``An Adaptive Multivariate Functional Control Chart''}

\author[1,2]{Fabio Centofanti\thanks{Corresponding author. e-mail: \texttt{fabio.centofanti@kuleuven.be}}}
\author[2]{Antonio Lepore}
\author[2]{Biagio Palumbo}

\affil[1]{Section of Statistics and Data Science, Department of Mathematics, KU Leuven, Belgium}
\affil[2]{Department of Industrial Engineering, University of Naples Federico II, Naples, Italy}

\setcounter{Maxaffil}{0}
\renewcommand\Affilfont{\itshape\small}
\date{}
\maketitle
} \fi

\if1\blind
{
  \bigskip
  \bigskip
  \bigskip
  \begin{center}
    {\LARGE\bf Supplementary Materials to ``An Adaptive Multivariate Functional Control Chart''}
\end{center}
  \medskip
} \fi

\vfill

\newpage
\spacingset{1.45} 
\appendix
\numberwithin{equation}{section}

\section{Details on Data Generation in the Simulation Study}
\label{sec_appB}
The data generation process is inspired by the works of \cite{centofanti2021functional,capezza2022robust}.
The compact domain $\mathcal{T}$ is  set, without loss of generality, equal to $\left[0,1\right]$ and the number components $p$ is set equal to 5.
 The aim is to generate discrete value realizations $\lbrace\bm{y}_{i,k}\rbrace_{k=1,\dots,p}$ corresponding to the functional observations $ \bm{X}_i $. 
 The IC functional observations $ \bm{X}_i $ are generated from a multivariate Gaussian random process with mean $\bm{m}=\left(m_1,\dots,m_p\right)^{T}$ equal to zero, i.e. $\bm{m}=\bm{0}$, and covariance function $\bm{G}=\lbrace G_{k_1k_2}\rbrace_{1\leq k_1,k_2 \leq p}$ with 
 \begin{equation}
 	G_{k_1k_2}\left(s,t\right)=\frac{0.01}{(8/\delta_{c})|k_1-k_2|+1}\rho(s-t,\delta_{c}) \quad s,t\in\mathcal{T},
 \end{equation}
 where $\rho$ is set as a multiple of the \textit{Bessel} correlation function  of the first kind \citep{abramowitz1964handbook} in Scenario 1, and the \textit{Gaussian} correlation function \citep{abrahamsen1997review} in Scenario 2. The function $\rho$ for Scenario 1 and Scenario 2 are reported in Table \ref{ta_corf}.
 \begin{table}
 	\caption{The function $\rho$  for the data generation process in Scenario 1 and Scenario 2 of the simulation study.}
 	\label{ta_corf}
 	
 	\centering
 	\resizebox{0.5\textwidth}{!}{
 		\begin{tabular}{cc}
 			\toprule
 			&$\rho(z,\delta_{c})$\\
 			\midrule
 			Scenario 1&$\left(-10\frac{(1-|z|)-1}{(1+(\delta_{c}-1)4)}+1\right)^{-1}\sum_{j=0}^{\infty}\frac{\left(-\left(|z|50/3\right)^{2}/4\right)^{j}}{j!\Gamma\left(j+1\right)}$\\[.35cm]
 				Scenario 2&$\exp\left[-\left(|z|40/\delta_{c}\right)^{2}\right]$\\
 			
 			\bottomrule
 	\end{tabular}}
 \end{table}
The parameter $\delta_{c}$ tunes the level of dependence within and between the multivariate functional characteristic components, and it is set equal to $1,2$ and $3$ for models D1, D2, and D3, respectively. 

To simulate departures from IC patterns, the OC observations are generated by simultaneously varying each component mean $m_k$, $k=1,\dots,p$ of the IC functional observations.
Specifically, Shift A is characterized by a mean function that differs from zero in the central part of the domain, in Shift B the mean function linearly decreases in the second half of the domain, in Shift C the mean function shift has a sinusoidal behaviour, whereas in Shift D quadratically increases.
Details on the shape of $m_k$ for each shift are given in Table \ref{ta_shiftmean}.
\begin{table}
	\caption{The function $m_k$ for each shift in the simulation study.}
	\label{ta_shiftmean}
	
	\centering
	\resizebox{0.5\textwidth}{!}{
		\begin{tabular}{cc}
			\toprule
			&$m_k(t)$\\
			\midrule
			Shift A&$\begin{cases}
				\frac{d0.07}{\left(0.75-0.5\right)^2}\left(t-0.5\right)^2-d0.07 &t\in\left[0.25,0.75 \right]\\
				0  & otherwise,
			\end{cases}$\\[.35cm]
		Shift B&$\begin{cases}
			\frac{-d0.09}{\left(1-0.5\right)}\left(t-0.5\right)&t\in\left[0.5,1 \right]\\
			0  & otherwise,
		\end{cases}$\\[.35cm]
			Shift C&$d0.05\sin(t2\pi),\quad t\in\left[0,1 \right]$\\
				Shift D&$d0.12t^2-d0.06,\quad t\in\left[0,1 \right]$\\
			\bottomrule
	\end{tabular}}
\end{table}
To generate discrete value realizations $\lbrace\bm{y}_{i,k}\rbrace_{k=1,\dots,p}$, each component of $ \bm{X}_i$, both under IC and OC condition,  is observed at 100 equally spaced time points in the domain $\left[0,1\right]$ and is contaminated with a normally distributed error with zero mean and variance $\sigma_e^2=0.1^2$.

\newpage
%
%
%
%
\section{Additional Simulation Results}
In this section, we present additional simulations in Scenario 2 mentioned in Section 3. Data are generated with a covariance structure obtained through the \textit{Gaussian} correlation function \citep{abramowitz1964handbook}.  

\begin{figure}[h]
	\caption{Mean FAR ($ d=0 $) or TDR ($ d\neq 0 $) achieved by AMFCC, MFCC\textsubscript{09}, MFCC\textsubscript{08},  MFCC\textsubscript{07}, MCC, and DCC for each dependence level (D1, D2 and D3), OC condition (Shift A, Shift B, Shift C, and Shift D) as a function of the severity level $d$  in Scenario 2.}
	
	\label{fi_results_2}
	
	\centering
	\hspace{-2.1cm}
	\begin{tabular}{cM{0.24\textwidth}M{0.24\textwidth}M{0.24\textwidth}M{0.24\textwidth}}
		\textbf{\footnotesize{D1}}&\includegraphics[width=.25\textwidth]{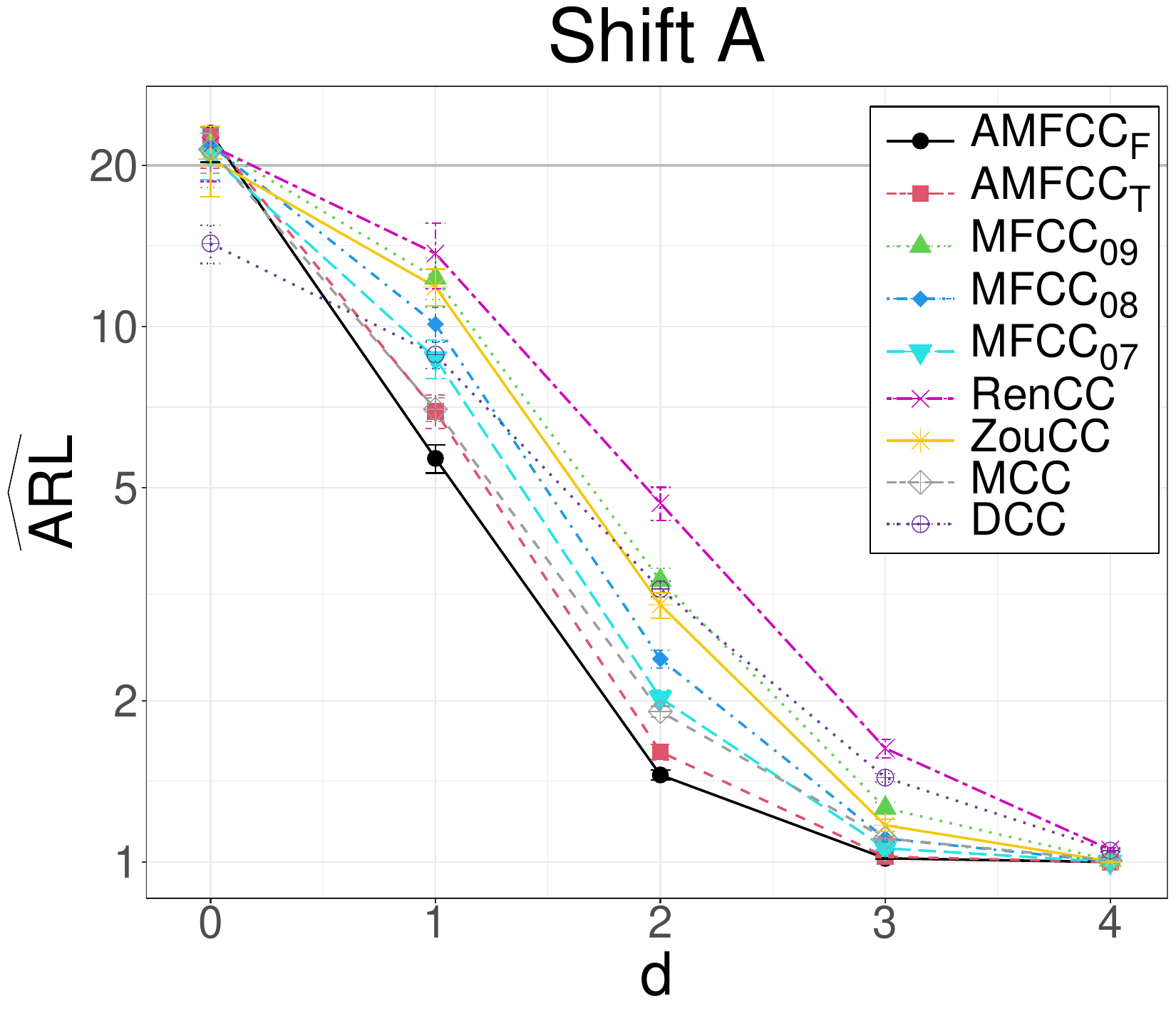}&\includegraphics[width=.25\textwidth]{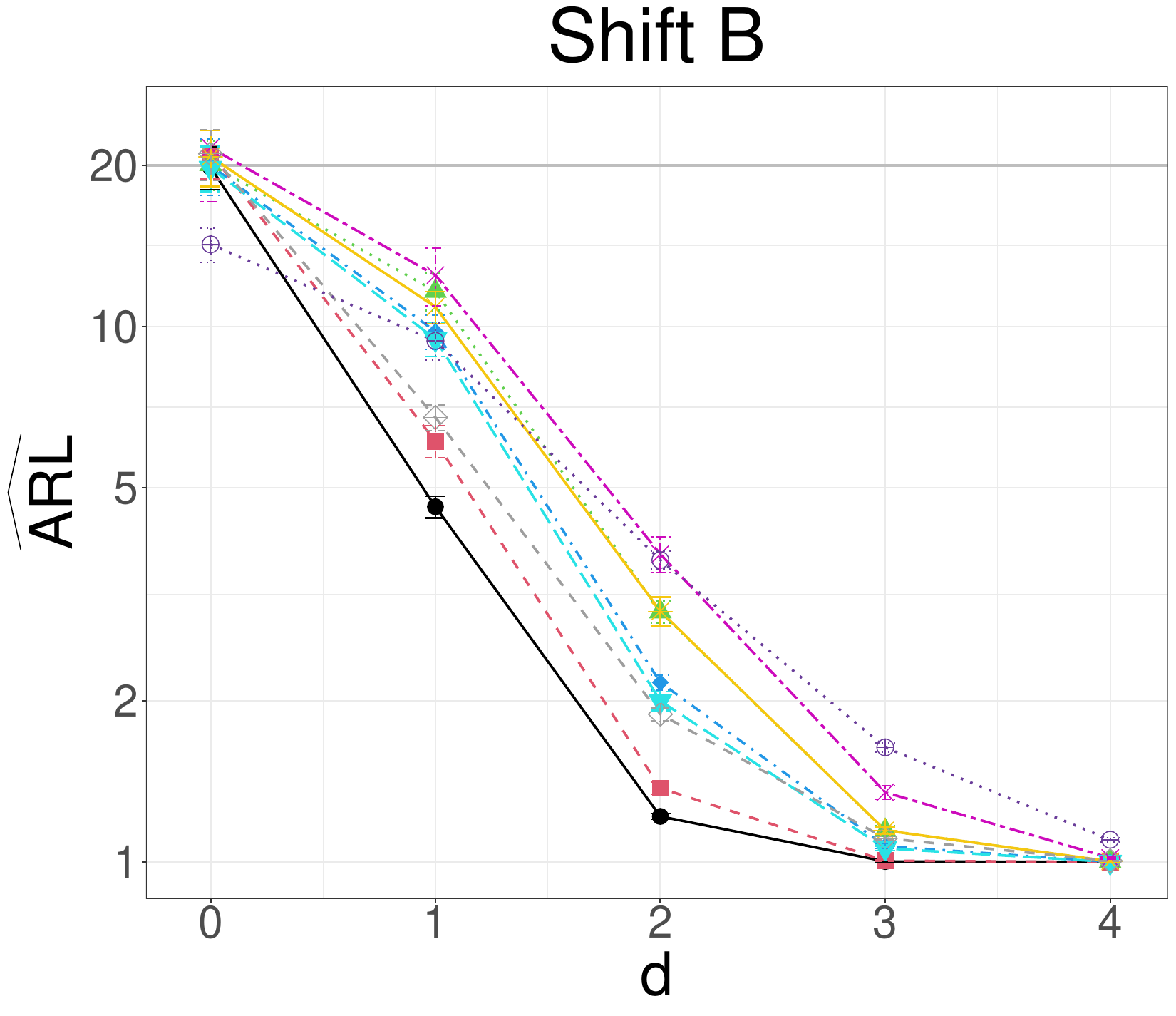}&\includegraphics[width=.25\textwidth]{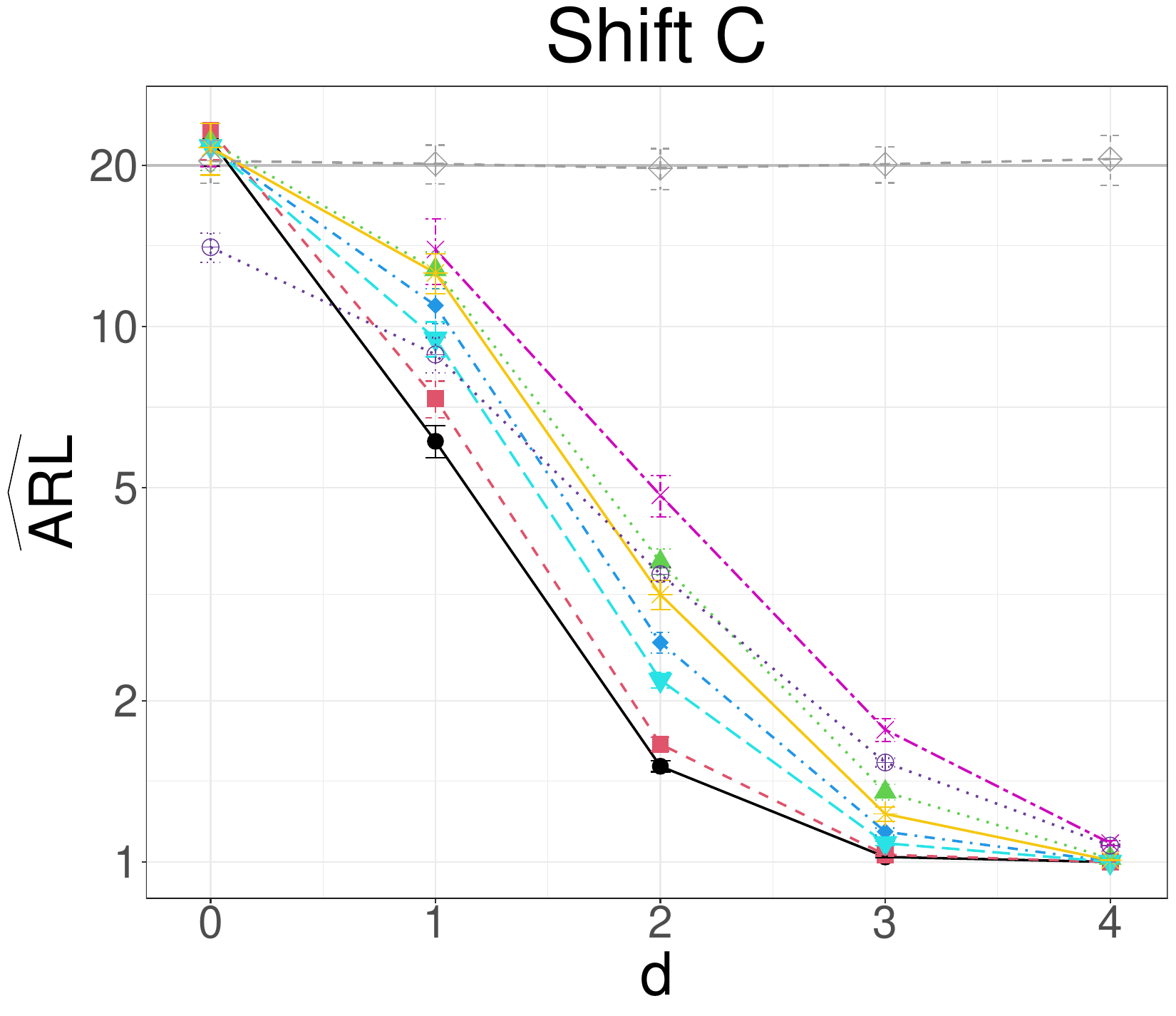}&\includegraphics[width=.25\textwidth]{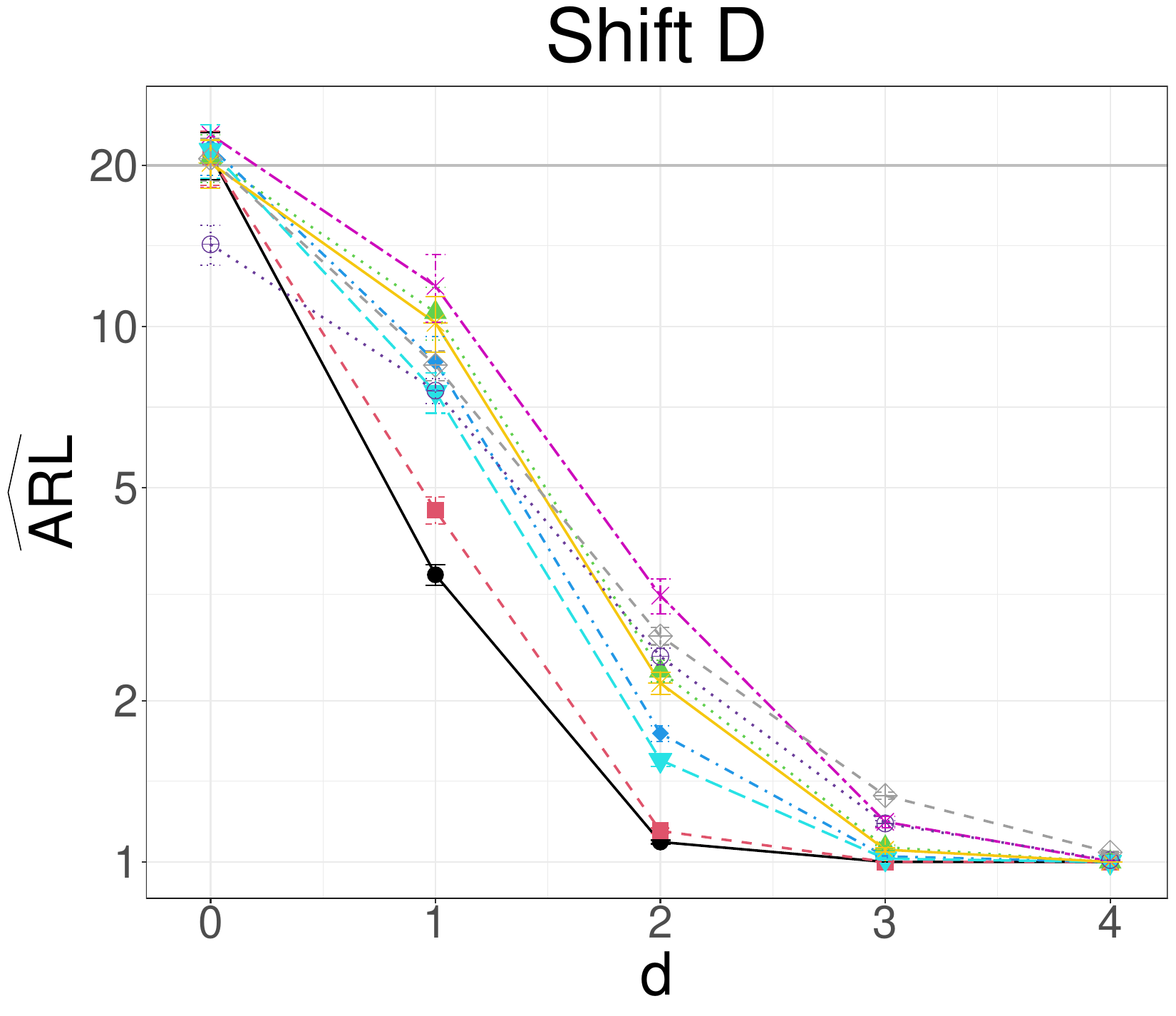}\\
		\textbf{\footnotesize{D2}}&\includegraphics[width=.25\textwidth]{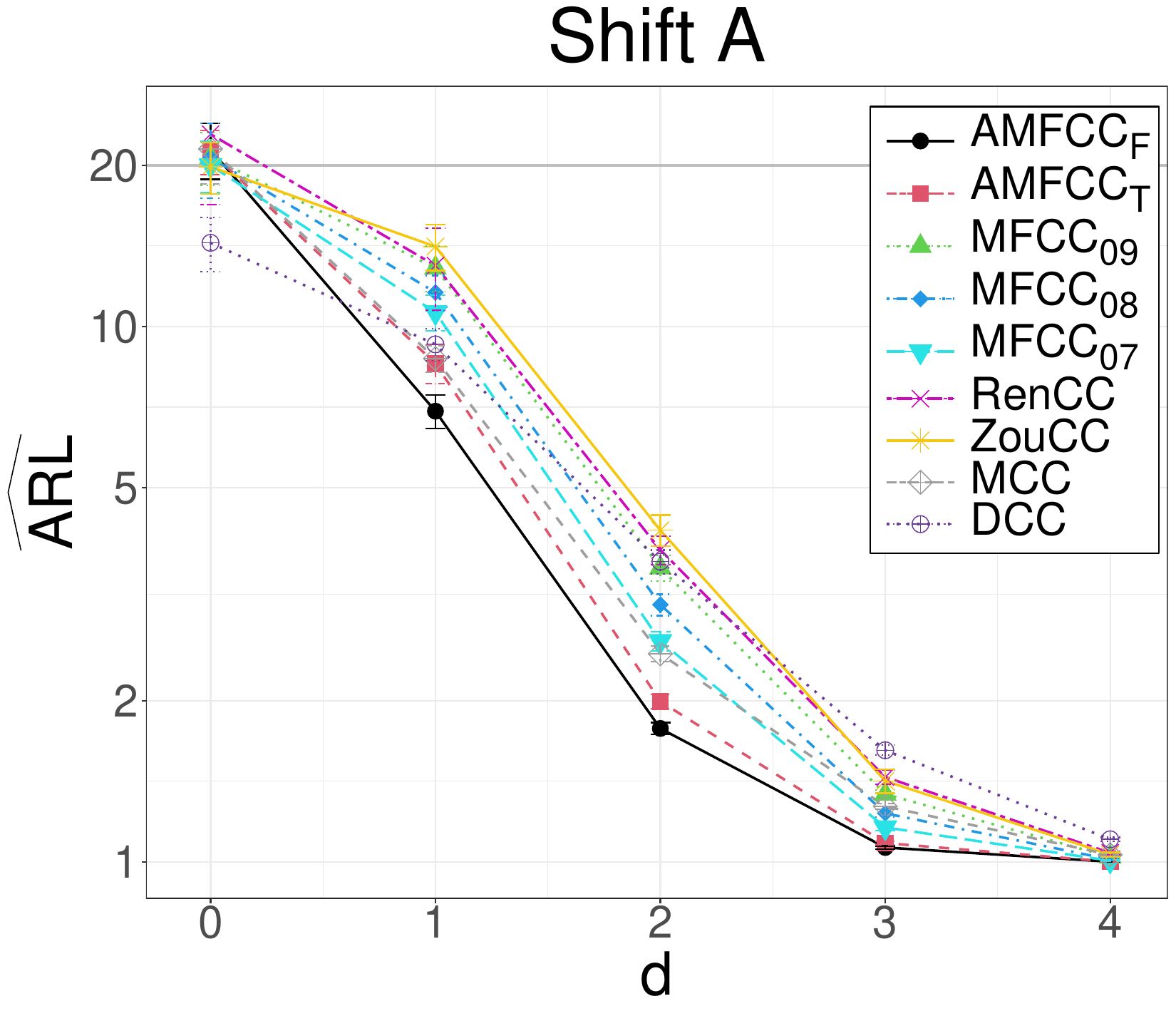}&\includegraphics[width=.25\textwidth]{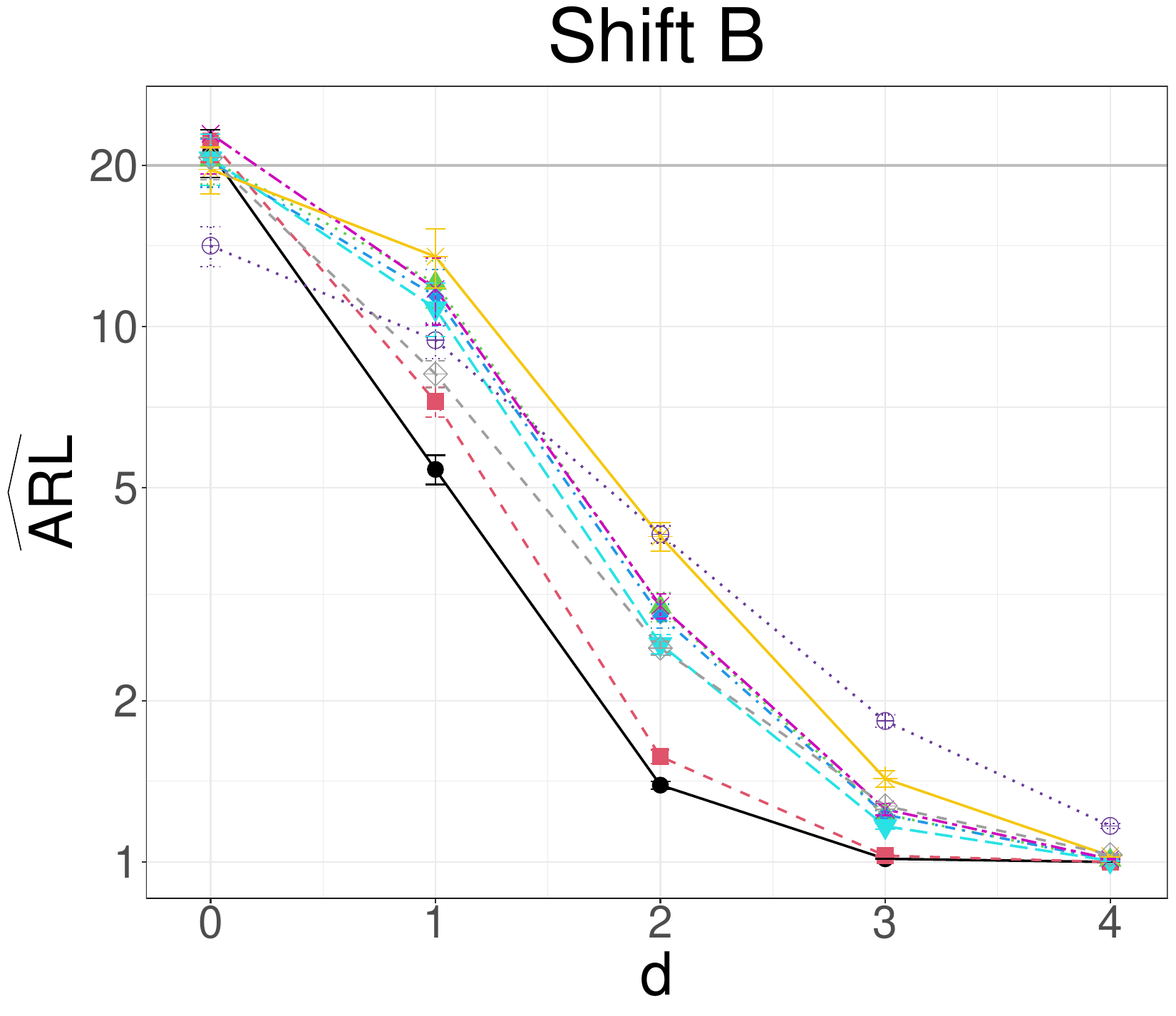}&\includegraphics[width=.25\textwidth]{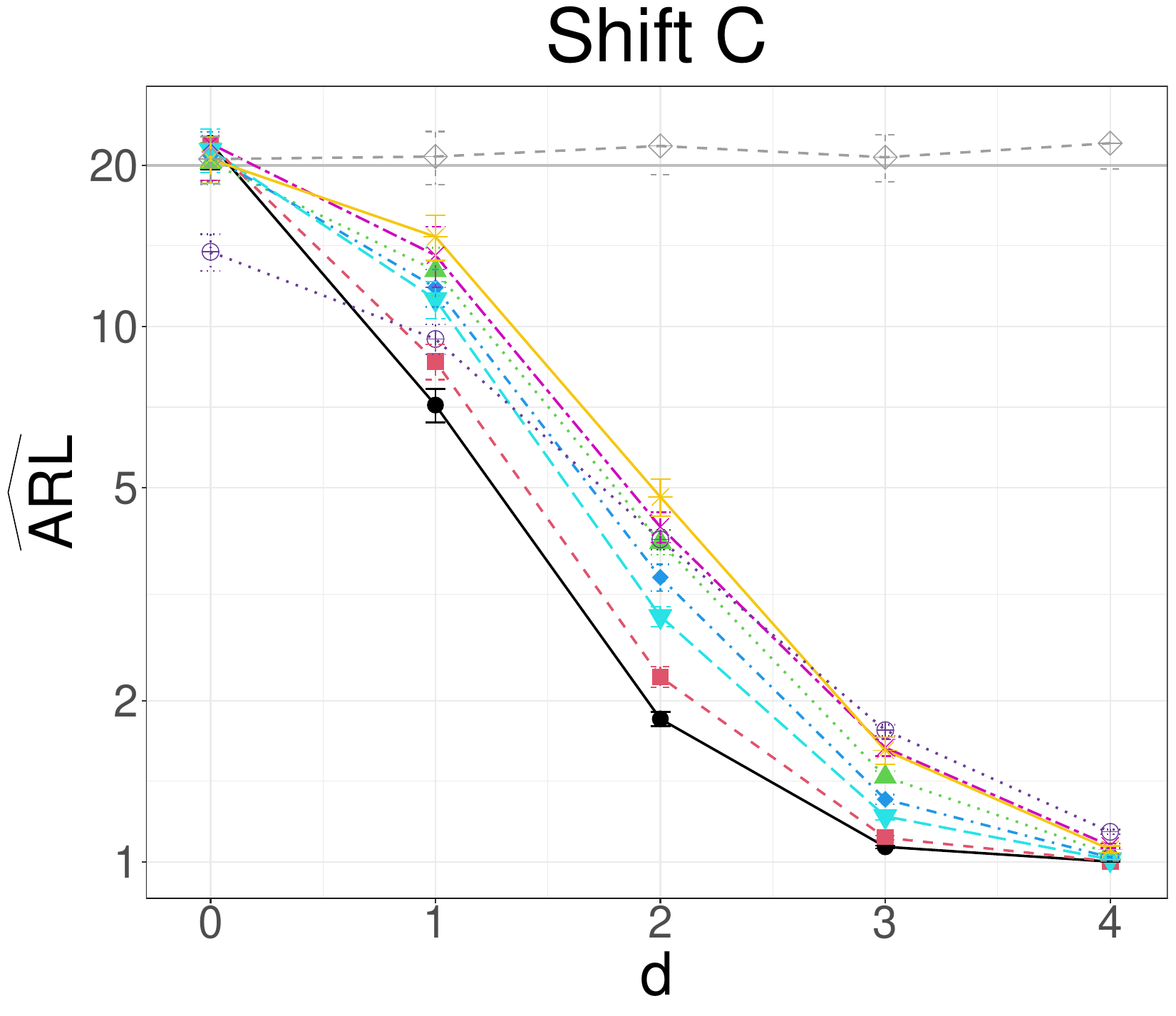}&\includegraphics[width=.25\textwidth]{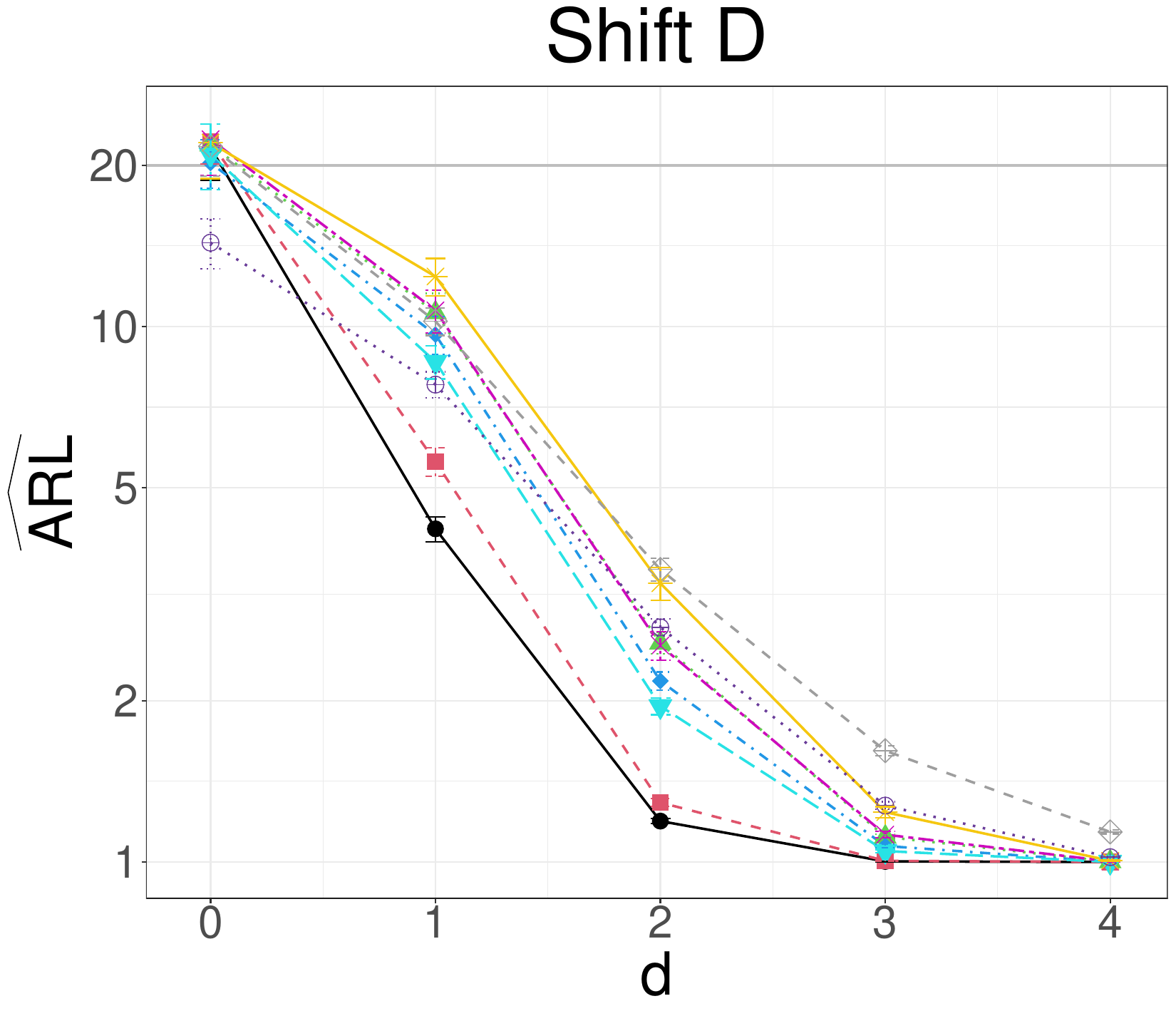}\\
		\textbf{\footnotesize{D3}}&\includegraphics[width=.25\textwidth]{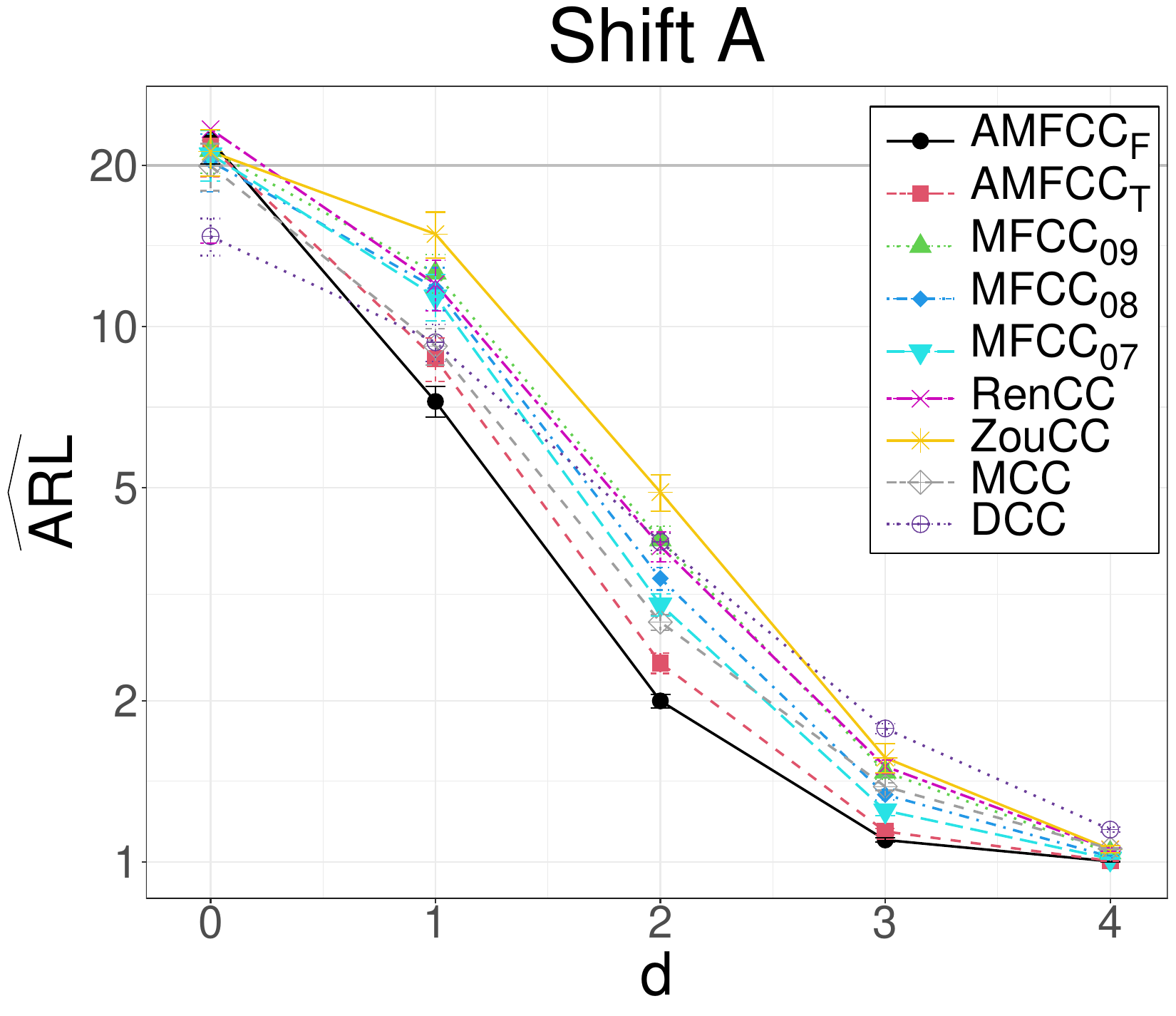}&\includegraphics[width=.25\textwidth]{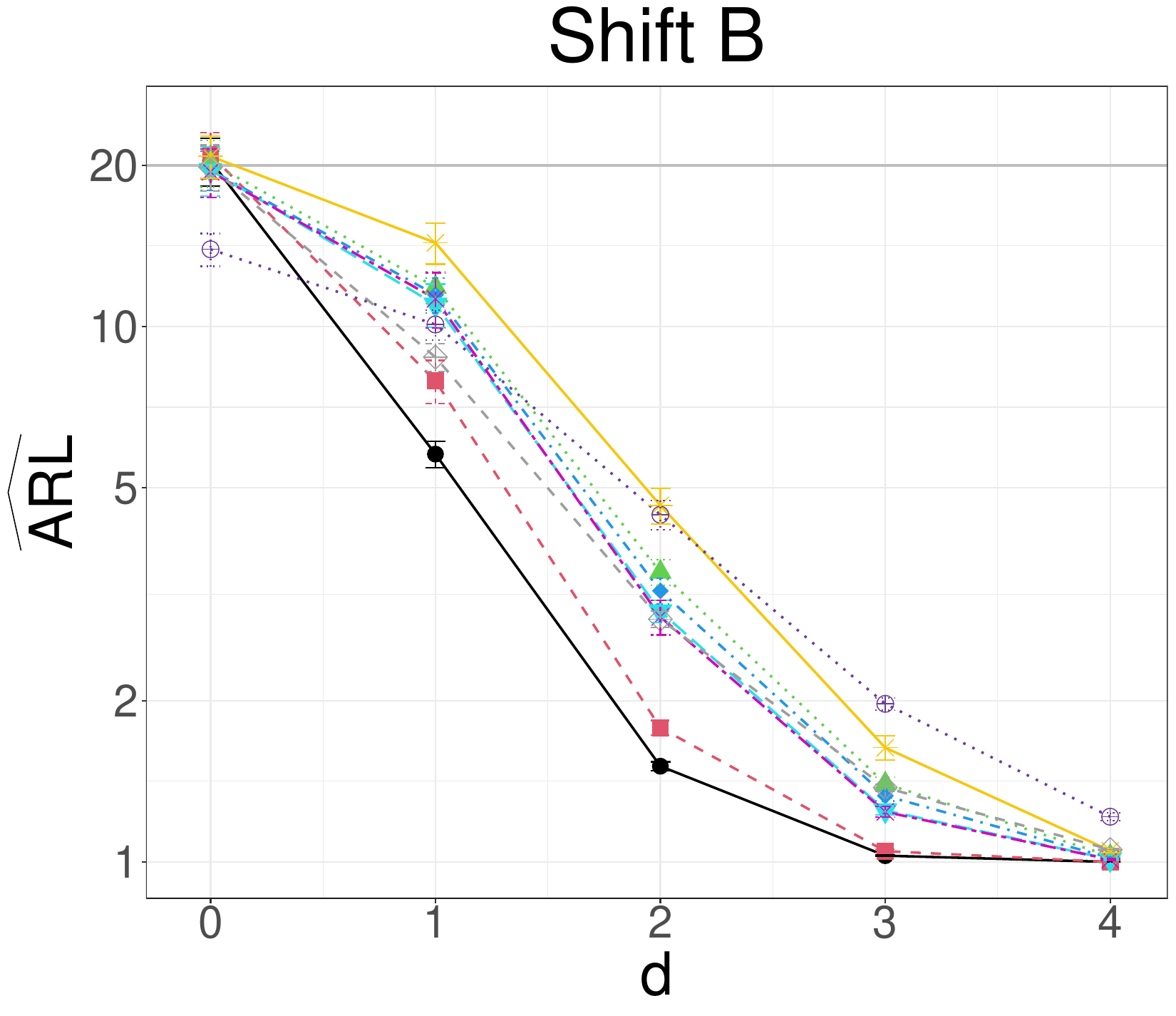}&\includegraphics[width=.25\textwidth]{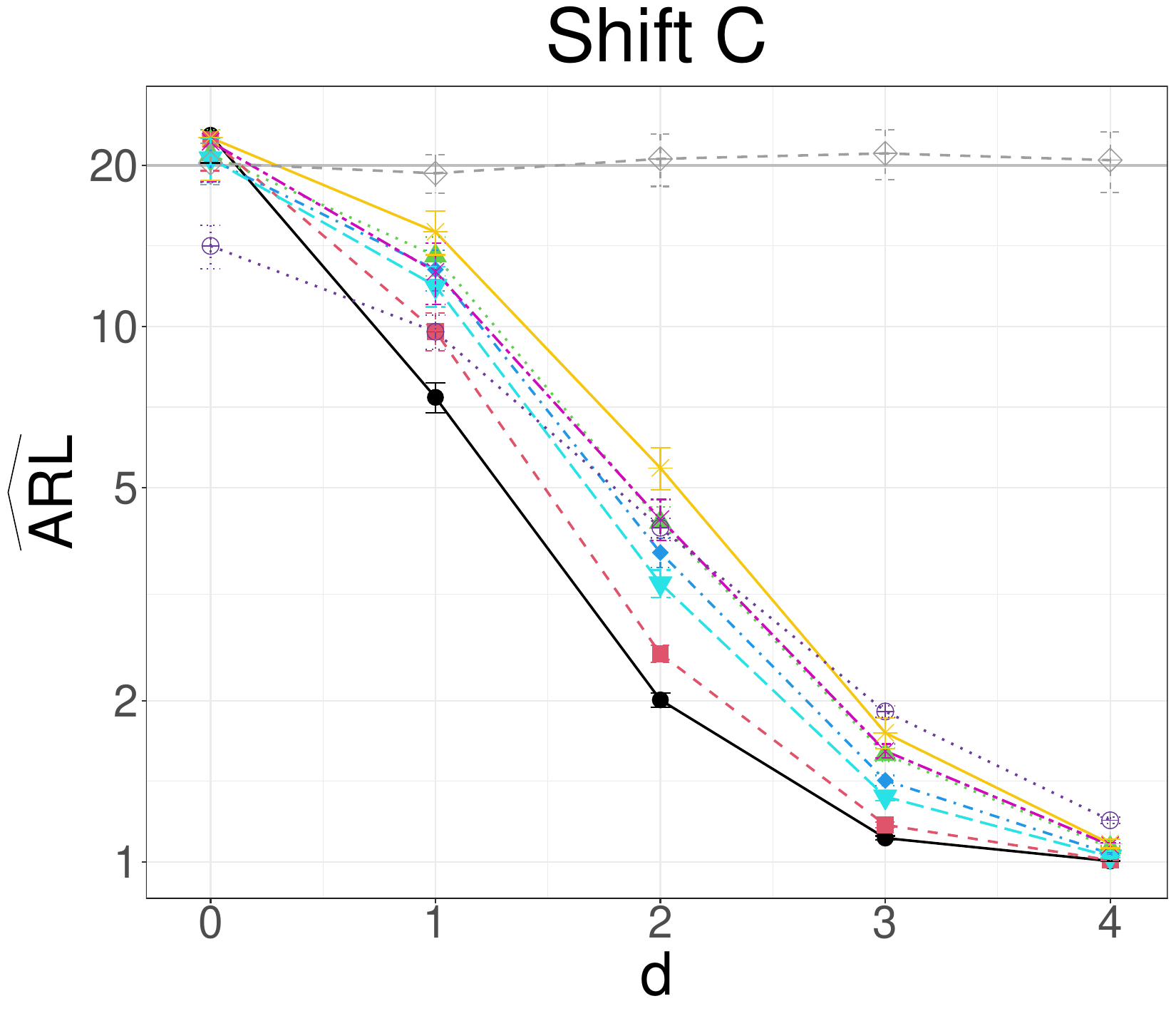}&\includegraphics[width=.25\textwidth]{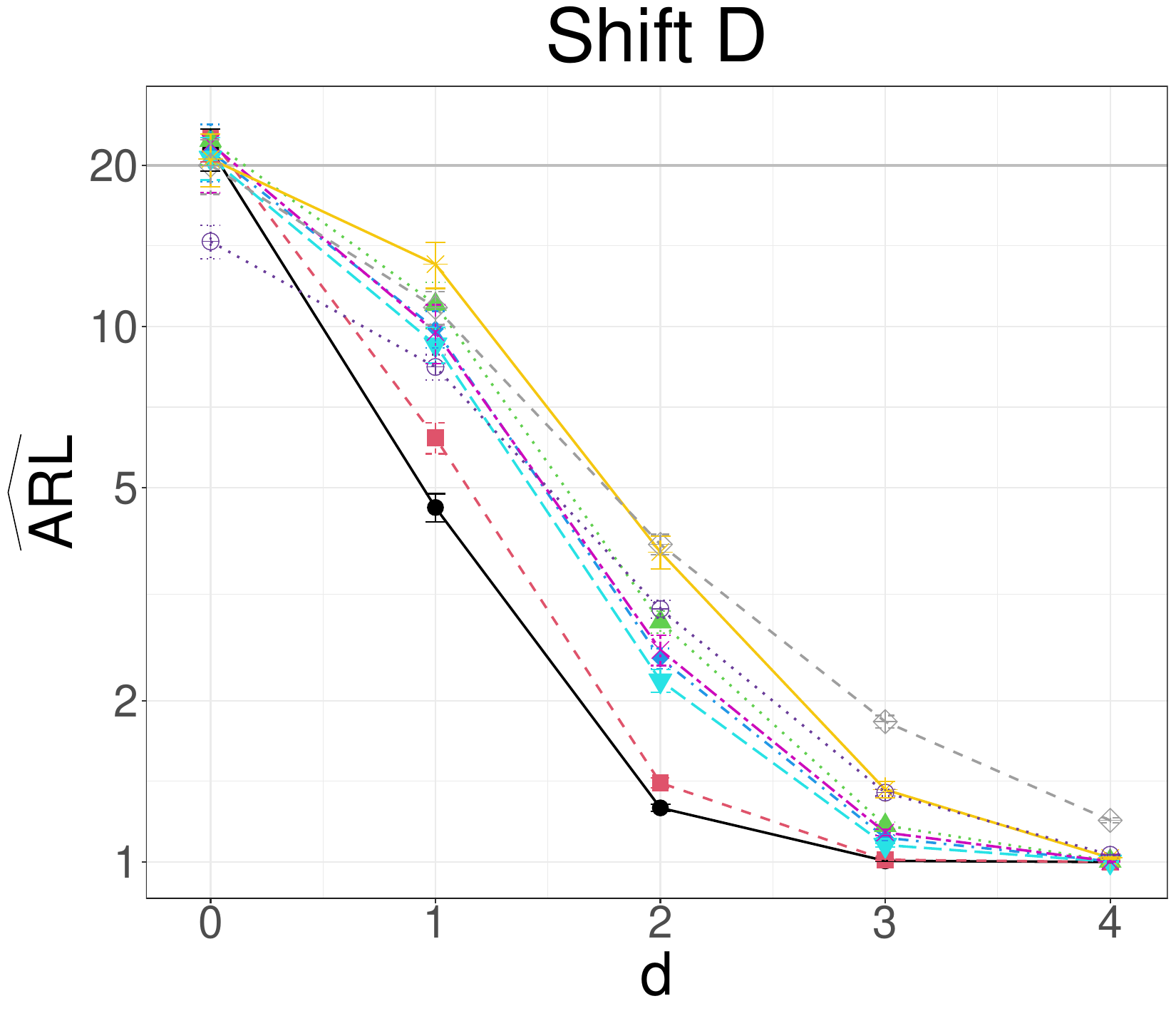}\\
		
	\end{tabular}
	\vspace{-.5cm}
\end{figure}

Figure  \ref{fi_results_2} displays the mean FAR ($ d=0 $) or TDR ($ d \neq 0 $) as a function of the severity level $d$ for each dependence level (D1, D2 and D3), and OC condition (Shift A, Shift B, Shift C, and Shift D). Results are very similar to those in Scenario 1.

The proposed AMFCC\textsubscript{F} and  AMFCC\textsubscript{T}, which are implemented as described in Section 3, clearly outperform all the competing methods.
By comparing Figure 1 and Figure 2, results appear as not influenced by the different covariance structure.
The MFCC methods show overall similar performance, even though MFCC\textsubscript{07} seems to perform slightly better than MFCC\textsubscript{09}, and MFCC\textsubscript{08}. The MCC shows a better performance for Shift A and  Shift B whereas the DCC  achieves very low TDR.

\begin{figure}[h]
	\caption{Mean cFAR ($ d=0 $) or cTDR ($ d \neq 0 $) achieved by AMFCC\textsubscript{F},  AMFCC\textsubscript{T}, MFCC\textsubscript{09}, MFCC\textsubscript{08}, and  MFCC\textsubscript{07} for each dependence level (D1, D2 and D3), OC condition (Shift A, Shift B, Shift C, and Shift D) as a function of the severity level $d$  in Scenario 2.}
	
	\label{fi_results_vs2}
	
	\centering
	\hspace{-2.1cm}
	\begin{tabular}{cM{0.24\textwidth}M{0.24\textwidth}M{0.24\textwidth}M{0.24\textwidth}}
		\textbf{\footnotesize{D1}}&\includegraphics[width=.25\textwidth]{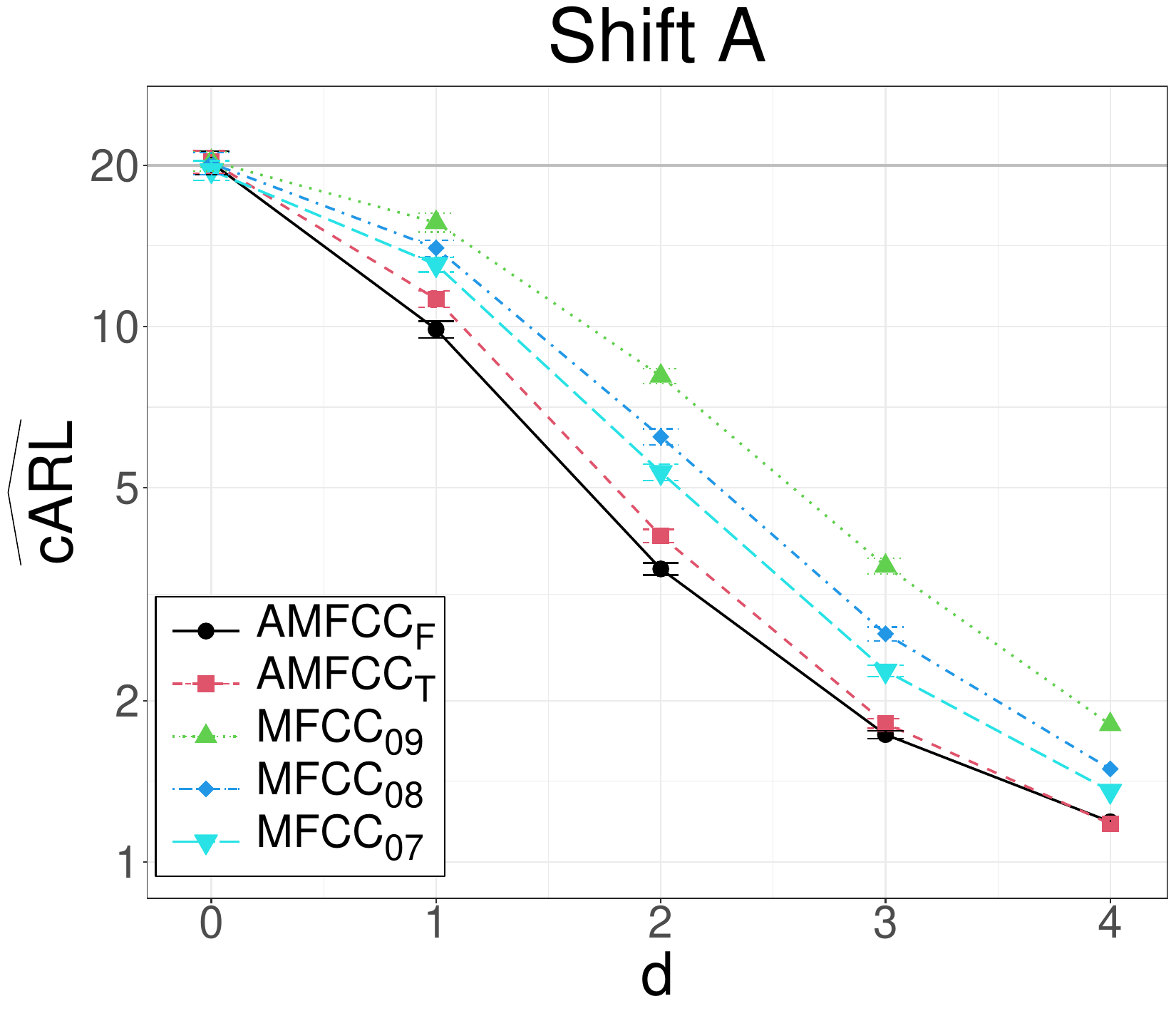}&\includegraphics[width=.25\textwidth]{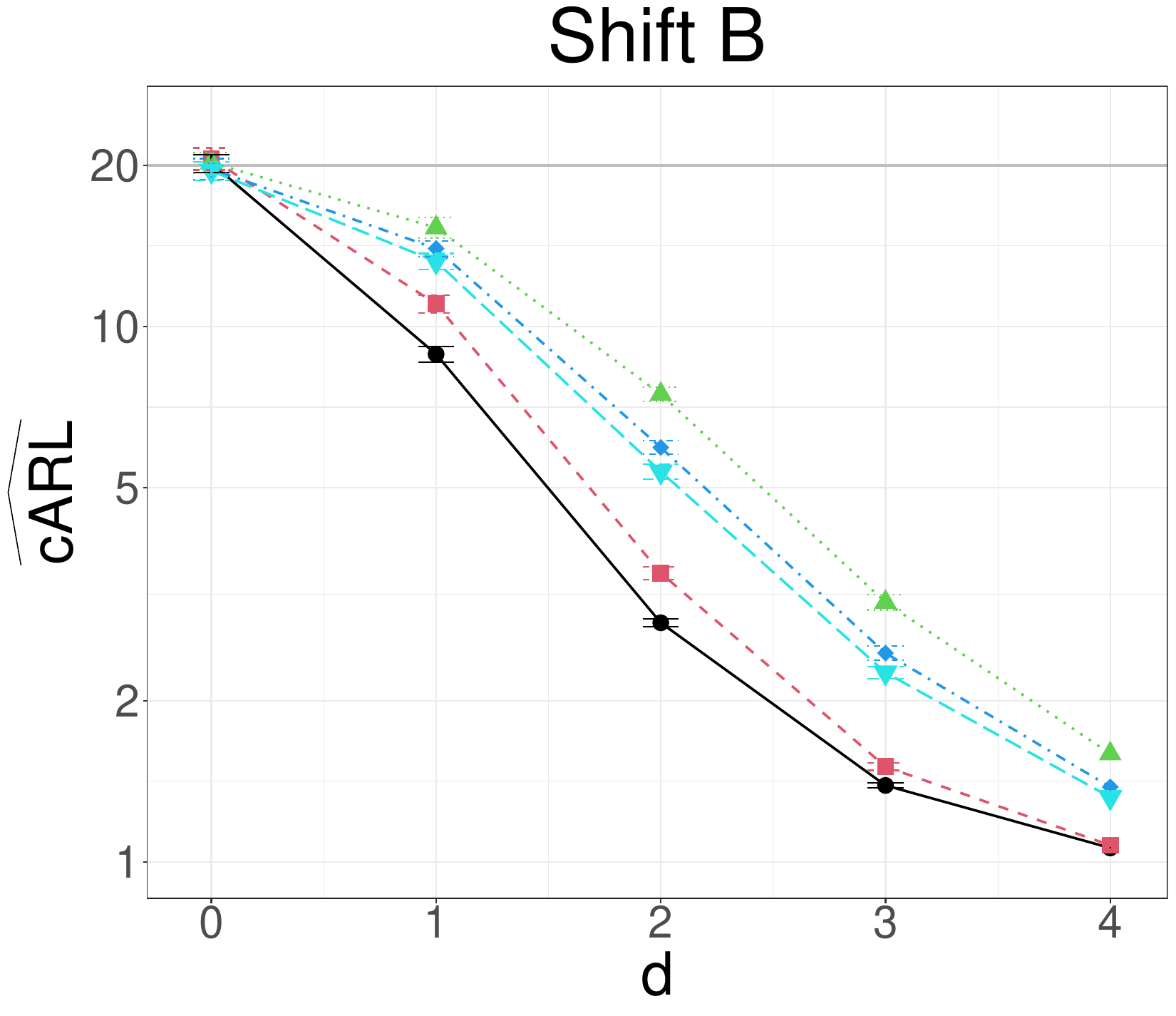}&\includegraphics[width=.25\textwidth]{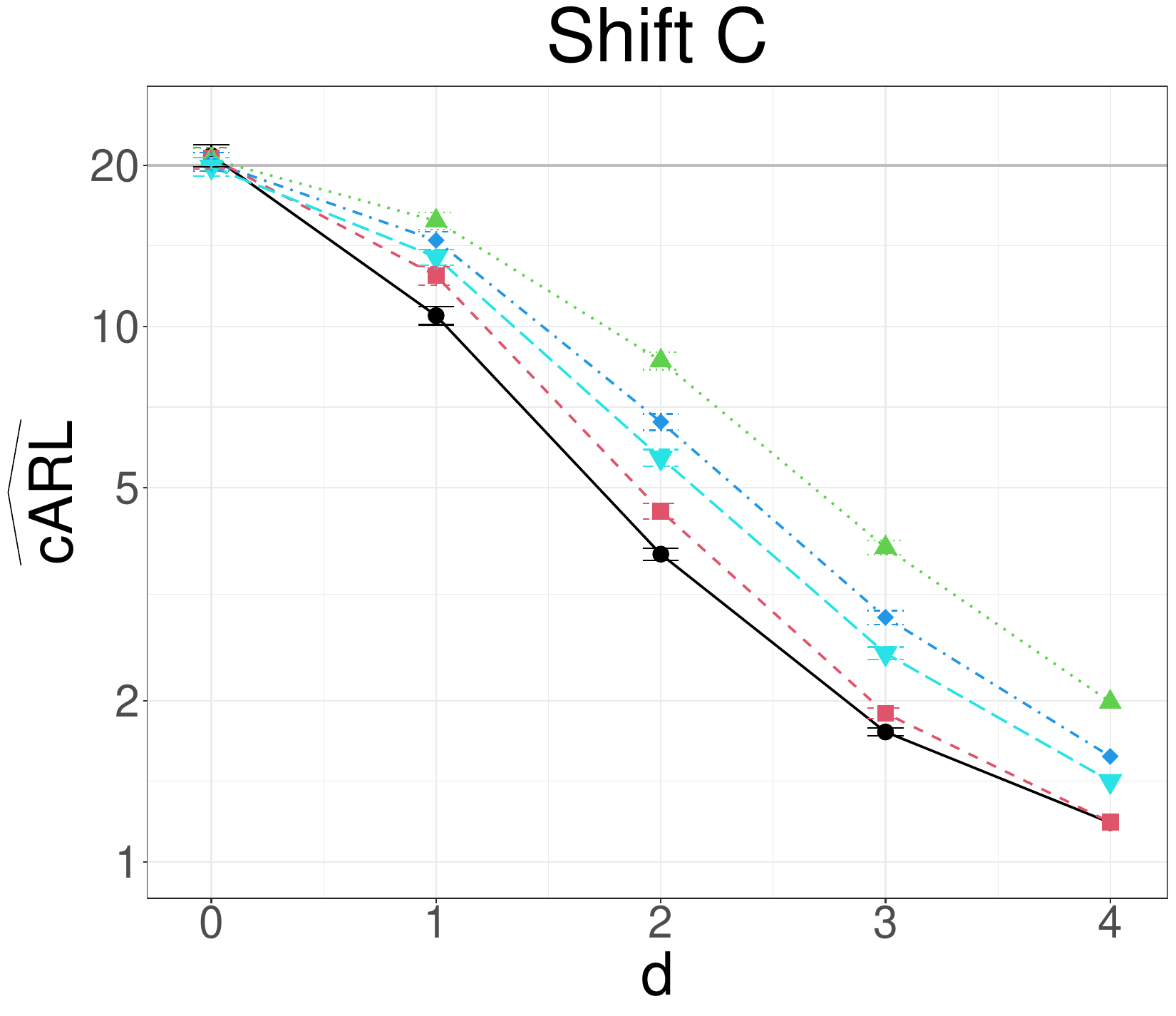}&\includegraphics[width=.25\textwidth]{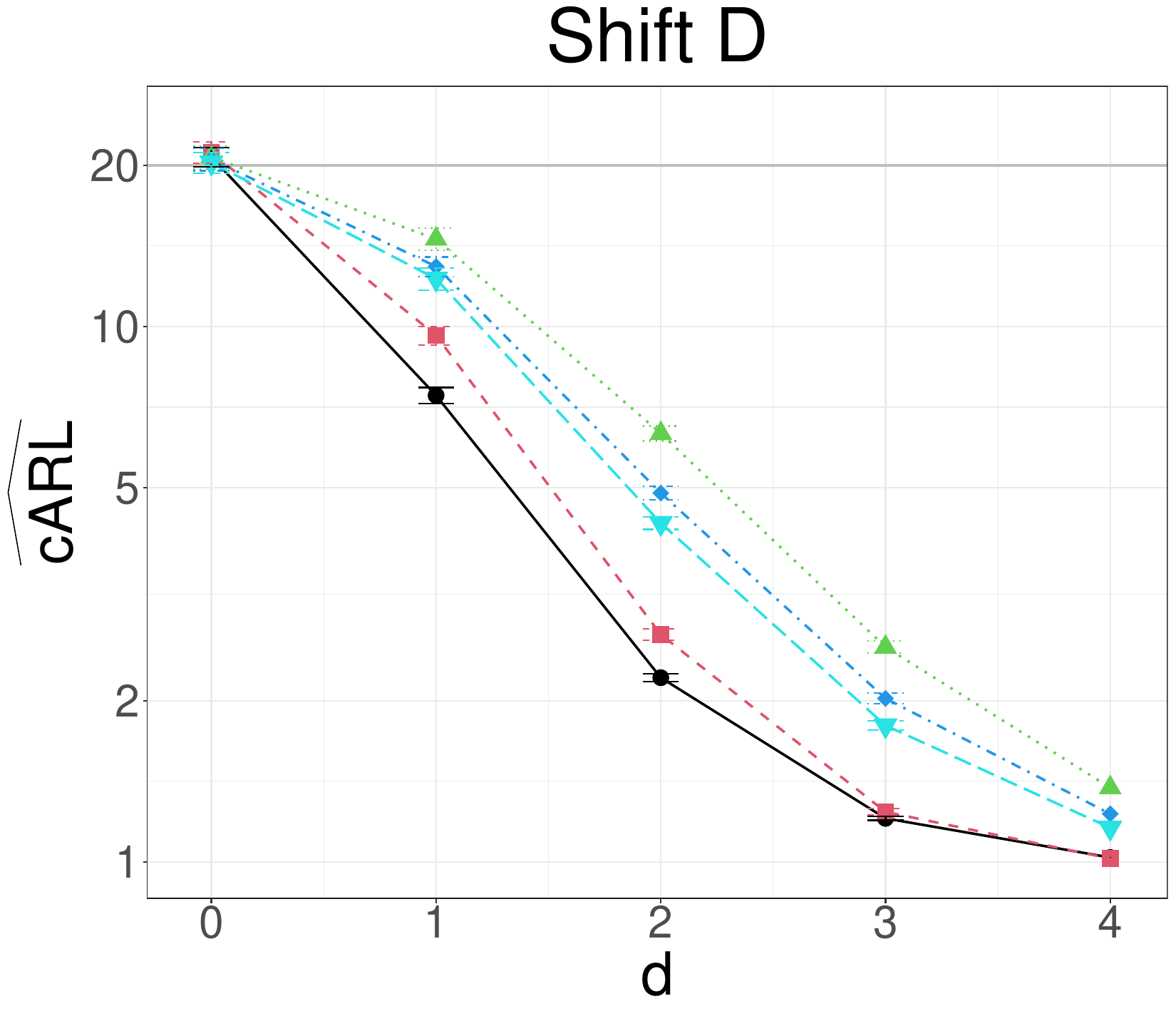}\\
		\textbf{\footnotesize{D2}}&\includegraphics[width=.25\textwidth]{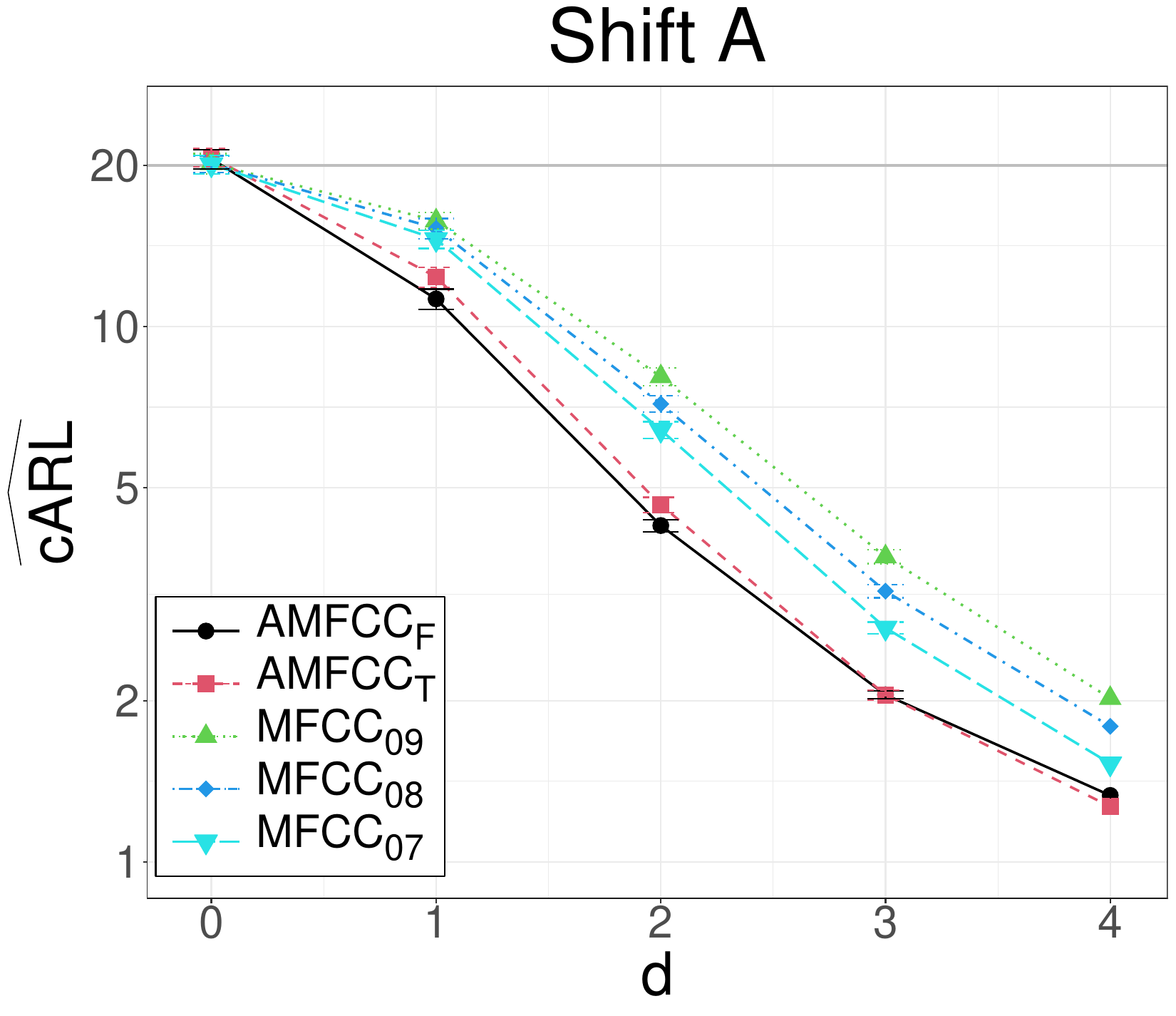}&\includegraphics[width=.25\textwidth]{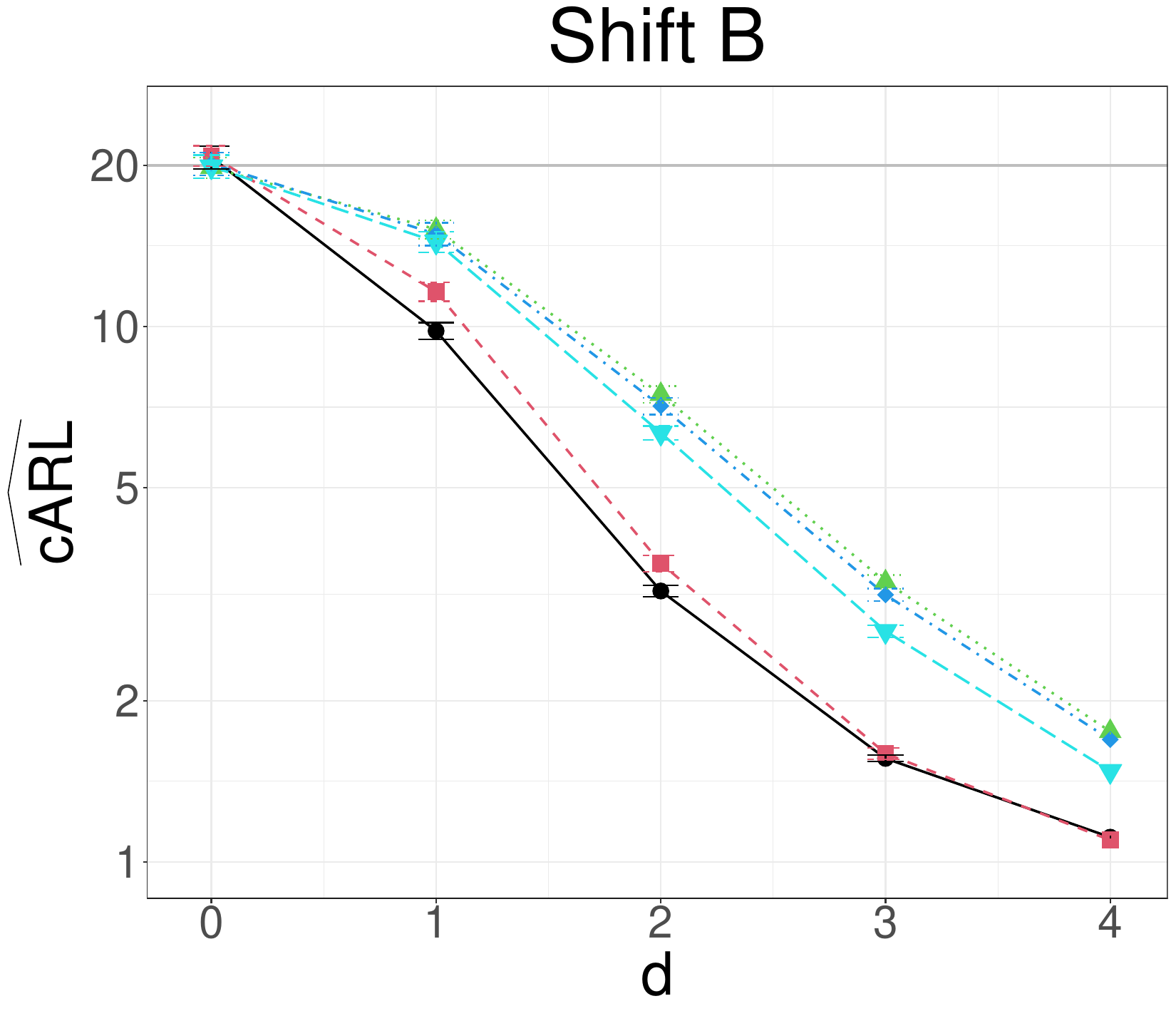}&\includegraphics[width=.25\textwidth]{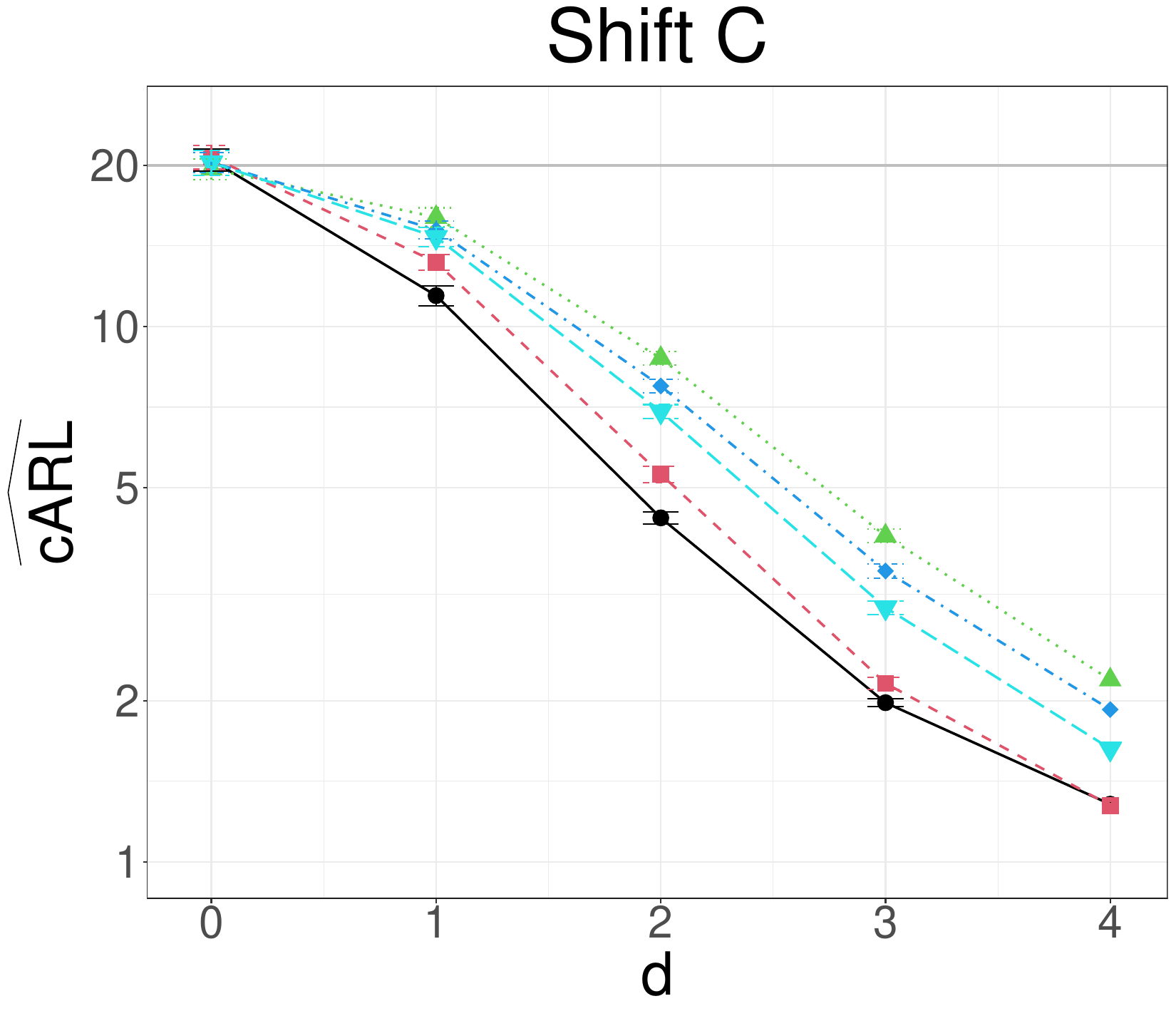}&\includegraphics[width=.25\textwidth]{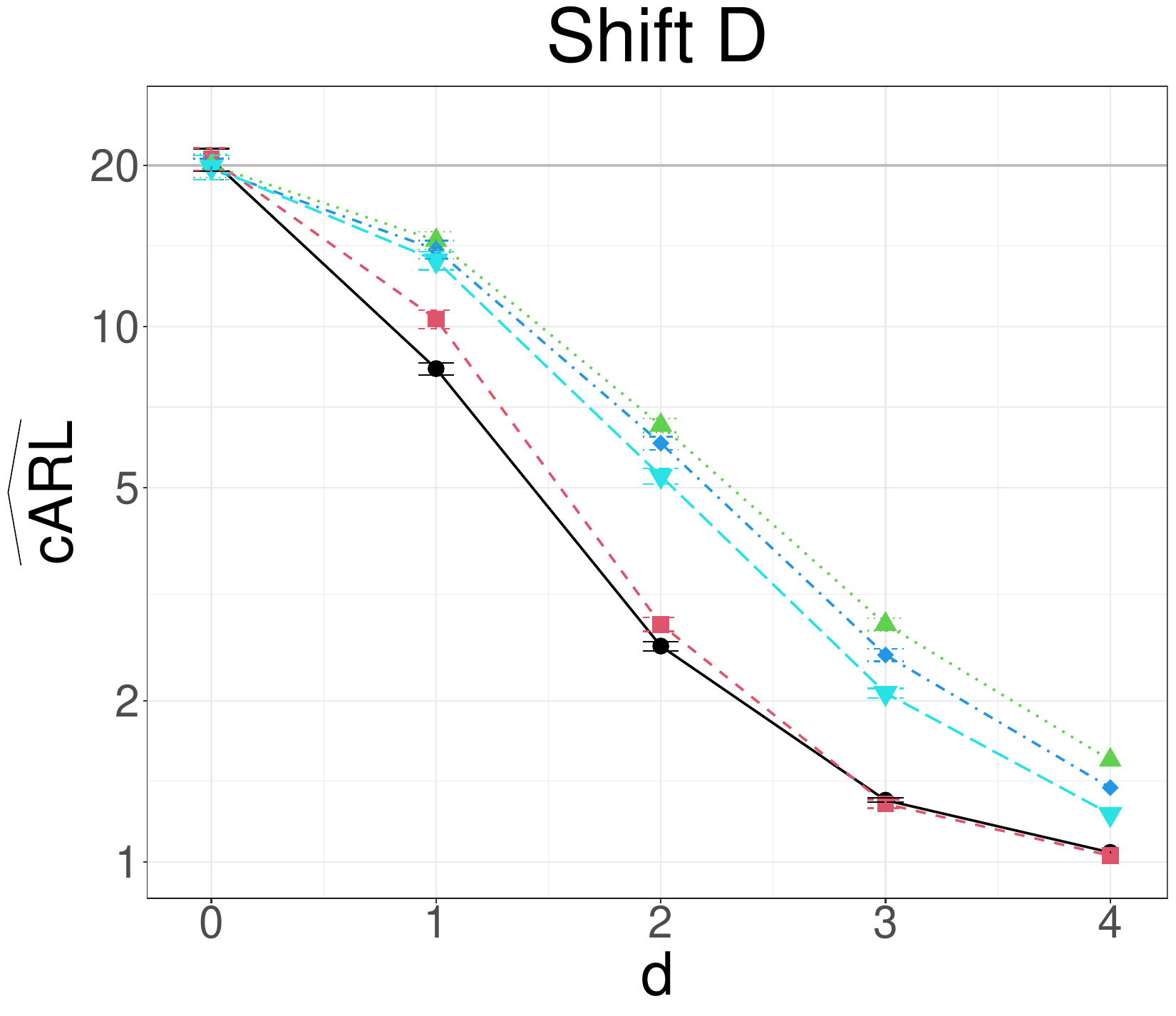}\\
		\textbf{\footnotesize{D3}}&\includegraphics[width=.25\textwidth]{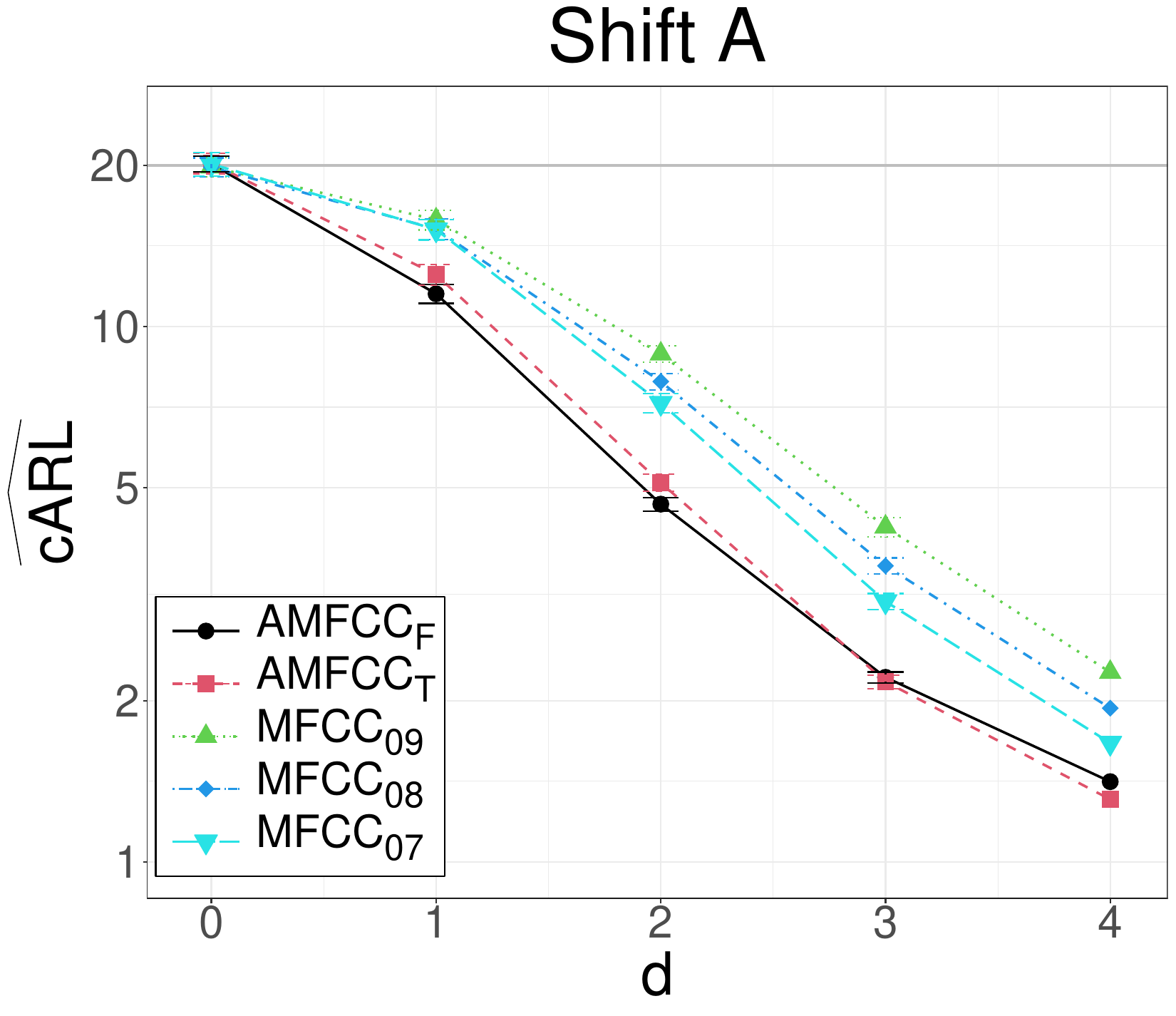}&\includegraphics[width=.25\textwidth]{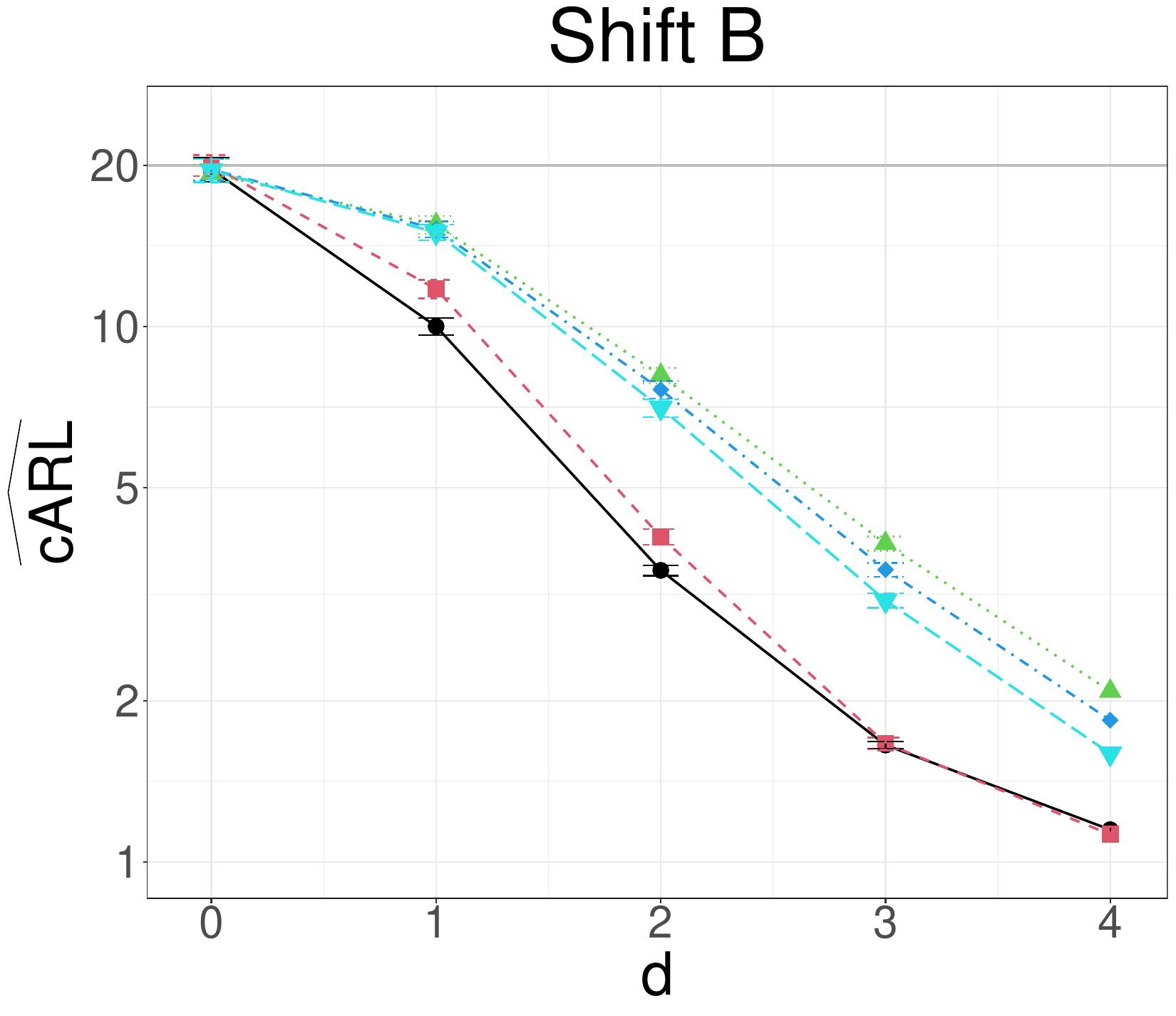}&\includegraphics[width=.25\textwidth]{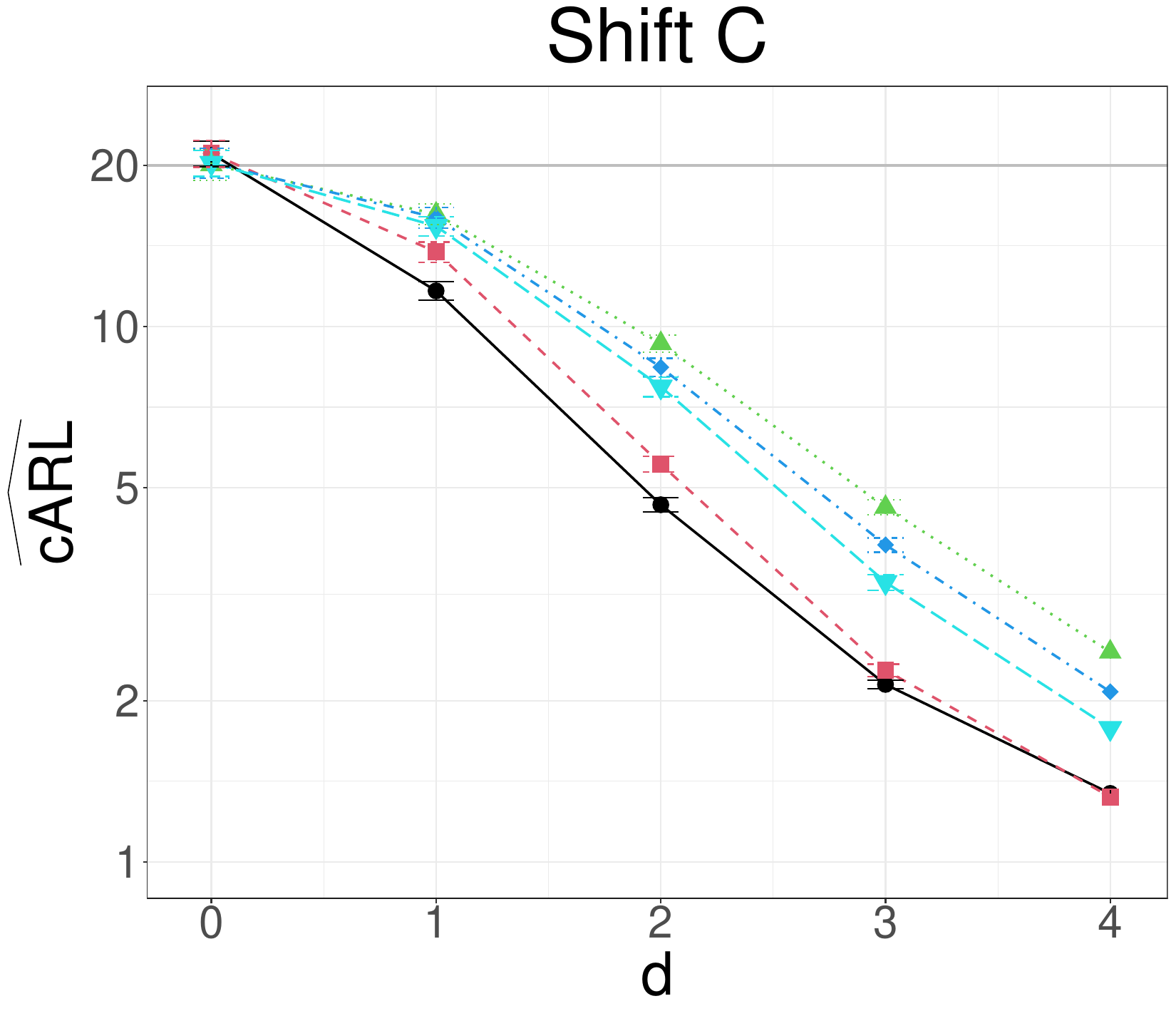}&\includegraphics[width=.25\textwidth]{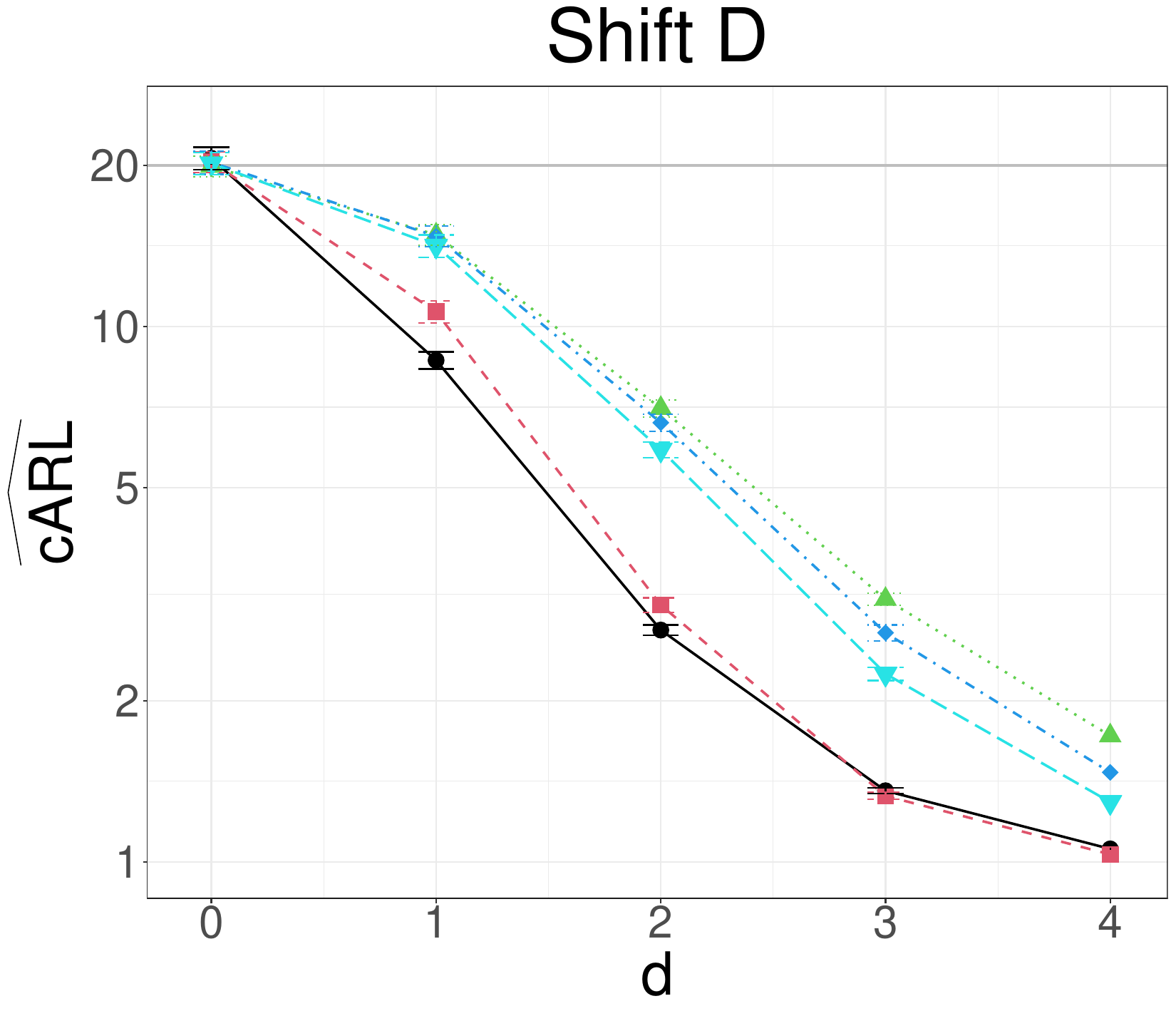}\\
		
	\end{tabular}
	\vspace{-.5cm}
\end{figure}

In addition, Figure \ref{fi_results_vs2}  displays the mean cFAR ($ d=0 $) or cTDR ($ d \neq 0 $)  as a function of the severity level $d$ for each dependence level (D1, D2 and D3), OC condition (Shift A, Shift B, Shift C, and Shift D). 
From this figure, the proposed diagnostic procedures  AMFCC\textsubscript{F} and  AMFCC\textsubscript{T} clearly appear to largely outperform  the competing approaches in identifying components responsible for the OC condition, and agree with results 
displayed for Scenario 1.

\section{Additional Plots in the Real-case Study.}
\begin{figure}[h]
	\caption{The 354 IC observations of the tuning set in the real-case study.}
	
	\label{fi_csvar}
	
	\centering
		\begin{tabular}{cc}
		\includegraphics[width=.4\textwidth]{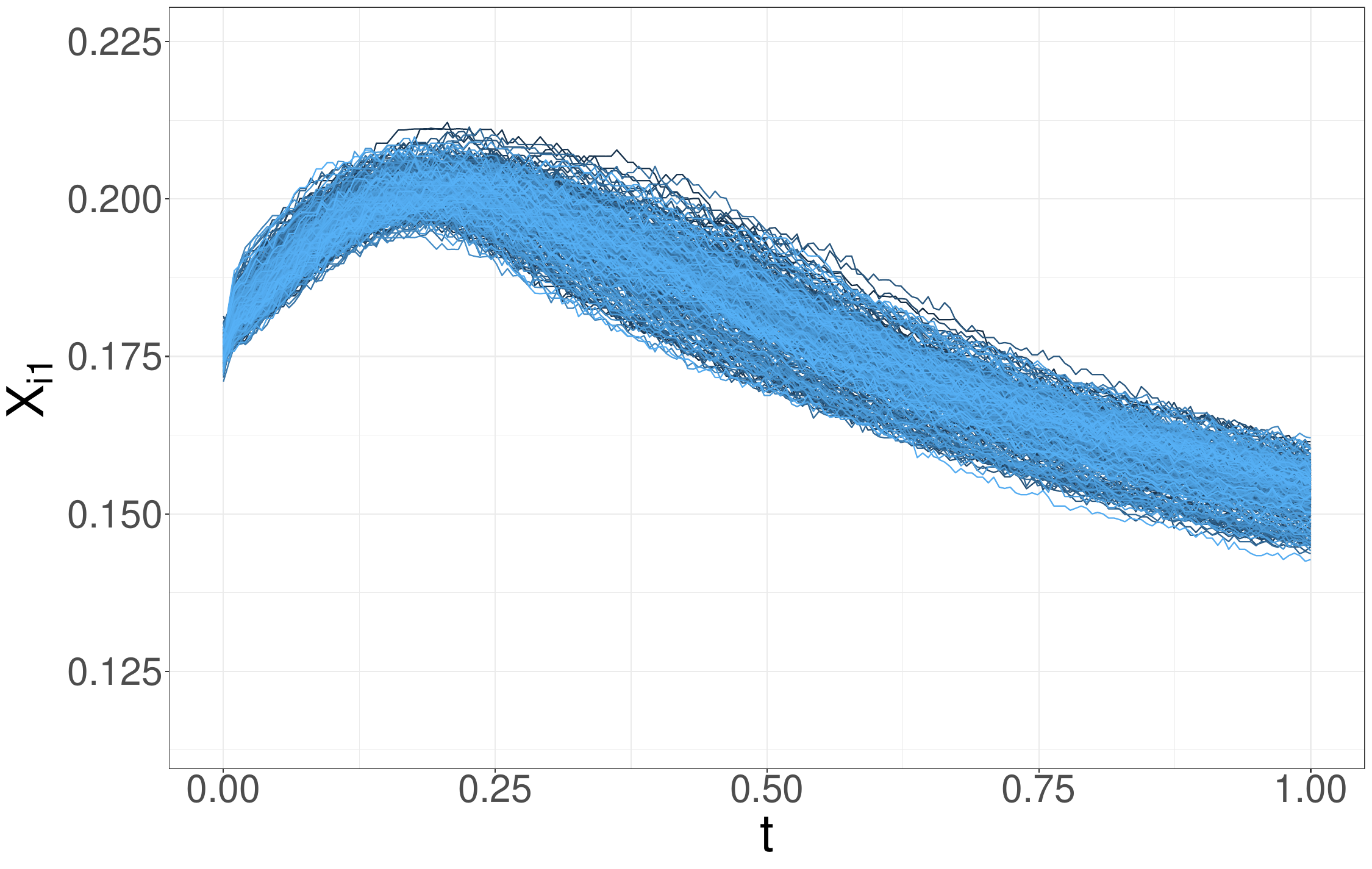}&\includegraphics[width=0.4\textwidth]{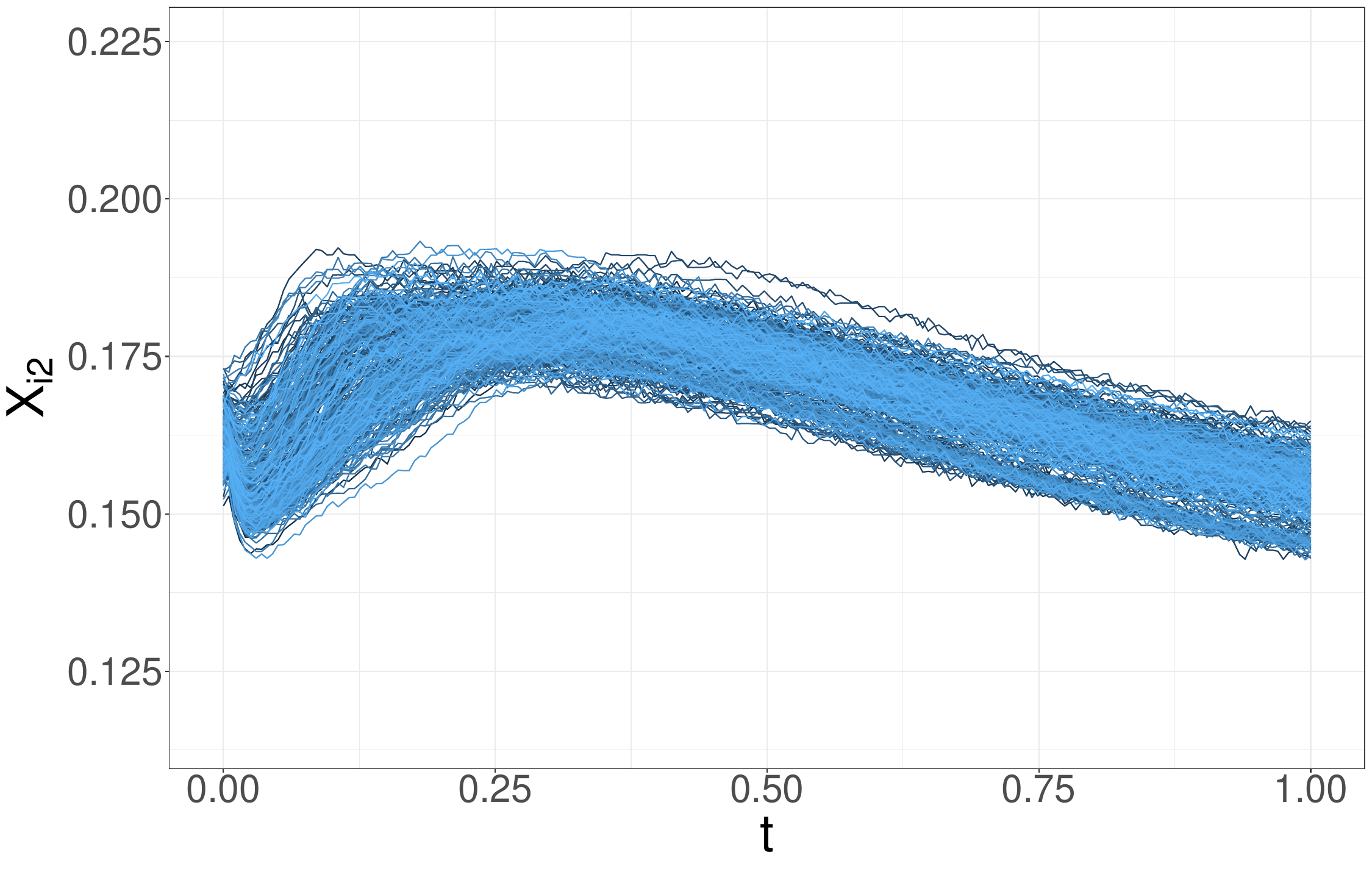}\\\includegraphics[width=0.4\textwidth]{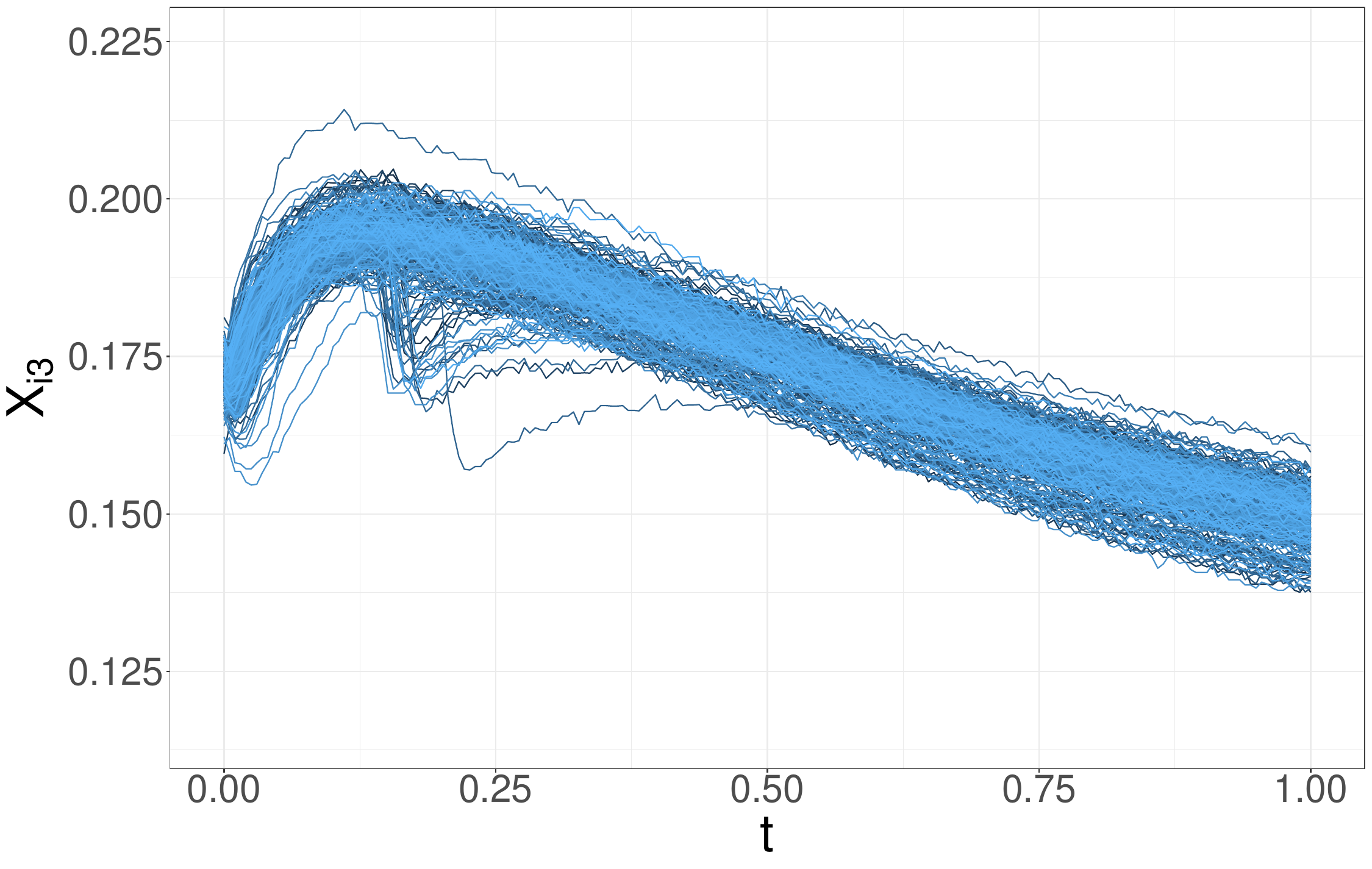}&\includegraphics[width=0.4\textwidth]{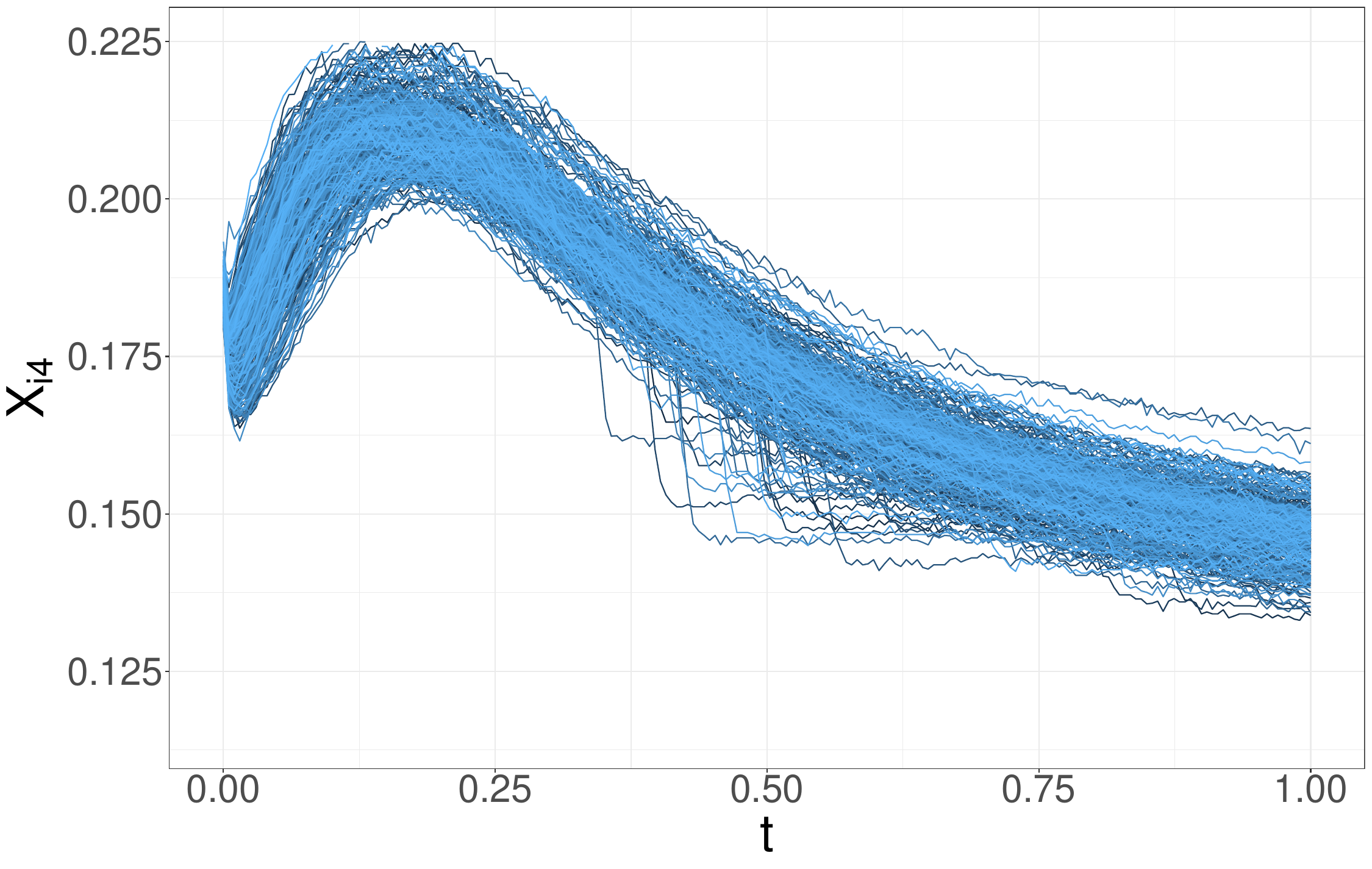}\\\includegraphics[width=0.4\textwidth]{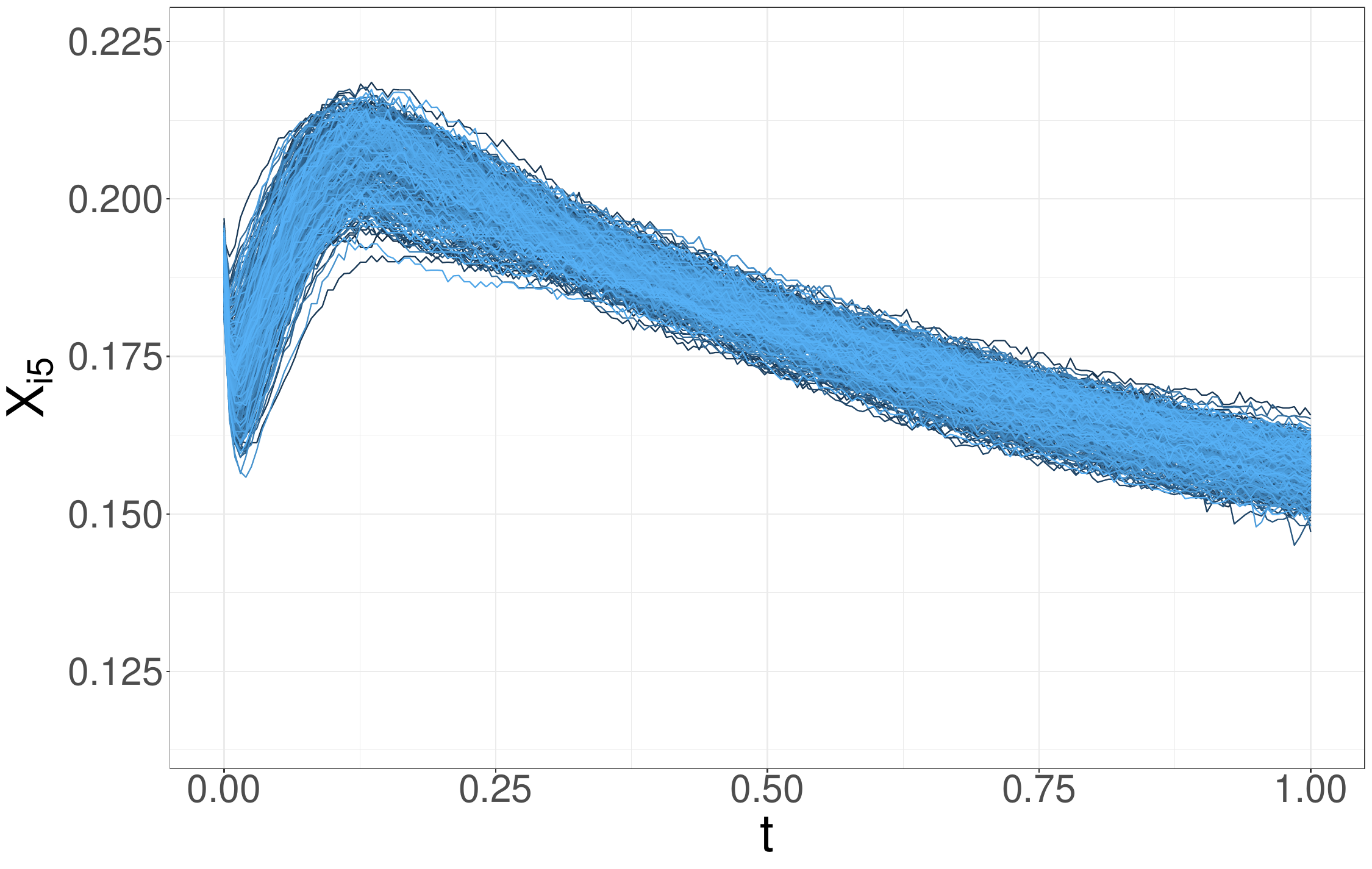}&
		\includegraphics[width=0.4\textwidth]{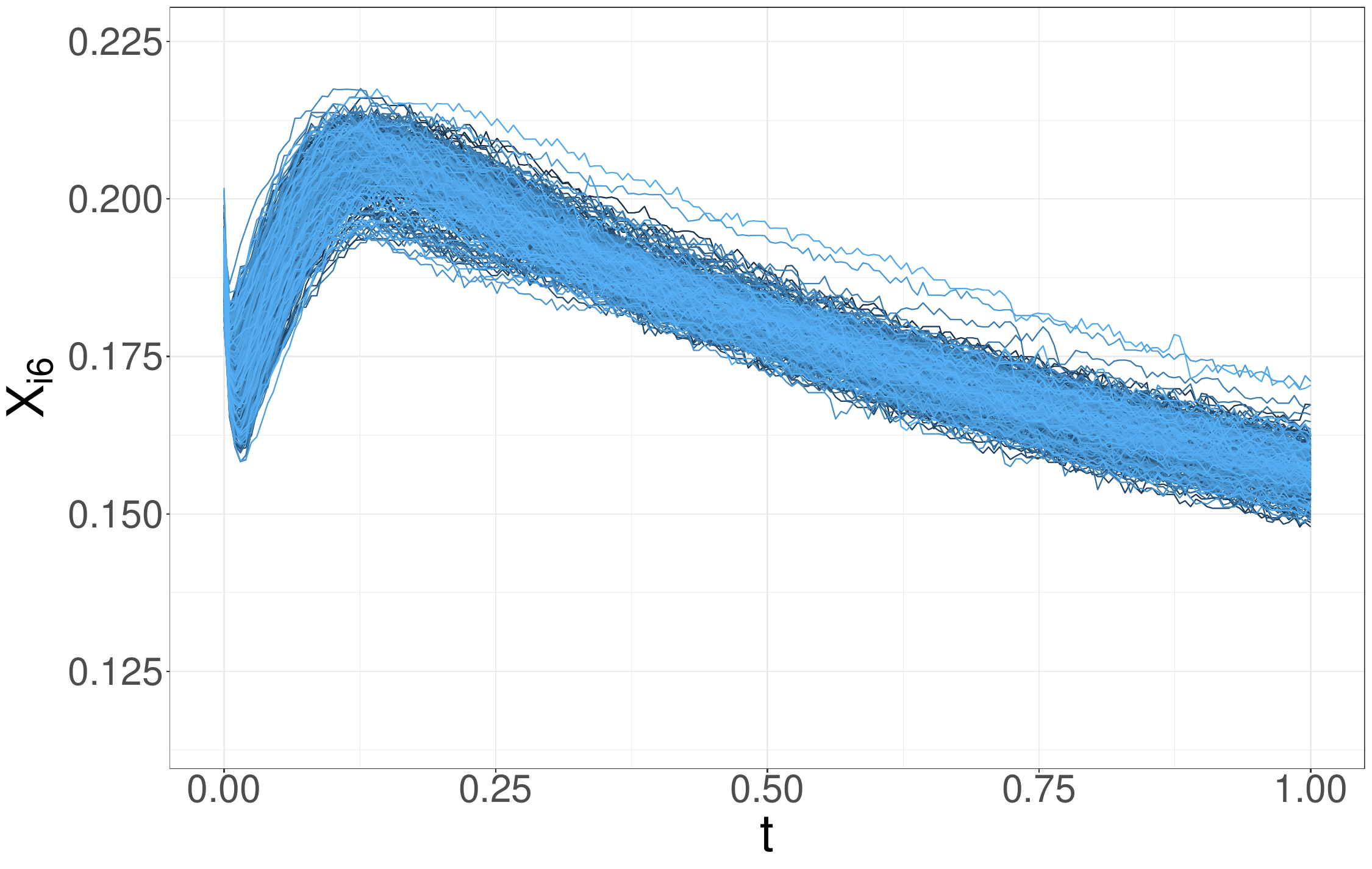}\\\includegraphics[width=0.4\textwidth]{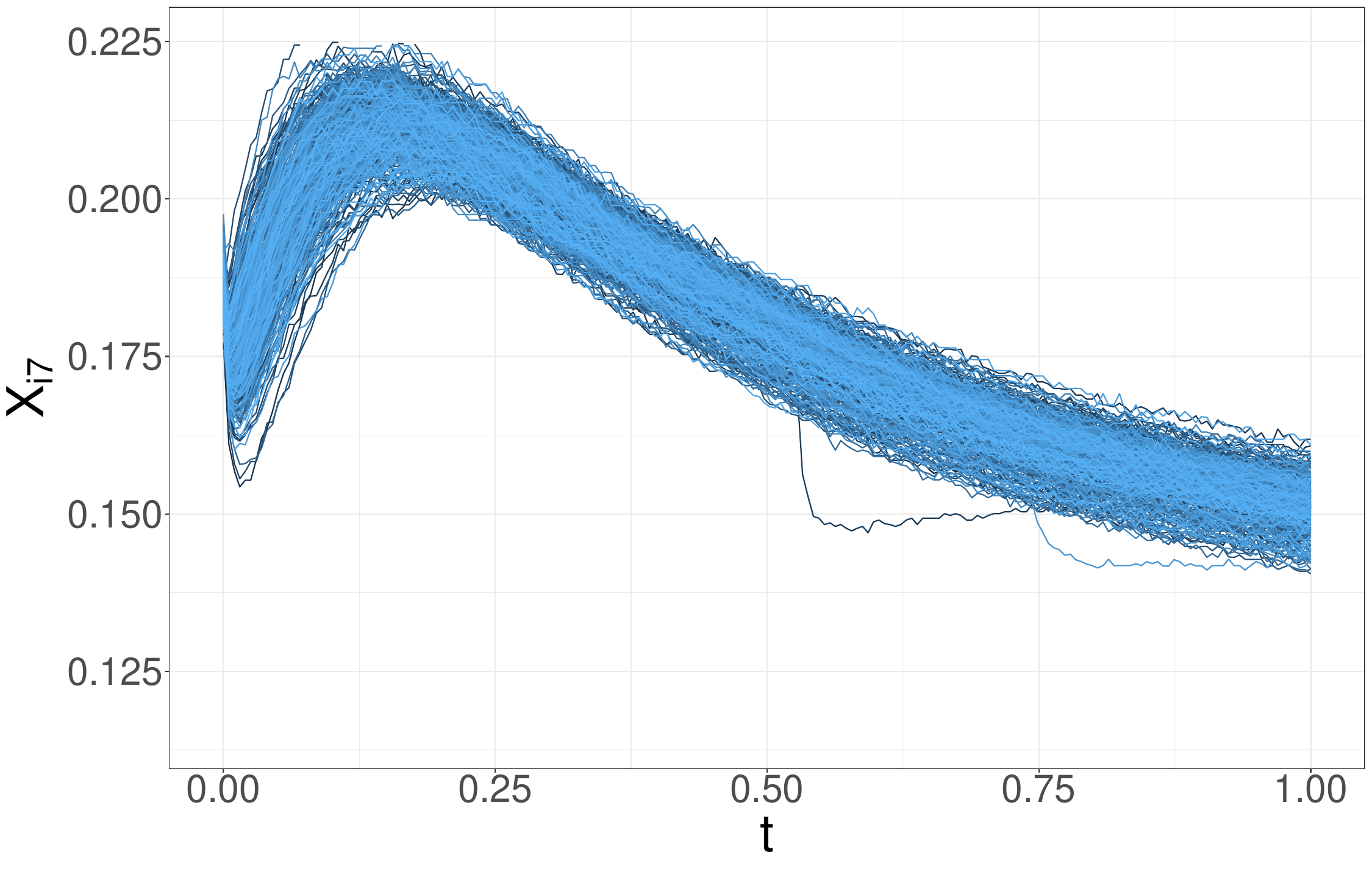}&\includegraphics[width=0.4\textwidth]{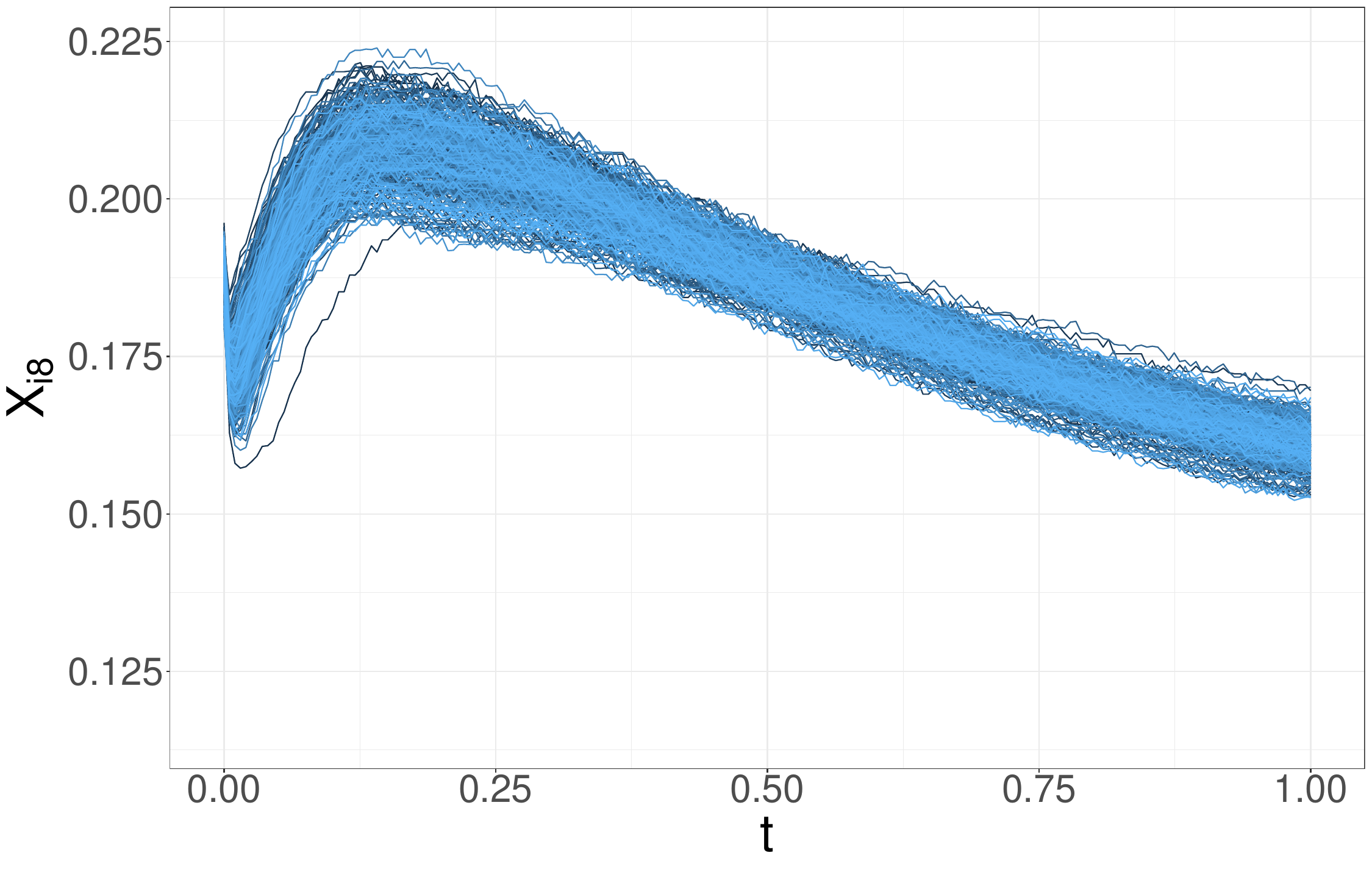}\\\includegraphics[width=0.4\textwidth]{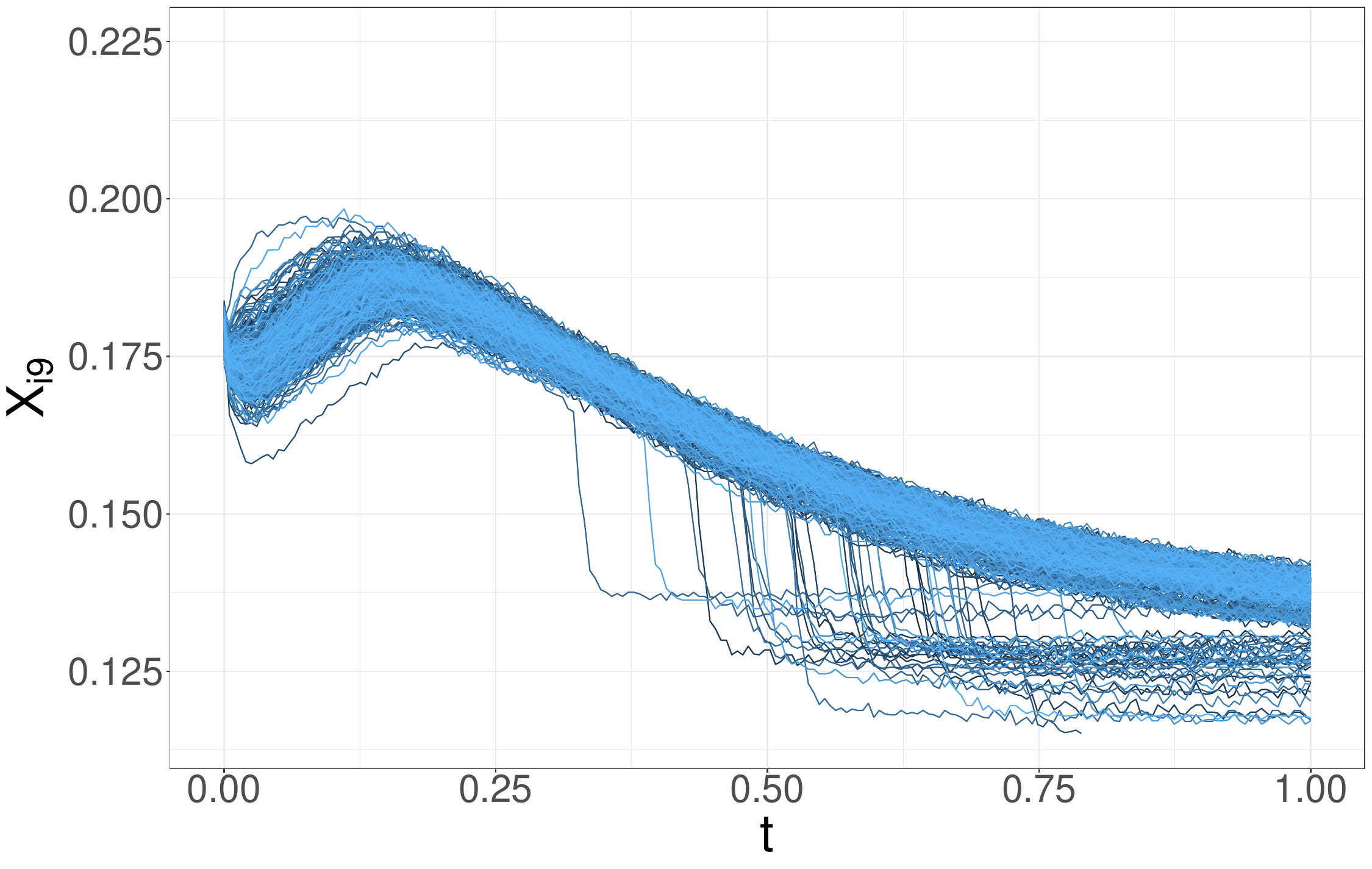}&\includegraphics[width=0.4\textwidth]{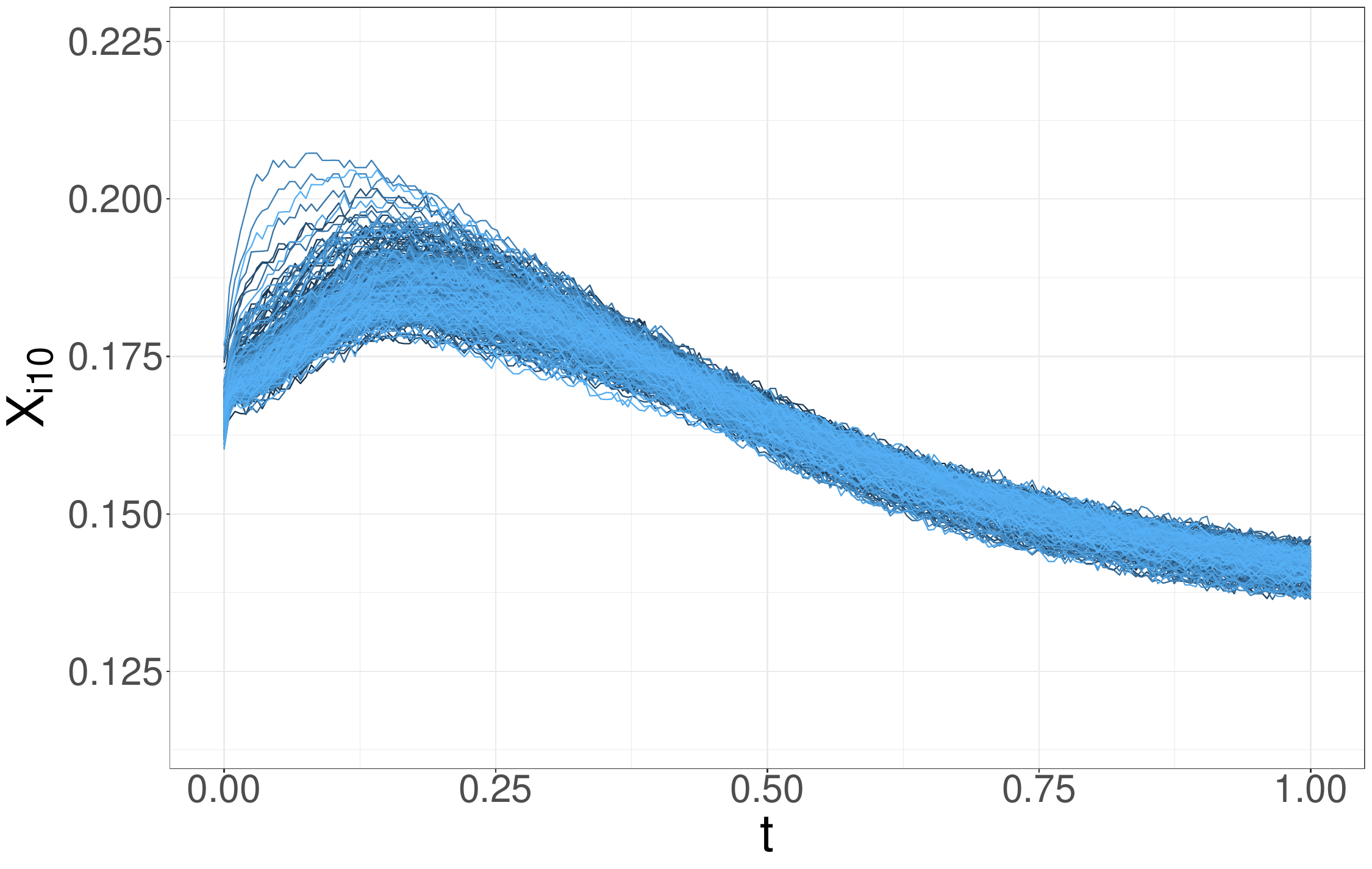}\\
\end{tabular}
	
	\vspace{-.5cm}
\end{figure}

 Figure \ref{fi_csvar} shows the 354 IC observations that form the tuning set of the real-case study examined in Section 4.

\bibliographystyle{chicago}
{\small
\bibliography{References}}
%